\documentclass[fleqn,usenatbib,useAMS]{mnras}

\usepackage{graphicx}	
\usepackage{amsmath}	
\usepackage{multicol}        
\usepackage{bm}		
\usepackage{pdflscape}	

\usepackage[T1]{fontenc}
\usepackage{ae,aecompl}

\usepackage{newtxtext,newtxmath}

\graphicspath{{.}}  
\usepackage{xspace}

\makeatletter
\renewcommand{\ion}[2]{#1\,{\scshape\@roman{#2}}}
\makeatother

\newcommand{\lya}{Ly$\alpha$\xspace}
\newcommand{\ha}{H$\alpha$\xspace}
\newcommand{\hb}{H$\beta$\xspace}
\newcommand{\oiii}{[O~{\sc iii}]\xspace}
\newcommand{\oii}{[O~{\sc ii}]\xspace}
\newcommand{\nii}{[N~{\sc ii}]\xspace}
\newcommand{\heii}{He~{\sc ii}\xspace}
\newcommand{\hei}{He~{\sc i}\xspace}
\ExplSyntaxOn
\prop_new:N \g__affil_text_prop
\prop_new:N \g__affil_num_prop
\seq_new:N  \g__affil_used_seq
\int_new:N  \g__affil_count_int
\seq_new:N  \l__affil_keys_seq
\seq_new:N  \l__affil_out_seq
\cs_generate_variant:Nn \prop_gput:Nnn { Nne }
\cs_generate_variant:Nn \seq_put_right:Nn { Ne }
\msg_new:nnn { autoaffil } { undefined }
  { Affiliation~key~'#1'~is~used~but~was~never~declared~with~\iow_char:N\\defaffil. }
\NewDocumentCommand \defaffil { m +m }
  { \prop_gput:Nnn \g__affil_text_prop {#1} {#2} }
\NewDocumentCommand \affil { m }
  {
    \seq_set_from_clist:Nn \l__affil_keys_seq {#1}
    \seq_clear:N \l__affil_out_seq
    \seq_map_inline:Nn \l__affil_keys_seq
      {
        \prop_if_in:NnF \g__affil_text_prop {##1}
          { \msg_error:nnn { autoaffil } { undefined } {##1} }
        \prop_if_in:NnF \g__affil_num_prop {##1}
          {
            \int_gincr:N \g__affil_count_int
            \prop_gput:Nne \g__affil_num_prop {##1} { \int_use:N \g__affil_count_int }
            \seq_gput_right:Nn \g__affil_used_seq {##1}
          }
        \seq_put_right:Ne \l__affil_out_seq
          { \exp_not:N \hyperlink { affil:##1 } { \prop_item:Nn \g__affil_num_prop {##1} } }
      }
    \textsuperscript { \seq_use:Nn \l__affil_out_seq { , } }
  }
\NewDocumentCommand \printaffils { }
  {
    \seq_map_inline:Nn \g__affil_used_seq
      {
        \hypertarget { affil:##1 } { \textsuperscript { \prop_item:Nn \g__affil_num_prop {##1} } }
        \prop_item:Nn \g__affil_text_prop {##1} \par
      }
  }
\ExplSyntaxOff

\RequirePackage{orcidlink}
\newcommand{\orcidsymb}[2]{\mbox{#1$^{\mbox{\orcidlink{#2}}}$}}

\newcommand{\corrmark}{$^{\star}$}
\newcommand{\corremail}[1]{\par\corrmark E-mail: \href{mailto:#1}{#1}}

\title[LATED Methodology]{LATED: Ly$\alpha$-anchored photometric selection of candidate metal-free and extremely metal-poor star formation from the end of reionisation to cosmic noon}

\author[M. Li et al.]{
\parbox{\textwidth}{\raggedright
\orcidsymb{Mingyu Li}{0000-0001-6251-649X}\affil{THU,KICC,Cavendish}\corrmark,
\orcidsymb{Zheng Cai}{0000-0001-8467-6478}\affil{THU},
\orcidsymb{Roberto Maiolino}{0000-0002-4985-3819}\affil{KICC,Cavendish,UCL},
\orcidsymb{Fuyan Bian}{0000-0002-1620-0897}\affil{ESO,CASSACA},
\orcidsymb{Sijia Cai}{0009-0003-4133-0292}\affil{THU},
\orcidsymb{Francesco D'Eugenio}{0000-0003-2388-8172}\affil{KICC,Cavendish},
\orcidsymb{Qiao Duan}{0009-0009-8105-4564}\affil{KICC,Cavendish},
\orcidsymb{Eiichi Egami}{0000-0003-1344-9475}\affil{Steward},
\orcidsymb{Xiaohui Fan}{0000-0003-3310-0131}\affil{Steward},
\orcidsymb{Yuki Isobe}{0000-0001-7730-8634}\affil{KICC,Cavendish,Waseda},
\orcidsymb{Xihan Ji}{0000-0002-1660-9502}\affil{KICC,Cavendish},
\orcidsymb{Gareth C. Jones}{0000-0002-0267-9024}\affil{KICC,Cavendish},
\orcidsymb{Maria Koller}{0009-0000-1950-9112}\affil{KICC,Cavendish},
\orcidsymb{Xiaojing Lin}{0000-0001-6052-4234}\affil{THU},
\orcidsymb{Boyuan Liu}{0000-0002-4966-7450}\affil{Heidelberg},
\orcidsymb{Christopher C. Lovell}{0000-0001-7964-5933}\affil{KICC,IoA},
\orcidsymb{Kimihiko Nakajima}{0000-0003-2965-5070}\affil{Kanazawa},
\orcidsymb{Masami Ouchi}{0000-0002-1049-6658}\affil{NAOJ,ICRR,SOKENDAI,IPMU},
\orcidsymb{Robert G. Pascalau}{0000-0001-9820-5773}\affil{KICC,Cavendish},
\orcidsymb{Zijin Su}{0009-0004-3303-3754}\affil{THU,UCL},
\orcidsymb{Jan Scholtz}{0000-0001-6010-6809}\affil{KICC,Cavendish},
\orcidsymb{Fengwu Sun}{0000-0002-4622-6617}\affil{Westlake},
\orcidsymb{Sandro Tacchella}{0000-0002-8224-4505}\affil{KICC,Cavendish},
\orcidsymb{Hannah \"Ubler}{0000-0003-4891-0794}\affil{MPE},
\orcidsymb{Yunjing Wu}{0000-0003-0111-8249}\affil{IPMU,CD3},
\orcidsymb{Fujiang Yu}{0000-0002-3489-6381}\affil{THU}, and 
\orcidsymb{Zijian Zhang}{0000-0002-2420-5022}\affil{KIAA,PKU,KICC,Cavendish}
}
\vspace{0.4cm}\\
\parbox{\textwidth}{Affiliations are listed at the end of the paper.\corremail{lmytime@hotmail.com}}
}

\defaffil{THU}{Department of Astronomy, Tsinghua University, Beijing 100084, People’s Republic of China}
\defaffil{KICC}{Kavli Institute for Cosmology, University of Cambridge, Madingley Road, Cambridge, CB3 0HA, UK}
\defaffil{Cavendish}{Cavendish Laboratory - Astrophysics Group, University of Cambridge, 19 JJ Thomson Avenue, Cambridge, CB3 0HE, UK}
\defaffil{UCL}{Department of Physics and Astronomy, University College London, Gower Street, London WC1E 6BT, UK}
\defaffil{ESO}{European Southern Observatory, Alonso de C\'ordova 3107, Casilla 19001, Vitacura, Santiago 19, Chile}
\defaffil{CASSACA}{Chinese Academy of Sciences South America Center for Astronomy, National Astronomical Observatories, CAS, Beijing 100101, People’s Republic of China}
\defaffil{Edinburgh}{Institute for Astronomy, University of Edinburgh, Royal Observatory, Blackford Hill, Edinburgh EH9 3HJ, UK}
\defaffil{NOIRLab}{NSF's National Optical-Infrared Astronomy Research Laboratory, 950 N. Cherry Avenue, Tucson, AZ 85719, USA}
\defaffil{Steward}{Steward Observatory, University of Arizona, 933 North Cherry Avenue, Tucson, AZ 85721, USA}
\defaffil{NRAO}{National Radio Astronomy Observatory, 520 Edgemont Road, Charlottesville, VA 22903, USA}
\defaffil{Waseda}{Waseda Research Institute for Science and Engineering, Faculty of Science and Engineering, Waseda University, 3-4-1, Okubo, Shinjuku, Tokyo 169-8555, Japan}
\defaffil{Heidelberg}{Universit\"at Heidelberg, Zentrum f\"ur Astronomie, Institut f\"ur Theoretische Astrophysik, Albert-Ueberle-Str. 2, D-69120 Heidelberg, Germany}
\defaffil{IoA}{Institute of Astronomy, University of Cambridge, Madingley Road, Cambridge, CB3 0HA, UK}
\defaffil{Kanazawa}{Institute of Liberal Arts and Science, Kanazawa University, Kakuma-machi, Kanazawa, Ishikawa 920-1192, Japan}
\defaffil{NAOJ}{National Astronomical Observatory of Japan, 2-21-1 Osawa, Mitaka, Tokyo 181-8588, Japan}
\defaffil{ICRR}{Institute for Cosmic Ray Research, The University of Tokyo, 5-1-5 Kashiwanoha, Kashiwa, Chiba 277-8582, Japan}
\defaffil{SOKENDAI}{Department of Astronomical Science, SOKENDAI (The Graduate University for Advanced Studies), 2-21-1 Osawa, Mitaka, Tokyo 181-8588, Japan}
\defaffil{IPMU}{Kavli Institute for the Physics and Mathematics of the Universe (WPI), The University of Tokyo Institutes for Advanced Study, The University of Tokyo, Kashiwa, Chiba 277-8583, Japan}
\defaffil{UCSC}{Department of Astronomy and Astrophysics, University of California, Santa Cruz, 1156 High Street, Santa Cruz, CA 95064, USA}
\defaffil{Westlake}{Department of Astronomy, School of Science, Westlake University, Hangzhou, Zhejiang 310030, People’s Republic of China}
\defaffil{CfA}{Center for Astrophysics $|$ Harvard \& Smithsonian, 60 Garden St., Cambridge, MA 02138, USA}
\defaffil{MPE}{Max-Planck-Institut f\"ur extraterrestrische Physik (MPE), Gie{\ss}enbachstra{\ss}e 1, 85748 Garching, Germany}
\defaffil{CD3}{Center for Data-Driven Discovery, Kavli IPMU (WPI), UTIAS, The University of Tokyo, Kashiwa, Chiba 277-8583, Japan}
\defaffil{KIAA}{Kavli Institute for Astronomy and Astrophysics, Peking University, Beijing 100871, People’s Republic of China}
\defaffil{PKU}{Department of Astronomy, School of Physics, Peking University, Beijing 100871, People’s Republic of China}

\date{}
\pubyear{2026}

\begin{document}
\label{firstpage}
\pagerange{\pageref{firstpage}--\pageref{lastpage}}
\maketitle

\begin{abstract}
Cosmological simulations allow a low-level tail of Population~III (Pop~III) star formation to persist to $z=2$--6.
Spectroscopic confirmation is expensive, so an efficient photometric pre-selection is needed.
We present LATED (Lyman-Alpha Tomography of Extremely Metal-poor Domains), which selects metal-free and extremely metal-poorcandidates from a Ly$\alpha$-emitter parent sample using strong-line diagnostics.
The method requires three bands and two colours, $x=m_{\rm OIII}-m_{{\rm H}\alpha}$ and $y=m_{{\rm H}\alpha}-m_{\rm cont}$, which trace oxygen abundance and the H$\alpha$ equivalent width, respectively.
Requiring that the filters simultaneously contain [O III]+H$\beta$ and H$\alpha$, together with a Ly$\alpha$ parent selection, defines five windows spanning $z=1.92$--6.60 (four JWST/NIRCam, one Roman/WFI).
The criteria, $x\geq x_{\rm min}(z)$ and $y\leq y_{\rm max}(z)$, are set by the per-redshift extrema of a forward-modelled Pop~III template locus.
Ordinary metal-enriched star-forming and AGN templates fall outside the selection region, and possible contaminants such as little red dots are flagged by their multi-band colours.
To check contamination empirically, we apply LATED to 1126 JADES spectroscopic galaxies, and no source is selected in the four NIRCam windows.
We release a Python package which converts the same photometry into R3=[O III]/H$\beta$ as a measurement or upper limit, reproducing JADES spectroscopy with small 0.12 dex scatter.
Applied to 85 archival MUSE Ly$\alpha$ emitters in Abell 2744, LATED recovers the confirmed extremely metal-poor galaxy AMORE6 and reveals three new candidates at $z=3$--5.
Photometry alone cannot establish a metal-free nature.
LATED delivers prioritised candidates for spectroscopic follow-up, providing a scalable route toward a systematic census of late-time Pop~III star formation.
\end{abstract}



\section{Introduction}
\label{sec:intro}

Population~III (Pop~III) stars, the first stellar generation to form from primordial chemical elements by Big Bang nucleosynthesis, occupy a foundational place in models of cosmic structure formation \citep[for reviews see][]{Bromm2004ARA&A..42...79B,Klessen2023ARA&A..61...65K,Venditti2026OJAp....967811V}.
They are expected to mostly form in dark-matter minihaloes of $\sim10^{5}$--$10^{6}\,M_{\sun}$ at $z\sim20$--30, where molecular hydrogen is the main available coolant; the resulting absence of metal and dust cooling keeps the gas warm and the Jeans mass high, suppressing fragmentation into small stars and favouring the formation of very massive stars \citep{Abel2002Sci...295...93A,Bromm2002ApJ...564...23B}.
Their initial mass function (IMF) is consequently thought to be top-heavy, reaching tens to hundreds of solar masses, but it remains one of the least constrained quantities in galaxy formation theory.
Stellar archaeology supplies an external prior: the abundance patterns of carbon-enhanced metal-poor halo stars favour enrichment by core-collapse and faint Pop~III supernovae with characteristic progenitor masses of only $\sim$10--20\,$M_{\sun}$ \citep{Ishigaki2018ApJ...857...46I,Koutsouridou2023MNRAS.525..190K}, well below the most extreme theoretical expectations.
As the first sources of ionising photons and of heavy elements, Pop~III stars are expected to initiate cosmic reionisation and the chemical enrichment of the intergalactic medium, while their massive remnants are a leading candidate for the seeds of the first supermassive black holes \citep{Tumlinson2000ApJ...528L..65T,Schneider2002ApJ...571...30S}.
Once star-forming gas is enriched beyond a critical metallicity of order $10^{-3.5}\,Z_{\sun}$, metal-line and dust cooling permit fragmentation to lower masses and star formation transitions to the metal-poor Population~II mode \citep{Bromm2001MNRAS.328..969B,Schneider2002ApJ...571...30S}.
The expected observational signatures of Pop~III star formation follow from the hard ionising spectra of hot, metal-free atmospheres: strong hydrogen recombination lines with very high equivalent widths (EWs), nebular \heii emission requiring photons above 54.4\,eV, a nebular-continuum-dominated spectrum redward of \lya, and, by construction, weak or absent metal lines \citep{Schaerer2002A&A...382...28S,Schaerer2003A&A...397..527S,Raiter2010A&A...523A..64R,Inoue2011MNRAS.415.2920I,Zackrisson2011ApJ...740...13Z, Nakajima2022MNRAS.513.5134N}.

Most observational effort has concentrated on the highest accessible redshifts, where Pop~III star formation is expected to peak.
However, the transition from Pop~III to metal-enriched (Pop~II) star formation probably is not synchronous across the Universe.
Metal enrichment proceeds inhomogeneously: galactic winds pollute the circumgalactic environment of early galaxies efficiently, but underdense regions and low-mass haloes that have not yet experienced star formation can retain chemically pristine gas to much later epochs \citep{Scannapieco2003ApJ...589...35S,Tornatore2007MNRAS.382..945T,Trenti2009ApJ...700.1672T,Pallottini2014MNRAS.440.2498P}.
Some cosmological simulations predict a low-level tail of Pop~III star formation extending to $z\sim2$--6, with rates that depend sensitively on the efficiency of metal mixing, the strength of the ionising background, and the critical metallicity for the Pop~III/II transition \citep{Jaacks2018MNRAS.475.4396J,Sarmento2019ApJ...871..206S,Liu2020MNRAS.497.2839L,Venditti2023MNRAS.522.3809V,Zier2025MNRAS.544..410Z}.
\citet{Tornatore2007MNRAS.382..945T} find Pop~III star formation continuing to $z\approx2.5$ and explicitly encourage `deep searches for pristine star formation at moderate ($2<z<5$) redshifts', while model-to-model termination redshifts span $z\approx5$ to effectively $z=0$ depending on the treatment of metal mixing \citep{Liu2020MNRAS.497.2839L}.
The predicted sites could be characteristically displaced from enriched regions, residing instead at galaxy peripheries, pristine satellites, and isolated haloes \citep{Trenti2009ApJ...700.1672T,Venditti2023MNRAS.522.3809V}.

If this tail exists, the most accessible laboratories for Pop~III-like star formation may lie not at cosmic dawn but from the end of reionisation to cosmic noon, where smaller cosmological dimming makes detection and characterisation much easier without gravitational lensing, and where rest-frame optical diagnostics fall into well-charted near-infrared windows by JWST.

The JWST era has sharpened both the motivation and the difficulty of this search.
Spectroscopy has uncovered candidate pristine and extremely metal-poor systems throughout the reionisation era: a \heii\,$\lambda1640$-emitting clump with no detected metal lines in the halo of GN-z11 at $z=10.6$ \citep{Maiolino2024A&A...687A..67M,Maiolino2026arXiv260320362M,Ubler2026arXiv260320360U}; an extremely metal-poor galaxy at $z=8.271$ from the JWST EXCELS survey \citep{Cullen2025MNRAS.540.2176C}; a gravitationally lensed, ultra-compact, very-low-metallicity \lya emitter behind Abell~370 at $z\approx8$ \citep{Willott2025ApJ...988...26W}; and the lensed, stellar-mass-starved system LAP1 at $z=6.6$, whose component LAP1-B has a measured oxygen abundance of only $0.42\pm0.18$ per cent solar \citep{Vanzella2023A&A...678A.173V,Nakajima2026Natur.653..363N}.
Toward lower redshift, the same signatures appear in \hb-detected, \oiii-undetected candidates such as AMORE6 at $z=5.7$ \citep{Morishita2025arXiv250710521M} and in a very low metallicity, star-forming complex at $z=4.19$ \citep{Vanzella2026A&A...705L..12V}.
Within the strongly lensed \lya emitter A370-z6LAE-1 at $z\approx5.9$ \citep{Claeyssens2022A&A...666A..78C}, a compact, \oiii-weak clump only $\sim$1\,kpc from enriched neighbours has been proposed as a metal-poor pocket or infalling satellite \citep{Fujimoto2025arXiv251211790F}, directly exhibiting the pristine-beside-enriched geometry the simulations predict.
The host galaxy of a strongly lensed, metal-poor Type II supernova at $z = 5.13$ is suggested to be extremely metal-poor \citep{Asada2026arXiv260714355A}.
Crucially, at cosmic noon, the \lya emitter MPG-CR3 at $z=3.19$ has rest-frame EW(\lya)\,$=822\pm101$\,\AA{}, EW(\ha)\,$=2814\pm327$\,\AA{}, and no detected \oiii, suggesting $12+\log({\rm O/H})<6.52$ \citep{Cai2025ApJ...993L..52C}, and it is joined by a second extremely metal-poor galaxy at $z=3.65$ \citep{Cai2026RAA....26k5005C}.
Some of the most metal-poor systems found by JWST at high and intermediate redshifts are little red dots (LRDs) or other classes of AGN \citep{Maiolino2026QSO1met,Ivey2026Cliff,Tripodi2025,Caputi2026pseudoLRD}, and others were found serendipitously as satellites of more massive systems \citep[e.g.][]{Koller2026MNRAS.551g1206K}.
At the same time, large JWST/NIRSpec and NIRCam WFSS surveys show that the metallicity floor of confirmed galaxy samples remains near $12+\log({\rm O/H})\approx7$ \citep{Nakajima2023ApJS..269...33N,Curti2024A&A...684A..75C,Hsiao2025arXiv250503873H}, i.e. a few per cent solar: spectroscopically confirmed, unambiguously metal-\emph{free} star formation remains elusive.

History also counsels caution about photometric claims.
The pattern predates CR7: unusually strong \lya \citep{Malhotra2002ApJ...565L..71M,Yamada2005PASJ...57..881Y} and very blue $z\sim7$ UV slopes \citep{Bouwens2010ApJ...708L..69B} were each initially regarded as possible metal-free signatures and later explained by young but enriched populations with nebular effects.
CR7 was advanced as a Pop~III candidate on the basis of strong \lya, tentative \heii, and an IRAC excess \citep{Sobral2015ApJ...808..139S}, then demoted when deeper data revealed the excess to be strong \oiii$+$\hb from metal-enriched gas \citep{Bowler2017MNRAS.469..448B,Sobral2019MNRAS.482.2422S}, an interpretation since confirmed by JWST, which resolves CR7 into a metal-enriched, multi-component merging system \citep{Kiyota2025ApJ...995..150K,Marconcini2025A&A...699A.154M}.
The same failure mode recurred in the JWST era: the NIRCam-selected Pop~III candidate GLIMPSE-16043 \citep{Fujimoto2025ApJ...989...46F} was refuted when NIRSpec detected \oiii with \oiii$\lambda5007$/\hb\,$=1.78\pm0.18$ at $z=6.2$ \citep{Fujimoto2025arXiv251211790F}.
In both cases the failure lay on the same axis: the inferred absence of \oiii proved wrong. 
Therefore, any credible photometric search should treat \oiii weakness as a hypothesis to be tested spectroscopically, never as a result.
This is now the stated consensus of the field: the central question is no longer only whether Pop~III stars can be detected, but how their signatures can be separated robustly from contaminants and from coeval metal-enriched star formation \citep{Venditti2026OJAp....967811V,Rusta2025ApJ...989L..32R}.

The central practical problem is scale.
Blind spectroscopic discovery of extremely metal-poor systems is expensive: \heii confirmation can require tens to hundreds of NIRSpec hours per target at high redshift \citep{Trussler2023MNRAS.525.5328T,Maiolino2026arXiv260320362M}, multi-object campaigns could geometrically miss most peripheral Pop~III clumps \citep{Venditti2024ApJ...973L..12V}, and predicted number densities are too low for blind lensed spectroscopy under most IMF assumptions \citep{Vikaeus2022MNRAS.512.3030V}.
What is needed is a photometric pre-selection framework that (i) starts from a population already known to be young, actively ionising, and relatively dust-free; and (ii) applies metallicity-sensitive photometric diagnostics over wide areas using existing and planned imaging.

Photometric pre-selection of metal-poor galaxies has been established at low redshift.
The EMPRESS survey selects candidates below $0.1\,Z_{\sun}$ from $\sim$500\,deg$^2$ of Subaru/Hyper Suprime-Cam imaging with a machine-learning classifier trained on model photometry, whose completeness (86 per cent) and purity (46 per cent) are measured against SDSS spectroscopic metallicities \citep{Kojima2020ApJ...898..142K}.
\citet{Nishigaki2023ApJ...952...11N} extend the search to $z\sim4$--5 with broadband colour excesses in deep JWST/NIRCam imaging, finding 17 candidates, some with \oiii weak enough to suggest $Z\approx0$--$0.01\,Z_{\sun}$.
At $z\simeq6$--7, \citet{Fujimoto2025ApJ...989...46F} design a NIRCam selection for Pop~III galaxies, validate it with completeness and contamination simulations, and apply it to $\simeq$500\,arcmin$^2$ of JWST legacy fields, deriving the first observational constraints on the Pop~III UV luminosity function.
After NIRSpec refuted GLIMPSE-16043, \citet{Fujimoto2025arXiv251211790F} revise the criteria to exclude extreme Balmer-jump objects and, adding five lensing-cluster fields, estimate a Pop~III star-formation-rate density of $\approx10^{-6}$--$10^{-4}\,M_{\sun}\,{\rm yr^{-1}\,cMpc^{-3}}$ at $z=6$--7.
\citet{Trussler2026MNRAS.550g1360T} further demonstrate the power of JWST NIRCam medium-band imaging in identifying \ha-strong, \oiii-weak extremely metal-poor candidates over a wider redshift range ($2.5<z<6.5$) from JADES fields.

A complementary route, which we develop here, anchors the selection to \lya emission, adding the redshift slice, a youth and low-dust prior, and, through wide-area surveys, survey-scale statistics.
\lya emitters (LAEs) are a natural starting point.
Strong \lya emission preferentially selects young, low-mass, low-dust, actively star-forming systems \citep[see review in][]{Ouchi2020ARA&A..58..617O}, and decades of narrow-band (NB), intermediate-band (IB), and integral-field-unit (IFU) surveys have produced large LAE samples with well-understood selection functions at $2\lesssim z\lesssim7$ \citep[e.g.][]{Rhoads2000ApJ...545L..85R,Gronwall2007ApJ...667...79G,Ouchi2008ApJS..176..301O,Sobral2018MNRAS.476.4725S,Urrutia2019A&A...624A.141U,Bacon2023A&A...670A...4B,Gebhardt2021ApJ...923..217G}.
Theoretical work predicts that Pop~III star formation enhances \lya emission even beyond Case~B expectations \citep{Raiter2010A&A...523A..64R}, and that surviving pockets of pristine gas may manifest as \lya-bright clumps in galaxy outskirts \citep{Mas-Ribas2016ApJ...833...65M}.
However, \lya alone cannot distinguish a chemically primitive system from an ordinary young metal-enriched starburst: most LAEs are metal-enriched \citep{Nakajima2013ApJ...769....3N,Trainor2016ApJ...832..171T,Maseda2023ApJ...956...11M}.
The discriminating information lies in the rest-frame optical, where the ratio of hydrogen recombination lines (especially \ha and \hb) to the strongest metal lines (especially \oiii\,$\lambda\lambda4959,5007$) varies by orders of magnitude between metal-free and metal-enriched gas \citep{Inoue2011MNRAS.415.2920I,Nakajima2022MNRAS.513.5134N}.

After reionisation these rest-frame optical lines fall into near-infrared bands now routinely imaged by JWST/NIRCam.
A galaxy with strong \ha boosts the band containing \ha relative to neighbouring continuum bands; a galaxy with strong \oiii$+$\hb boosts the corresponding bluer band.
The photometric contrast between the two excesses is a metallicity-sensitive diagnostic that can be computed for every LAE with adequate imaging, at no additional observational cost where archival data exist.

We therefore introduce LATED (Lyman-Alpha Tomography of Extremely Metal-poor Domains), a framework that combines \lya-emitter parent selection with rest-frame optical photometric line diagnostics to isolate \ha-strong, \oiii-weak/absent candidate Pop~III sources for spectroscopic follow-up.
The usable redshift range is set by filter feasibility: it extends from $z\approx1.9$ where \lya enters the ground-based near-ultraviolet window, to $z\approx6.6$ where \ha leaves the reddest JWST/NIRCam band.

This paper, the first in the series, defines the LATED methodology in general form.
Section~\ref{sec:concept} sets out the survey concept and the two pillars (tomographic \lya parent selection and extremely metal-poor domain identification).
Section~\ref{sec:algorithm} presents the general selection algorithm: the rest-frame optical photometric diagnostics and the EW(\ha)--\oiii/\hb (We refer to this as EW(\ha)--O/H in this paper.) diagram at the method's core, together with the selection recipe itself (Section~\ref{sec:diagnostics}); the construction of the redshift windows in which the required bands exist, including the parent \lya selection and the continuum-band rule (Section~\ref{sec:windows}); the Monte-Carlo mock that realises each window and fixes its selection criteria from the forward-model library (Section~\ref{sec:models}); and the mock and model uncertainties, recovery, and leakage (Section~\ref{sec:mockunc}).
Section~\ref{sec:performance} quantifies detectability, required depth, and the contaminants and failure modes, and compares the selection with published photometric searches.
Section~\ref{sec:recovery} presents the photometric line-ratio inference and the candidate ranking it supports.
Section~\ref{sec:a2744} demonstrates the complete workflow on archival MUSE and JWST data in the Abell~2744 field.
Section~\ref{sec:future} discusses the selection-completeness budget a survey should supply, the statistical products a candidate sample supports, a call for community surveys, and the spectroscopic path to confirmation; Section~\ref{sec:summary} summarises.
Throughout this paper, we use AB magnitudes and quote rest-frame equivalent widths (EWs) unless otherwise stated.
Where needed, we adopt a flat $\Lambda$CDM cosmology with $H_0=70\,{\rm km\,s^{-1}\,Mpc^{-1}}$ and $\Omega_{\rm m}=0.3$.
We use \oiii to refer to \oiii\,$\lambda5007$ unless stated otherwise.

\begin{figure*}
\includegraphics[width=\linewidth]{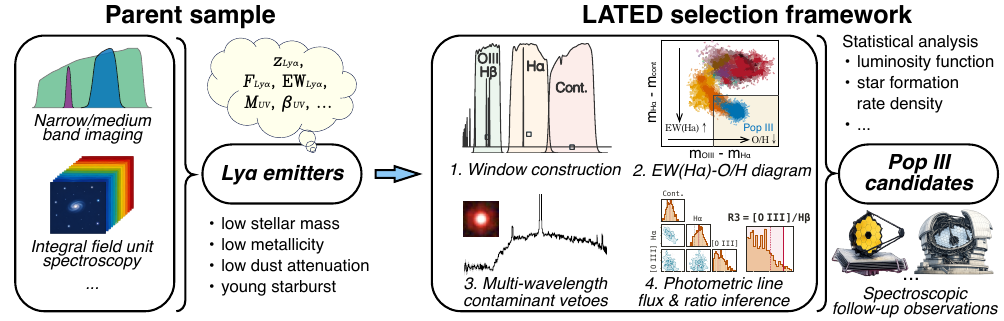}
\caption{The LATED selection workflow.
\emph{Left:} the parent sample is a tomographic \lya-emitter selection, built from narrow/medium-band imaging, integral-field spectroscopy, or other routes (Section~\ref{sec:parent}); each emitter retains its measured \lya properties ($z_{{\rm Ly}\alpha}$, $F_{{\rm Ly}\alpha}$, EW$_{{\rm Ly}\alpha}$, $M_{\rm UV}$, $\beta_{\rm UV}$), and the population is already biased toward low stellar mass, low metallicity, low dust attenuation, and young starbursts (Section~\ref{sec:emd}).
\emph{Right:} the LATED framework then applies four stages: (1) construction of the redshift windows in which the filter set holds the \oiii$+$\hb, \ha, and continuum bands (Section~\ref{sec:windows}); (2) the EW(\ha)--O/H diagram, $y=m_{\rm H\alpha}-m_{\rm cont}$ versus $x=m_{\rm OIII}-m_{\rm H\alpha}$, with redshift-dependent criteria derived from the model grid (Sections~\ref{sec:diagnostics}); (3) multi-wavelength contaminant vetoes, including the LRD flag; and (4) Line flux and ratio from photometry, realised in this paper as the photometric emission-line inference Python package \textsc{lated} (Section~\ref{sec:recovery}).
The output is a ranked Pop~III candidate sample feeding spectroscopic follow-up and per-slice statistical analyses such as luminosity functions and star-formation-rate densities.}
\label{fig:workflow}
\end{figure*}

\section{The LATED concept}
\label{sec:concept}

LATED stands for \emph{Lyman-Alpha Tomography of Extremely Metal-poor Domains}.
The name encodes the two pillars of the method: a redshift-sliced (\emph{tomographic}) \lya emitter selection, followed by photometric identification of \emph{extremely metal-poor} and even metal-free candidates within each slice.
The acronym is also meant to be read as a word, because the search is a belated one in two distinct senses.
The first is astrophysical: the star formation we target is itself belated, occurring in pockets of gas that stay pristine long after the first stars, so we look at $z\lesssim7$ and down to $z\approx2$ rather than at the $z>10$ frontier where Pop~III searches are usually pitched.
The second is observational: the parent samples this method needs were assembled years ago, so a fraction of the candidates has been sitting in published LAE catalogues for a decade or more, waiting for the rest-frame optical bands that only JWST and \textit{Roman} supply to turn them into photometric selections and, from there, into spectroscopic targets.
Figure~\ref{fig:workflow} summarises the workflow: a \lya-selected parent sample, carrying its measured \lya properties, enters a four-stage selection framework (window construction, the EW(\ha)--O/H diagram, contaminant vetoes, and photometric emission-line inference) that returns a ranked Pop~III candidate sample for spectroscopic follow-up and per-slice statistics.

\subsection{Tomography of Ly\texorpdfstring{$\alpha$}{a} emitters}
\label{sec:tomography}

LAE selection is intrinsically tomographic.
A narrow- or intermediate-band filter isolates \lya within a thin redshift slice ($\Delta z\approx0.05$--0.3 depending on filter width); a battery of adjacent filters, such as the Subaru/Suprime-Cam IA set in COSMOS \citep{Taniguchi2007ApJS..172....9T,Taniguchi2015PASJ...67..104T,Sobral2018MNRAS.476.4725S} or the DECam/IBIS medium bands \citep{Ebina2026JCAP...03..019E}, maps the LAE population in contiguous slices across cosmological redshift.
IFU surveys such as MUSE provide continuous \lya tomography over their full bandpass \citep[e.g.,][]{Urrutia2019A&A...624A.141U,Bacon2023A&A...670A...4B}, and wide-field spectroscopic experiments such as HETDEX extend this to hundreds of square degrees \citep{Gebhardt2021ApJ...923..217G}.
Each \lya slice can then be cross-matched against rest-frame optical imaging whose filters sample \ha and \oiii$+$\hb at the corresponding redshift.
The tomographic structure matters for two reasons.
First, the known, narrow redshift of each slice removes the dominant degeneracy in photometric redshift and line diagnostics, which line is in which filter, so that band excesses can be converted into specific line EWs.
Second, slicing enables differential statistics: number densities and luminosity functions of extremely metal-poor candidates can be measured per slice and compared across redshift.

\subsection{Extremely metal-poor domains}
\label{sec:emd}

We deliberately use `extremely metal-poor' rather than `Pop~III'.
As our results below show, photometry cannot establish that a stellar population is metal-free; it can only establish that the nebular metal lines are weak relative to the hydrogen lines, within the uncertainties of the colour measurements and models.
The populations selected by LATED therefore span a continuum: genuinely primordial (Pop~III) star formation, if it exists at these redshifts; extremely metal-poor (`Pop~III-like') galaxies (EMPGs) below $\sim0.01\,Z_{\sun}$; and chemically young systems with a few times $10^{-2}\,Z_{\sun}$ undergoing their first major star-formation and enrichment episodes.
We refer to the selected objects throughout as \emph{candidate} extremely metal-poor, Pop~III candidates, or Pop~III-like systems, and reserve stronger language for spectroscopic confirmation.

We also use the term `domains' for the selected entities.
The method is sensitive to any region dominated by \lya and \ha emission within the matching aperture: entire compact galaxies, star-forming clumps within larger galaxies, satellite or outskirt pockets of chemically primitive gas \citep{Mas-Ribas2016ApJ...833...65M,Vanzella2023A&A...678A.173V}, or extended nebular regions.


\section{The general LATED selection algorithm}
\label{sec:algorithm}

We assemble the framework into a practical recipe.
This section presents, in order: \lya emitters as parent sample (Section~\ref{sec:laeparent}); the selection algorithm and the rest-frame optical photometric diagnostics of extremely metal-poor and metal-free star formation that motivate it, with the EW-O/H diagram at its core (Section~\ref{sec:diagnostics}); the construction of the redshift windows in which the bands the algorithm needs exist (Section~\ref{sec:windows}); the selection criteria, built on the forward-model library (Section~\ref{sec:models}); and the mock and model uncertainties (Section~\ref{sec:mockunc}).

\subsection{Ly\texorpdfstring{$\alpha$}{a} emitters as parent sample}
\label{sec:laeparent}\label{sec:parent}

We begin from an LAE parent sample, and the choice is deliberate rather than one of convenience.
\lya is the strongest intrinsic feature of young, actively star-forming galaxies, and both its production and its escape favour precisely the systems we target: a large \lya equivalent width demands a hard, copious ionising continuum with little dust, the hallmark of young, low-metallicity, high-specific-SFR populations \citep{Ouchi2020ARA&A..58..617O}.
Metal-free and extremely metal-poor populations are predicted to be the most extreme \lya emitters of all, because their hot, massive stars and, for genuinely metal-free gas, the absence of metal cooling drive the intrinsic rest-frame EW(\lya) to many hundreds and up to $\sim$1500--3000\,\AA{}, far above the few hundred \AA{} of normal star-forming galaxies \citep{Schaerer2002A&A...382...28S,Raiter2010A&A...523A..64R,Inoue2011MNRAS.415.2920I}.
To first order, therefore, the Pop~III-like population LATED seeks is a \emph{subset} of the LAE population.

Two practical advantages follow from starting here.
First, a \lya detection delivers a redshift for every source, from the narrow- or medium-band slice that isolates the line or directly from IFU spectroscopy, which pins which broadband filters contain \ha and \oiii$+$\hb and turns band excesses into specific line measurements; without this redshift the photometric line diagnostics would be badly degenerate, and we quantify in Section~\ref{sec:contaminants} how much of that degeneracy survives when the redshift is only photometric.
Second, \lya is detectable to faint limits over very wide fields, from the ground in optical narrow- and medium-band imaging and, increasingly, over large volumes with IFU and wide-field spectroscopy, so that LAE catalogues already reach thousands to millions of sources and will keep growing over the coming decade.
The imaging side is scaling now: ODIN is tiling $\sim$100\,deg$^2$ in three custom narrow bands at $z=2.4$, $3.1$ and $4.5$ \citep{Lee2024ApJ...962...36L}, IBIS is tiling $z=2.26$--3.41 gaplessly in five medium bands for DESI Run 2 \citep{Ebina2026JCAP...03..019E}, and the HSC-Niji medium bands extend gapless coverage to $z\approx7$.
The spectroscopic side follows in the next decade with the Stage-V facilities Spec-S5, MUST and WST \citep{Besuner2025arXiv250307923B,Zhao2024arXiv241107970Z,Mainieri2024arXiv240305398M}, and with BlueMUSE opening the $z<2.9$ range that MUSE cannot reach.
Space observatories include the CSST wide survey, whose $v$ and near-UV bands select bright \lya below $z\approx2.3$ over its full footprint \citep{Luo2026arXiv260919105L}.
That scale is what makes a statistical census of an intrinsically rare population feasible.

The choice still carries costs.
\lya is a resonant line, so its escape is set by radiative transfer through neutral hydrogen and dust and is intrinsically stochastic: a young metal-poor galaxy can be a weak or absent LAE along a given sightline when its H\,\textsc{i} column or dust content is high, so an LAE-parent selection can be incomplete and biased toward low-column, low-dust sightlines \citep{Verhamme2006A&A...460..397V,Hayes2015PASA...32...27H,Dijkstra2014PASA...31...40D,Ouchi2020ARA&A..58..617O}.
The same physics worsens with redshift, as the increasingly neutral intergalactic medium of the reionisation era scatters \lya out of the line of sight and depresses the observed \lya fraction beyond $z\approx6$ \citep{Stark2013ApJ...763..129S,Ouchi2020ARA&A..58..617O}, precisely the epoch in which Pop~III is most expected, so the highest-redshift windows are the most affected.
\lya can also be powered by an AGN rather than by young stars, an ambiguity resolved downstream by the rest-optical diagnostics and the AGN vetoes (Section~\ref{sec:contaminants}).
Most fundamentally, the overwhelming majority of LAEs are chemically evolved star-forming galaxies rather than metal-free systems, so LAE selection is a \emph{necessary but not sufficient} condition: it defines the parent sample, and it is the rest-optical EW(\ha)--O/H step (Section~\ref{sec:diagnostics}) that isolates the extremely metal-poor candidates from within it.
We adopt the LAE parent because its incompleteness acts along quantifiable axes, \lya escape and IGM transmission, which can be folded into the selection function, while the interlopers it admits are removed by the photometric diagnostics that follow.

LATED is independent of the method used to construct the parent LAE sample.
The requirements are only that (i) the \lya selection function is characterised, (ii) the redshift of each source is constrained to its \lya slice, and (iii) \lya fluxes and EWs (or limits) are available for scoring.
The standard LAE selection routes are well studied and their methodologies are established in the literature, so we do not re-derive them here.
Narrow- and intermediate-band colour-excess selection, with its excess-significance and EW thresholds, is set out in \citet{Bunker1995MNRAS.273..513B}, \citet{Rhoads2000ApJ...545L..85R}, \citet{Gronwall2007ApJ...667...79G}, \citet{Ouchi2008ApJS..176..301O}, \citet{Sobral2013MNRAS.428.1128S}, \citet{Konno2016ApJ...823...20K}, \citet{Sobral2018MNRAS.476.4725S} and \citet{Shibuya2018PASJ...70S..14S}; IFU and wide-field spectroscopic detection, where the line is measured directly and the selection function is set by line flux and profile rather than by a colour excess, is set out in \citet{Wisotzki2016A&A...587A..98W}, \citet{Herenz2017A&A...606A..12H}, \citet{Urrutia2019A&A...624A.141U} and \citet{Gebhardt2021ApJ...923..217G}.

\subsection{Rest-frame optical photometric diagnostics}
\label{sec:diagnostics}

Here, we investigate the diagnostics of extremely metal-poor and metal-free star formation, the observables needed, and the recipe that turns them into a candidate sample.

\subsubsection{Key signature of Pop~III-like star formation}
\label{sec:linefeature}

We surveyed the full library of metal-free and extremely-metal-poor populations assembled for this work (e.g., \citealt{Schaerer2003A&A...397..527S,Raiter2010A&A...523A..64R,Zackrisson2011ApJ...740...13Z, Inoue2011MNRAS.415.2920I,  Nakajima2022MNRAS.513.5134N}; see Section~\ref{sec:models}), and one conclusion is robust across all of them: a Pop~III-like star formation is selected by a strong \ha together with a weak or absent \oiii.
The reason these two lines suffice is that they are the dominant features of the rest-optical spectrum.
\ha and the \oiii\,$\lambda\lambda4959,5007+$\hb complex are the strongest rest-optical emission lines; the other optical lines (\oii, \nii, [S~{\sc ii}], [Ne~{\sc iii}], \hei) are intrinsically faint, and the metal lines among them weaken further as metallicity drops, so the \ha band and the \oiii$+$\hb band already carry essentially all of the line information a chemically primitive system offers to broadband photometry.

The first signature is a very high EW(\ha).
Every metal-free family predicts rest-frame EW(\ha)\,$\gtrsim1900$\,\AA{} for ongoing ($\leq10$\,Myr) Pop~III star formation, rising to $\gtrsim3200$\,\AA{} in the most extreme cases, and young extremely metal-poor populations reach comparable values.
Ordinary star-forming galaxies at these redshifts have EW(\ha)\,$\sim100$--500\,\AA{} \citep[e.g.][]{Sobral2013MNRAS.428.1128S}.
The \ha excess therefore separates very young, chemically primitive systems from the bulk population, though not by itself from extreme emission-line galaxies (EELGs), which the second diagnostic will separates.
At very low metallicity the \nii and [S~{\sc ii}] contamination of the \ha band is negligible \citep[e.g.,][]{Sanders2024ApJ...962...24S,Isobe2026arXiv260611345I}.

The second signature is weak or absent \oiii, which inverts the usual emission-line selection logic.
\oiii\,$\lambda5007$ is normally the strongest rest-optical line in star-forming galaxies at these redshifts, and `normal' low-metallicity galaxies ($Z\sim0.1\,Z_{\sun}$) are typically \emph{strong} \oiii emitters, because the \oiii/\hb ratio peaks at such metallicity, where the gas is metal-poor but hot enough to excite \oiii strongly \citep[e.g.][]{Nakajima2022ApJS..262....3N,Nakajima2022MNRAS.513.5134N}.
Below $12+\log({\rm O/H})\approx7$, \oiii collapses under typical conditions.
For example, the record-holding local extremely metal-poor galaxy J0811$+$4730 ($12+\log({\rm O/H})=6.98$) has \oiii\,$\lambda5007$/\hb\,$=1.61$ \citep{Izotov2018MNRAS.473.1956I}.
In extremely metal-poor gas ($Z\sim10^{-5}$--$10^{-4}$, about 0.07--0.7 per cent solar), the \oiii/\hb ratio spans 0.1--1 across all conditions \citep{Nakajima2022MNRAS.513.5134N}.
And truly primordial gas would show \oiii/\hb\,$<0.1$, identified by \citet{Inoue2011MNRAS.415.2920I} as the most robust single criterion for $Z<10^{-3}\,Z_{\sun}$.
A genuine metal-free population is therefore \ha-strong and \oiii-weak/absent \emph{together}.
This is the core of LATED selection.

In photometry, the \oiii$+$\hb band excess of a Pop~III-like system is small but \emph{not} zero, because \hb itself is strong (e.g., rest EW $\sim300$--540\,\AA{} in the \citealt{Inoue2011MNRAS.415.2920I} models), and the nebular continuum contributes additional flux redward of the Balmer jump.
We note that weak \oiii must therefore be interpreted jointly with the \ha excess, the \lya line detection, the UV continuum shape, and the contaminant vetoes, never in isolation.
In particular, metal-\emph{rich} galaxies may also show weak \oiii in low excitation conditions, and the photometric degeneracy between the metal-poor and metal-rich ends of the \oiii/\hb relation is broken in LATED by the \lya pre-selection, the blue UV continuum requirement, and the very high EW(\ha) threshold, which metal-rich populations do not reach (Section~\ref{sec:contrast}).

Photometrically, a strong emission line inside a filter raises the measured mean flux density above the continuum level.
Writing $\langle f_\nu^{\rm LB}\rangle$ for the observed mean flux density in the line band and $\langle f_\nu^{\rm cont}\rangle$ for the continuum flux density at the same wavelength, estimated by interpolation between adjacent line-free bands, the line flux is, to first order,
\begin{equation}
F_{\rm line} \simeq \Delta\lambda_{\rm eff}\,
\left(\langle f_\nu^{\rm LB}\rangle-\langle f_\nu^{\rm cont}\rangle\right)\frac{c}{\lambda_{\rm eff}^{2}},
\label{eq:lineflux}
\end{equation}
where $\lambda_{\rm eff}$ is the pivot wavelength of the line band and $\Delta\lambda_{\rm eff}=\int T(\lambda)\,{\rm d}\lambda / T(\lambda_{\rm line})$ is its effective width evaluated at the line wavelength; the equivalent $f_\lambda$ expression follows directly.
The observed-frame EW is ${\rm EW}_{\rm obs}=F_{\rm line}/f_\lambda^{\rm cont}$ and the rest-frame EW is ${\rm EW}_0={\rm EW}_{\rm obs}/(1+z)$.
In magnitudes, the band excess is
\begin{equation}
\Delta m \;=\; m_{\rm cont}-m_{\rm LB} \;=\; 2.5\log_{10}\!\left(1+\frac{{\rm EW}_{\rm obs}}{\Delta\lambda_{\rm eff}}\right),
\label{eq:colourew}
\end{equation}
so that a rest-frame ${\rm EW}_0({\rm H}\alpha)\approx2000$\,\AA{} at $z\approx3$ produces an excess of order one magnitude in the NIRCam wide band F277W.
Equations~(\ref{eq:lineflux})--(\ref{eq:colourew}) are adequate for survey design and for intuition; production measurements must propagate the source spectrum through the full filter transmission curves, as we do for all numerical results in this paper, because line position within the band, filter shape, and secondary lines each shift the conversion at the tens-of-per-cent level.

\subsubsection{Observables and selection algorithm}
\label{sec:observables}

LATED is built on a compact set of observables.
The parent measurement is the \lya flux and equivalent width, from NB/MB imaging surveys or IFU spectroscopic surveys, which also fixes the Ly$\alpha$ slice redshift.
To it we add the two rest-optical line excesses, each the magnitude boost of a line-bearing band over the interpolated continuum: the \ha excess, convertible into an inferred EW(\ha), and the \oiii$+$\hb excess or its absence, measured the same way.
Their difference, the \ha--\oiii contrast, traces the \ha/\oiii flux ratio and hence the oxygen abundance, that is, the tracer of metallicity.
Continuum constraints (the UV slope, the continuum luminosity, and any accessible hydrogen Balmer or Paschen jumps) and, where available, auxiliary diagnostics (high-ionisation recombination lines from primordial gas, such as \heii\,$\lambda1640$ and \heii\,$\lambda4686$ in existing spectra, source morphology, and multi-wavelength AGN and dust vetoes) complete the set.

The selection algorithm assembles these observables into an operational recipe.
We begin from the LAE catalogue, constructed or imported with its selection function, \lya fluxes and EWs, and slice redshifts (Section~\ref{sec:parent}), and apply quality masks for bright-star haloes, detector artefacts, image edges, and known instrumental features in both the \lya-selection and the rest-frame optical imaging.
Each LAE is cross-matched to the multi-wavelength photometric catalogues, recording matching radii and flagging multiple matches for de-blending, and is required to have a robust ($\mathrm{S/N}\geq5$) detection in the \ha-band.
The continuum band itself may be \emph{undetected}, since a faint continuum is the signature of the highest-EW targets \citep[e.g.][]{Nakajima2026Natur.653..363N, Maiolino2026arXiv260320362M,Ubler2026arXiv260320360U}: it is then carried as an upper limit and the \ha excess becomes a conservative \emph{lower} limit, whereas an additional cut on the continuum S/N would discard exactly those highest-EW objects.

From the detected bands we estimate the line excesses by local continuum interpolation across the line-free bands, and apply the window's selection criteria (Section~\ref{sec:models}).
Surviving sources are screened for contaminants: low-$z$ emission-line galaxies whose spectroscopic redshift is inconsistent with the \lya slice, AGN (from X-ray, radio, mid-infrared, or variability indicators), stars and brown dwarfs (proper motion, morphology, and colours), dusty starbursts (red continuum slopes or far-infrared detections), blends (high-resolution morphology), and photometric artefacts (visual inspection).
The remaining candidates are ranked by their inferred \oiii/\hb, measurement or $2\sigma$ bound (Section~\ref{sec:ranking}), and prioritised for spectroscopic follow-up by rank and brightness (Section~\ref{sec:followup}).

At $z\lesssim6$, the key rest-frame optical lines move through the near-infrared filter sets of JWST/NIRCam; at $z\approx2$ they can move through \textit{Roman}/WFI and ground-based imagers.
For a source whose redshift is pinned by its \lya slice, the filters containing \ha, \oiii$+$\hb, and line-free continuum are known a priori, and band excesses become line measurements.

\subsubsection[The EW(\texorpdfstring{\ha}{Ha})--O/H diagram]{The EW(\texorpdfstring{\ha}{Ha})--\texorpdfstring{\oiii/\hb}{[OIII]/Hb} (EW(\texorpdfstring{\ha}{Ha})--O/H) diagram}
\label{sec:contrast}

The core of LATED selection is the EW(\ha)--\oiii/\hb diagram, hereafter the EW(\ha)--O/H diagram: \ha equivalent width against the \oiii/\hb ratio, the standard rest-optical proxy for gas-phase oxygen abundance, with EW(\ha) measuring the strength of \ha line relative to the continuum.
It is one of the simplest and cleanest model-population planes available at these redshifts: the Pop~III target population occupies an extreme corner, high EW(\ha) and \oiii/\hb\,$\to0$, separated from the strong-line population by more than an order of magnitude in either axis individually and by orders of magnitude jointly.
Every model family with metal-free or near-metal-free nebular gas (e.g., \citealt{Zackrisson2011ApJ...740...13Z}, \citealt{Schaerer2003A&A...397..527S}, \citealt{Raiter2010A&A...523A..64R}, \citealt{Inoue2011MNRAS.415.2920I}, \citealt{Nakajima2022MNRAS.513.5134N}) clusters in this corner despite quite different IMF and SFH assumptions, while metal-enriched starbursts, normal star-forming galaxies, dusty star-forming galaxies, AGN and quiescent populations populate disjoint regions far above the \oiii/\hb threshold.

LATED measures this diagram photometrically with a pair of magnitude colours.
With generic notation, not tied to any one filter set, the adopted colours are
\begin{align}
y &\;=\; m_{{\rm H}\alpha} - m_{\rm cont},
\label{eq:ydef}\\
x &\;=\; m_{{\rm OIII}} - m_{{\rm H}\alpha},
\label{eq:xdef}
\end{align}
where $m_{\rm cont}$ is the magnitude in a line-free continuum band redward of \oiii.
For example, in the NIRCam implementation of the $z\approx3$ window (Section~\ref{sec:windows}), these become $y=m_{\rm F277W}-m_{\rm F356W}$ and $x=m_{\rm F200W}-m_{\rm F277W}$.
These are the photometric axes of the EW(\ha)--O/H diagram: $y$ tracks EW(\ha) within each window and $x$ tracks (\oiii+\hb)/\ha.
Given proper assumptions on the continuum slope and Balmer line ratios (Section~\ref{sec:recovery}), these two photometric axes of the EW(\ha)--O/H diagram can then trace EW(\ha) and \oiii/\hb.
The mapping is not exact because the photometric colours are not the line ratios, but it is monotonic in the regime that matters and is calibrated against the full model grid in Section~\ref{sec:models}, where the redshift-dependent cuts are derived.

\begin{figure*}
\includegraphics[width=\linewidth]{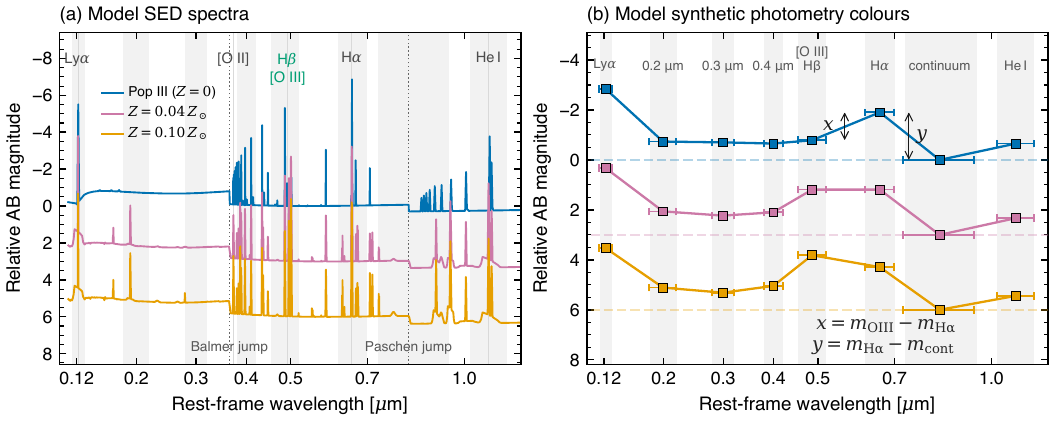}
\caption{The LATED selection principle, independent of any single window.
\emph{(a)} Rest-frame model SEDs from the rest FUV out to Paschen$\,\beta$: a Yggdrasil Pop~III model (extremely top-heavy Pop~III.1 IMF, 2\,Myr burst, $f_{\rm cov}=1$; \citealt{Zackrisson2011ApJ...740...13Z}), an FSPS boundary model ($Z=0.04\,Z_{\sun}$, 3\,Myr burst, $\log U=-1.5$) whose \oiii$+$\hb band carries exactly the same flux as its \ha band ($x=0$; see below), and an FSPS metal-enriched burst ($Z=0.1\,Z_{\sun}$, 3\,Myr, $\log U=-1.5$) \citep{Conroy2009ApJ...699..486C,Conroy2010ApJ...712..833C,Byler2017ApJ...840...44B}, in AB magnitude relative to the rest 5500\,\AA{} continuum and offset per class.
The defining signature is a strong \ha, a present \hb, a conspicuously \emph{weak} \oiii$\lambda5007$, the Balmer and Paschen jumps (dotted), and the \hei\,$\lambda10830$ line: metal-poor systems are \ha-strong but \oiii-weak; the enriched burst is the reverse.
Faint vertical bands mark the schematic windows of panel~(b).
\emph{(b)} The band photometry those SEDs produce in eight schematic rest-frame bands (a \lya band, FUV and NUV continuum bands, a rest-optical continuum band between \oii\,$\lambda3727$ and \hb, an \oiii$+$\hb band, an \ha band, a strong-line-free continuum band redward of \ha, and an \hei\,$\lambda10830$ band), for all three classes, each normalised to its own red continuum band and offset vertically for clarity, so that the \ha EW is directly traced by $y=m_{\rm H\alpha}-m_{\rm cont}$ and the O/H is the line-band contrast $x=m_{\rm OIII}-m_{\rm H\alpha}$ (equations~\ref{eq:ydef}--\ref{eq:xdef}).
Pop~III is \ha-strong and \oiii-weak ($x>0$); the boundary model sits exactly at $x=0$; and the $Z{=}0.1\,Z_{\sun}$ burst is \oiii-strong ($x<0$).
At each window redshift, the observed bands map onto these rest-frame positions; the concrete per-window realisations are in Section~\ref{sec:windowgallery}.}
\label{fig:sed}
\end{figure*}

Because brighter means numerically smaller magnitudes, strong \ha makes the \ha band bright and drives $y$ \emph{negative}, so the \ha-strength criterion is $y\leq y_{\rm max}(z)$; and when \ha dominates over \oiii$+$\hb, the \ha band is brighter than the \oiii band and $x$ is \emph{positive}, so the metal-deficiency criterion is $x\geq x_{\rm min}(z)$.
The pair is shown in Fig.~\ref{fig:sed} on three reference model classes: a Pop~III burst lands at $x>0,\, y<0$, the rejected metal-enriched burst at $x<0$, and the $Z=0.04\,Z_\odot$ boundary model exactly at $x\simeq0$.
We note the boundary condition $x=0$ here equates the two \emph{band} magnitudes, not the line fluxes: an AB band mean weights an emission line by its wavelength and by the inverse fractional band width, both of which favour \ha, so band equality requires the \oiii$\lambda\lambda4959,5007$$+$\hb complex to carry $\simeq$1.5 times the \ha line flux.
Equality of the line fluxes themselves would occur at $Z\approx0.01$--$0.02\,Z_{\sun}$ (e.g.\ from the empirical R3=\oiii$\lambda5007$/\hb calibration of \citealt{Isobe2026arXiv260611345I,Sanders2024ApJ...962...24S}).

\subsection{Window construction}
\label{sec:windows}

\begin{figure*}
\includegraphics[width=\textwidth]{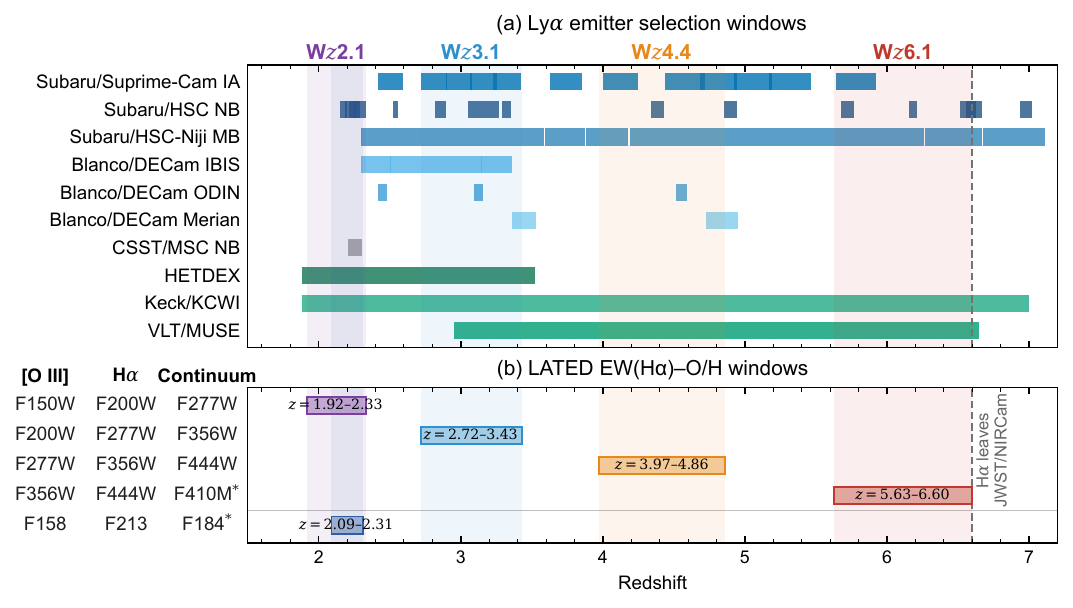}
\caption{Window construction.
\emph{Panel (a)}: the redshift interval over which each facility selects \lya, from the half-power edges of the real filter curves, or from the wavelength coverage for the spectroscopic routes.
\emph{Panel (b)}: the broadband EW(\ha)--O/H windows, \ha in the redder band and \oiii$+$\hb in the bluer, one row per window, with the three bands listed in the left margin and the adopted redshift range annotated on the bar.
Each bar carries its window's signature colour, which runs purple to red with redshift and is reused in every later figure keyed on window.
The shaded columns are the five adopted LATED windows of Table~\ref{tab:windows}, and the asterisk on W$z$2.2 and W$z$6.1 marks a continuum band lying between the two line bands rather than redward of \ha, which is what makes those two windows exploratory (Section~\ref{sec:lrdveto}).
The dashed line at $z\approx6.6$ is the derived limit of the method, where \ha leaves F444W, the reddest wide NIRCam band.
Medium-band, narrow-band, and mixed windows and lower-$z$ windows without ground-based Ly$\alpha$ coverage are not shown; the \textsc{Lated Explorer} provides them.}
\label{fig:windowcon}
\end{figure*}

The applicability of LATED to a given dataset is the product of two independent questions, which we deliberately separate: where can \lya parent samples be built, and where can \ha and \oiii$+$\hb be photometrically separated?
Their intersection defines a usable LATED window.
Figure~\ref{fig:windowcon} shows both layers, computed from the real filter transmission edges.

\subsubsection{Ly\texorpdfstring{$\alpha$}{a} parent-selection windows}
\label{sec:lyawin}

The parent stage does not depend on \emph{how} \lya is found: any survey that delivers a \lya flux, equivalent width, and slice redshift can seed a LATED selection.
Three techniques dominate, and we introduce each in turn together with the \lya redshifts it makes accessible.
Figure~\ref{fig:windowcon}(a) plots the coverage of the individual facilities from their wavelength coverage.

\textbf{Narrow-band imaging.}
Wide-field narrow-band (NB) imaging, with $R\sim50$--100 filters placed in the dark gaps between night-sky OH-airglow bands, has been the workhorse for assembling large, redshift-clean LAE samples since the pioneering blank-field Keck searches of \citet{Cowie1998AJ....115.1319C} and \citet{Hu1998ApJ...502L..99H}.
Dedicated campaigns then tiled \lya across cosmic time: the Large-Area Lyman-Alpha survey at $z\approx4.5$ \citep[LALA;][]{Rhoads2000ApJ...545L..85R}, MUSYC in the ECDF-S at $z\approx3.1$ and $2.1$ \citep{Gronwall2007ApJ...667...79G,Guaita2010ApJ...714..255G}, and the Subaru/Suprime-Cam SXDS survey at $z=3.1$, $3.7$ and $5.7$ \citep{Ouchi2008ApJS..176..301O}, reaching into reionisation with the NB921 $z=6.6$ sample of \citet{Ouchi2010ApJ...723..869O}.
The current generation scales this up by more than an order of magnitude: Subaru/Hyper Suprime-Cam SILVERRUSH at $z=5.7$ and $6.6$, extended to $z\approx2.2$--7.3 in its later narrow-band catalogues \citep{Ouchi2018PASJ...70S..13O,Shibuya2018PASJ...70S..14S}, the MAMMOTH-Subaru survey, which follows up the MAMMOTH overdensity fields \citep{Cai2016ApJ...833..135C} with Subaru/HSC NB387 and NB400 imaging to assemble $\sim$3300 LAEs at $z=2.2$ and $2.3$ over $\sim$12\,deg$^2$ \citep{Li2024ApJS..275...27L,Zhang2024ApJ...961...63Z}, the DECam LAGER survey at $z\approx7$ \citep{Zheng2017ApJ...842L..22Z,Hu2019ApJ...886...90H}, and the $\sim$100\,deg$^2$ ODIN survey, whose custom N419/N501/N673 filters select \lya at $z=2.4$, $3.1$ and $4.5$ \citep{Lee2024ApJ...962...36L,Firestone2024ApJ...974..217F}.
Emitters are selected from the NB colour excess, converted to line flux and rest-frame EW, with candidacy set by an excess-significance parameter and an EW floor \citep{Bunker1995MNRAS.273..513B,Sobral2013MNRAS.428.1128S}.
The net effect is that ground-based NB imaging samples \lya at a set of discrete slices fixed by which window each filter occupies.
For example, the Hyper Suprime-Cam (HSC) on the Subaru Telescope carries 18 NB filters\footnote{\url{https://subarutelescope.org/Instruments/HSC/sensitivity.html}}, covering \lya at $z=2.1$--7.3.

\textbf{Medium- and intermediate-band imaging.}
Broadening the filters to $R\sim20$--40 trades some EW sensitivity for contiguous redshift coverage\footnote{We note that the terms ``intermediate-band imaging'' and ``medium-band imaging'' are used interchangeably and refer to the same technique.}, so that a battery of adjacent medium bands acts as a wide-field, low-resolution integral-field spectrograph; the technique descends from the medium-band photometric-redshift surveys COMBO-17 \citep{Wolf2004A&A...421..913W} and ALHAMBRA \citep{Moles2008AJ....136.1325M}, though those targeted lower-redshift galaxies rather than \lya.
The twelve Subaru/Suprime-Cam intermediate bands imaged over COSMOS \citep{Taniguchi2007ApJS..172....9T,Taniguchi2015PASJ...67..104T} were the first such battery exploited systematically for \lya: SC4K treated the IA427--IA827 set, which tiles \lya over $z\approx2.4$--5.9, as an $R\approx20$--80 IFU to assemble $\sim$3900 emitters in sixteen redshift slices from $z\approx2$ to $6$ \citep{Sobral2018MNRAS.476.4725S}.
This slicing strategy is now scaling to wide areas: the DECam IBIS (Intermediate-Band Imaging Survey) tiles \lya gaplessly across $z=2.26$--3.41 in five medium bands (M411--M517) to feed DESI Run 2 \citep{Ebina2026JCAP...03..019E}, the DECam Merian survey images $\sim$850\,deg$^2$ of the HSC-SSP wide layer in two medium bands, N540 ($\lambda_{\rm c}=5400$\,\AA, $\Delta\lambda=210$\,\AA) and N708 ($7080$\,\AA, $275$\,\AA; \citealt{Luo2024MNRAS.530.4988L}).
Merian was designed to catch \ha and \oiii from $z\approx0.06$--0.10 dwarf galaxies rather than \lya, but the same two bands place \lya at $z=3.36$--3.53 (N540) and $z=4.73$--4.95 (N708), so the same imaging is a wide-area \lya parent route at those two redshifts.
The ongoing Subaru HSC Medium-Band survey, dubbed HSC-Niji, deploys sixteen medium bands, providing gapless coverage from $z=2.3$--7.1.

\textbf{IFU and wide-field spectroscopy.}
Integral-field and wide-field spectroscopic surveys detect \lya directly, so they need no colour-excess pre-selection and reach the faint population that NB and MB imaging misses.
Blind spectroscopy with VLT/MUSE (4800--9300\,\AA{}, so \lya over $z=2.95$--6.65, extendable to $z\approx2.83$) now yields hundreds to over a thousand emitters, from the Hubble Ultra Deep Field mosaic and 141-hr MXDF \citep[1308 LAEs at $2.8<z<6.7$;][]{Bacon2017A&A...608A...1B,Bacon2023A&A...670A...4B} to the wider MUSE-Wide survey \citep[479 LAEs at $2.9<z<6.3$;][]{Herenz2017A&A...606A..12H,Urrutia2019A&A...624A.141U} and lensing-cluster atlases that reach the intrinsically faintest luminosities \citep{Richard2021A&A...646A..83R,deLaVieuville2019A&A...628A...3D}.
Keck/KCWI with its KCRM red arm spans 3500--10800\,\AA{}, so \lya over $z=1.88$--7.88 \citep{Morrissey2018ApJ...864...93M}, while the planned BlueMUSE (3500--5800\,\AA{}, $z=1.88$--3.77) and the wide-field Magellan/LLAMAS (3500--9800\,\AA{}, $z=1.88$--7.06) extend or rebalance this coverage toward the blue.
At the wide-field end, the untargeted VIRUS/HETDEX survey blankets 540\,deg$^2$ at 3500--5500\,\AA{} to assemble of order $10^6$ \lya detections at $1.88<z<3.52$ \citep{Gebhardt2021ApJ...923..217G,MentuchCooper2023ApJ...943..177M}, and the Stage-V spectroscopic survey facilities such as Spec-S5, MUST and WST plan still much larger samples in the same range \citep{Besuner2025arXiv250307923B,Zhao2024arXiv241107970Z,Mainieri2024arXiv240305398M}.

\subsubsection{EW(\texorpdfstring{\ha}{Ha})--O/H windows}
\label{sec:optwin}

The second layer asks a purely geometric question: in which imager, and over what redshift interval, can the redshifted \ha and \oiii$+$\hb bands each be isolated while a strong-line-free continuum band remains available?
We answer it by an automated search over the real filter curves.

\textbf{Instruments and bands.}
Capturing the \ha line at $z\gtrsim2$ requires wavelength coverage beyond $\approx2\,\mu$m.
JWST/NIRCam is the primary facility: its wide bands F150W, F200W, F277W, F356W and F444W can place \oiii+\hb and \ha in adjacent filters from $z\approx1.9$ to $z\approx6.6$.
Its medium bands can further refine the coverage where they exist.
In addition to JWST/NIRCam, \textit{Roman}/WFI adds a window, its F213 band covering \ha at $z\approx2.2$.
By contrast, HST/WFC3 (limited to $\lesssim 1.7\,\mu\mathrm{m}$) and Euclid (limited to $\lesssim 2.0\,\mu\mathrm{m}$) cannot define a complete window of their own.
The legacy HST/NICMOS instrument extended coverage to $  \sim 2.5\,\mu\mathrm{m}  $ and could in principle observe \ha up to $  z \lesssim 2.8  $, but it has not been operational since 2008.
Current ground-based imaging at $\gtrsim 2\,\mu\mathrm{m}$ struggles to meet the depth requirements for our science case because of the bright thermal sky background.
For example, reaching a $5\sigma$ depth of $m_\mathrm{AB} \approx 27$ in $K_s$ band typically requires an impractical several hundred hours of integration on 8--10\,m telescopes such as Subaru, VLT, or Keck.
In contrast, \textit{Roman}/WFI can reach comparable depth in about 12 hours in the F213 band, and JWST reaches significantly greater depths within five minutes in the F200W band.

\textbf{The continuum-band rule.}
\label{sec:contrule}
The continuum band is preferred to be taken \emph{redward} of \ha and \emph{blueward} of \hei\,$\lambda10830$ whenever a suitable band exists in this window.
This placement brackets the stellar continuum while avoiding contamination from strong emission lines.
When no appropriate band is available in that interval, the continuum band is instead taken to be a strong-line-free filter lying between \oiii and \ha.
This selection strategy is designed to minimise contamination from red continuum sources, such as little red dots (Section~\ref{sec:lrdveto}).

\textbf{The window search.}
\label{sec:combosearch}
We built a pipeline to search EW(\ha)--O/H windows automatically.
The pipeline enumerates every (\oiii band, \ha band, continuum band) triplet drawn from the NIRCam bands and the \textit{Roman}/WFI bands, and keeps those that remain geometrically valid across a redshift interval.
Three conditions must hold throughout that interval: \ha lies inside its band; the full \oiii\,$\lambda\lambda4959,5007$+\hb complex lies inside its band, with \hb (the bluest line) setting the low-redshift edge and \oiii\,$\lambda5007$ (the reddest) the high-redshift edge; and the continuum band is free of \ha, \oiii, \hb, and \hei\,$\lambda10830$.
A line is \emph{inside} a band where it sits at or above half of the band's peak transmission ($\geq50$ per cent), and a band is \emph{free of} a line where the line falls below a tenth of the peak ($<10$ per cent).
The gap between these two thresholds is deliberate: a line at intermediate transmission is counted as cleanly captured by no band.
This matters where adjacent NIRCam wide bands overlap in wavelength (F277W/F356W near $3.13\,\mu$m, F356W/F444W near $3.93\,\mu$m), because a line landing in such an overlap sits near half power in \emph{both} filters and so is measured cleanly by neither, trimming the interval at the crossover.
Each surviving triplet returns the redshift interval over which it holds, and the union of these intervals is the set of usable EW(\ha)--O/H windows.

\subsubsection{Combined LATED windows}
\label{sec:combwin}

\begin{table*}
\caption{Redshift windows for LATED selection. The quoted range is the joint
window over which Ly$\alpha$ falls in at least one selection band, H$\alpha$ in the
designated H$\alpha$ band, and the full [O\,\textsc{iii}]\,$\lambda\lambda4959,5007+$H$\beta$
complex in the designated [O\,\textsc{iii}]+H$\beta$ band (half-power filter edges); the
bluest line, H$\beta$, sets the low-redshift edge. Within each window the
usable range for an individual object is set by its Ly$\alpha$ selection band.
The Ly$\alpha$ parent routes are listed one facility per row, with the bands of
that facility that fall in the window (Section~\ref{sec:combwin}); a band is
quoted where at least 30 per cent of its Ly$\alpha$ slice lies inside the window, and the
IFU and wide-field spectroscopic routes are quoted instead by the part of the
window they cover, since they measure the Ly$\alpha$ redshift directly. All three
selection bands belong to the facility named in the window column.
$^{\dagger}$Part of this band's Ly$\alpha$ slice falls outside the window, so the band
alone does not place a source inside it and a further redshift constraint is needed;
bands without the dagger are contained in the window to the 0.01 precision quoted here.
$^{*}$The continuum band lies \emph{between} the two line bands, sampling the rest-frame
optical continuum between [O\,\textsc{iii}]$+$H$\beta$ and H$\alpha$, rather than
redward of H$\alpha$ as the priority rule prefers; such a band could have
red-continuum interlopers at the colour stage, so these two windows are treated as
exploratory.}
\label{tab:windows}
\footnotesize
\setlength{\tabcolsep}{4pt}
\begin{tabular}{llllccc}
\hline
Window & $z$ range & Ly$\alpha$ facility & Ly$\alpha$ bands & [O\,\textsc{iii}]+H$\beta$ & H$\alpha$ & Continuum \\
\hline
NIRCam W$z$2.1 & $1.92$--$2.33$ & HSC NB & NB387, NB391, NB395, NB400 & F150W & F200W & F277W \\
 &  & HETDEX & $z=1.92$--$2.33$ &  &  &  \\
 &  & CSST/MSC NB & $v$ &  &  &  \\
\hline
\textit{Roman} W$z$2.2 & $2.09$--$2.31$ & HSC NB & NB387, NB391, NB395, NB400$^{\dagger}$ & F158 & F213 & F184$^{*}$ \\
 &  & HETDEX & $z=2.09$--$2.31$ &  &  &  \\
 &  & CSST/MSC NB & $v$ &  &  &  \\
\hline
NIRCam W$z$3.1 & $2.72$--$3.43$ & SC IA & IA464, IA484, IA505, IA527 & F200W & F277W & F356W \\
 &  & HSC NB & NB468, NB497, NB506, NB515, NB527 &  &  &  \\
 &  & HSC-Niji MB & MB465, MB490, MB516, MB543$^{\dagger}$ &  &  &  \\
 &  & IBIS & M464, M490, M517 &  &  &  \\
 &  & ODIN & N501 &  &  &  \\
 &  & Merian & N540$^{\dagger}$ &  &  &  \\
 &  & HETDEX & $z=2.72$--$3.43$ &  &  &  \\
 &  & MUSE & $z=2.95$--$3.43$ &  &  &  \\
\hline
NIRCam W$z$4.4 & $3.97$--$4.86$ & SC IA & IA624, IA679, IA709$^{\dagger}$ & F277W & F356W & F444W \\
 &  & HSC NB & NB656 &  &  &  \\
 &  & HSC-Niji MB & MB611$^{\dagger}$, MB648, MB685 &  &  &  \\
 &  & ODIN & N673 &  &  &  \\
 &  & Merian & N708$^{\dagger}$ &  &  &  \\
 &  & MUSE & $z=3.97$--$4.86$ &  &  &  \\
\hline
NIRCam W$z$6.1 & $5.63$--$6.60$ & SC IA & IA827 & F356W & F444W & F410M$^{*}$ \\
 &  & HSC NB & NB816, NB872, NB921$^{\dagger}$, NB926$^{\dagger}$ &  &  &  \\
 &  & HSC-Niji MB & MB811$^{\dagger}$, MB858, MB908$^{\dagger}$ &  &  &  \\
 &  & MUSE & $z=5.63$--$6.60$ &  &  &  \\
\hline
\end{tabular}
\end{table*}

\begin{figure*}
\includegraphics[width=\textwidth]{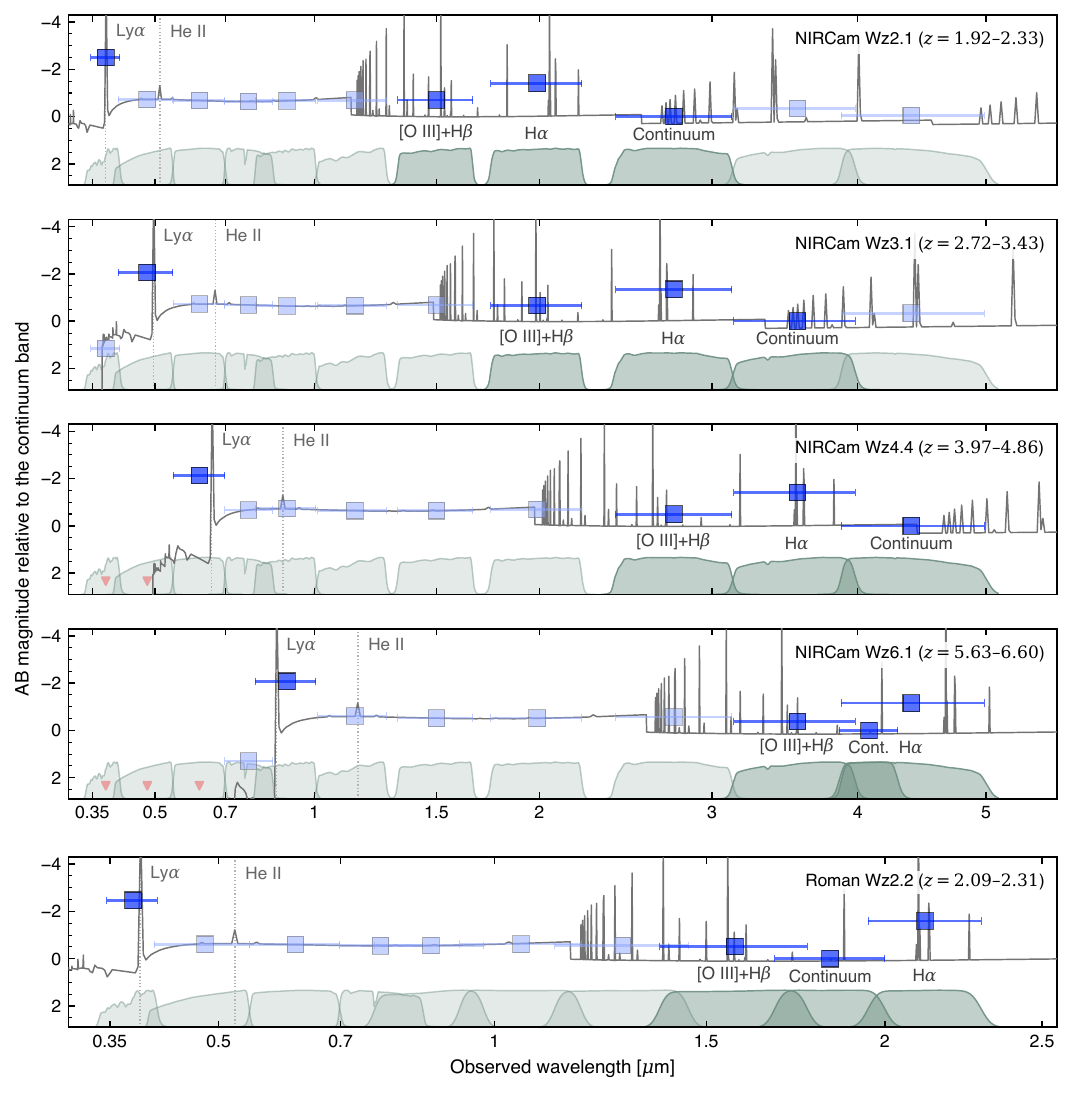}
\caption{The five LATED windows seen through the Pop~III template: the four NIRCam windows in redshift order, then the \textit{Roman} window W$z$2.2 on the bottom with narrower wavelength axis.
Each row shows the same Yggdrasil Pop~III model (moderately top-heavy Pop~III.2 IMF, 1\,Myr instantaneous burst, $f_{\rm cov}=1$, full \lya escape), whose \heii\,$\lambda1640$ emission is marked alongside \lya, redshifted to the window-centre redshift with mean IGM attenuation applied, in AB magnitude relative to the window's continuum band.
Squares are the synthetic photometry through representative optical bands (CFHT/MegaCam $u$; Subaru/HSC $g$, $r$, $i$) and the NIRCam wide bands F090W--F444W ($+$F410M for W$z$6.1), or the \textit{Roman}/WFI bands F087--F213 in the W$z$2.2 row; horizontal bars span each band's half-power range, and bands with no transmitted flux (blueward of the Lyman break) are drawn as red arrows at the panel floor.
Saturated blue marks the four bands the window actually leans on, the \lya parent band together with the [O\,\textsc{iii}]$+$\hb{} / \ha{} / continuum triplet, and pale blue the remaining bands, which are shown for context.
The filter transmission curves are shown in the bottom of each panel, with the window's [O\,\textsc{iii}]$+$\hb / \ha / continuum triplet emphasised and labelled on the curves.
Rising redshift walks \lya through the optical bands, dropping the bands blueward of the Lyman break one by one, while \oiii$+$\hb and \ha walk out along the NIRCam ladder, where the window construction of Table~\ref{tab:windows} encodes exactly this geometry.}
\label{fig:windowseds}
\end{figure*}

Five combined windows are revealed, with four built on JWST/NIRCam and one on \textit{Roman}/WFI, presented in Fig.~\ref{fig:windowcon} and listed in Table~\ref{tab:windows}.
Each window is labelled W$z$\texttt{x.y}, the suffix \texttt{x.y} being its approximate central redshift (so W$z$3.1 is centred near $z\simeq3.1$).
Figure~\ref{fig:windowseds} previews them all at once: the Pop~III spectrum template placed at each window's central redshift, showing how \lya walks through the optical bands while \oiii$+$\hb and \ha walk out along the NIRCam (and \textit{Roman}) filter ladder from $z=1.92$ to $z=6.60$.

Below we describe them in turn, grouped by facility.
Table~\ref{tab:windows} lists the \lya parent bands of each window, one facility per row, and we quote the same bands here.
We list a band if at least 30 per cent of its \lya slice falls inside the window, and we dagger it if part of that slice falls outside.
A daggered band cannot be used on its own: it selects emitters over a redshift interval that straddles a window edge, so an emitter it selects may lie outside the window, where the band's selection assumptions no longer hold.
Imaging alone cannot resolve that ambiguity, because a narrow- or medium-band detection constrains the redshift only to the width of its own slice.
A further spectroscopic constraint does resolve it, and the wide-field spectroscopic surveys now coming online supply exactly that: DESI and DESI Run 2 \citep{Schlegel2022arXiv220903585S}, Subaru/PFS \citep{Takada2014PASJ...66R...1T,Greene2022arXiv220614908G}, and the Stage-V facilities such as Spec-S5 \citep{Besuner2025arXiv250307923B} will deliver redshifts over the same footprints as the imaging surveys above.
Where such a redshift is available, a daggered band becomes usable on the same footing as a fully contained one.

\textbf{NIRCam W$z$2.1}  ($z=1.92$--2.33).
\oiii falls in F150W and \ha in F200W, with the red-side continuum in F277W and \hei in F356W.
\lya parents can be drawn across the full range from the HETDEX spectroscopic survey, and over its upper part from the HSC narrow bands (NB387, NB391, NB395, NB400) and the CSST $v$ band.
Blue IFUs like KCWI can cover this window.
The bluest medium bands, IBIS M411 and HSC MB413, both reach \lya only at $z\approx2.30$--2.51, so no more than a sixth of either slice lies inside the window and neither is listed as a parent route here.
Every band that is listed is contained in the window: NB400 alone runs past the upper edge, and only by $\Delta z=0.001$, within the precision to which that edge is quoted.
W$z$2.1 is the lowest-redshift NIRCam window, extending the selection down to cosmic noon.
Its blue edge $z=1.92$ is set by the continuum rather than a line band, below which \hei\,$\lambda10830$ enters F277W.
A still lower-redshift window with \ha in F115W or F150W is geometrically possible, but there \lya shifts into the near-UV, beyond the reach of ground-based surveys, so we do not adopt it.

\textbf{NIRCam W$z$3.1} ($z=2.72$--3.43).
\oiii falls in F200W and \ha in F277W, with the red-side continuum in F356W and \hei in F444W.
\lya parents can be selected from IFU spectroscopic surveys, HETDEX and KCWI across the full range and MUSE over its $z\gtrsim2.95$ coverage (or $z\gtrsim2.83$ coverage in blue extended mode), and from the HSC narrow bands (NB468, NB497, NB506, NB515, NB527), the ODIN narrow band N501, and the HSC (MB465, MB490, MB516) and IBIS (M464, M490, M517) medium bands.
This is the best-fed window of the five: all of these bands are contained in it, so none of them needs an external redshift.
Two further bands reach the window's upper edge without being contained in it, the HSC medium band MB543 ($z=3.36$--3.59, a third of it inside) and Merian N540 ($z=3.36$--3.53, two fifths inside), and Merian brings $\sim$850\,deg$^2$ of imaging with it.
That MUSE reaches only the upper part of the window does not matter, in contrast to the medium-band case: an IFU measures the \lya redshift precisely, so every emitter it detects is placed inside the window unambiguously.
In addition, an archival \lya-emitter parent sample is available from the SC4K Suprime-Cam intermediate bands (IA464, IA484, IA505, IA527; \citealt{Sobral2018MNRAS.476.4725S}), whose slices tile the full \lya redshift range of the window.

\textbf{NIRCam W$z$4.4} ($z=3.97$--4.86).
\oiii falls in F277W and \ha in F356W, with the red-side continuum in F444W and \hei in MIRI F560W below $z=4.70$.
\lya parents: the Suprime-Cam intermediate bands (IA624, IA679), the HSC medium bands (MB648, MB685), the HSC narrow band NB656, and the ODIN narrow band N673, together with, decisively for deep fields, the IFU spectroscopic surveys using MUSE and KCWI.
All of these are contained in the window.
Three more bands overlap it without being contained: MB611 ($z=3.88$--4.18) and IA709 ($z=4.69$--4.95) hang over the lower and upper edges respectively, each about two thirds inside, and Merian N708 ($z=4.73$--4.95) adds three fifths of its slice, and with it wide-area coverage of the top of the window.
The upper edge $z=4.86$ is set by the F356W/F444W filter overlap ($3.93\,\mu$m): above it \ha is split between the F356W \ha band and the F444W continuum, corrupting both, so the window stops there rather than at the geometric $z=5.07$ where \ha formally leaves F356W.
The lower edge $z=3.97$ is set by the \oiii$+$\hb band: below it \hb, the bluest line of the complex, drops out of the F277W half-power range, so the band no longer captures the whole \oiii$+$\hb flux (the \oiii\,$\lambda5007$-only coverage would reach $z=3.83$).
We note that \citet{Nishigaki2023ApJ...952...11N} search for extremely metal-poor galaxies at $z\approx4$--5 with JWST, within the redshift range this window spans; their colour cuts are compared with the LATED boundaries directly in Section~\ref{sec:selcompare}.

\textbf{NIRCam W$z$6.1} ($z=5.63$--6.60).
\oiii falls in F356W, \ha in F444W, and \hei in MIRI F770W, with the continuum in the \emph{medium} band \textbf{F410M}, which over this window samples the rest-optical continuum between \oiii$+$\hb and \ha ($\sim$5800\,\AA{} rest at $z=6$).
This is the configuration \citet{Fujimoto2025ApJ...989...46F,Fujimoto2025arXiv251211790F} use for their Pop~III photometric selection.
\lya parents: the Suprime-Cam IA827 band, the HSC narrow bands (NB816, NB872, NB921, NB926), the HSC medium bands (MB811, MB858, MB908), and the IFU spectroscopic surveys MUSE and KCWI, which span the whole window.
This window is the most dependent on partial coverage of the five.
The bands contained in it (IA827, NB816, NB872, MB858) between them reach only $z=6.26$, so the top of the window is served entirely by bands that straddle its upper edge: MB908 ($z=6.27$--6.67), NB921 ($z=6.52$--6.63) and NB926 ($z=6.56$--6.67).
MB811 ($z=5.49$--5.87) straddles the lower edge in the same way.
An external redshift therefore matters more here than in any other window, and above $z\approx6.3$ imaging alone cannot place a source inside W$z$6.1 at all.
The lower edge $z=5.63$ is where \ha clears F410M, and the upper edge $z=6.60$ is the wavelength limit of the method, where \ha leaves the reddest NIRCam wide band F444W.
Although F410M is a medium band, it is one of the most widely observed: because it captures the \oiii$+$\hb complex at $z\approx7$, a key reionisation-era diagnostic, it is carried, usually as the single medium band, by many otherwise broad-band NIRCam programs such as CEERS, PRIMER and UNCOVER \citep{Finkelstein2025ApJ...983L...4F,Donnan2024MNRAS.533.3222D,Bezanson2024ApJ...974...92B}.
Since W$z$6.1 needs only F410M alongside the ubiquitous F356W and F444W wide bands, it is available over far more sky than the deep multi-medium-band fields, if still less than a pure wide-band window.
Because F410M lies \emph{blueward} of \ha (F444W being the reddest wide band), it is hard to reject red-continuum interlopers at the colour stage, so the window is treated as exploratory and calls for additional post-selection diagnostics to reach a pure sample.
Because that requirement depends on how many bands a field carries, W$z$6.1 is released with two sets of criteria: a default, conservative one for fields holding only the three selection bands, and an extended one for fields with enough bands to run the multi-band contamination flag, which supplies that downstream diagnostic.
A geometric cost compounds this near the upper edge: across the window the F444W transmission at \ha falls from near its peak to about half, so a genuine metal-poor source's \ha-band excess weakens toward $z\approx6.6$ and the window's discriminating power degrades, which also drives the unusually strong redshift variation of its derived criteria (Section~\ref{sec:selcrit}).

The redshift gaps between the windows are real, set by the coverage gaps and overlaps of NIRCam bands.
In the gap between W$z$4.4 and W$z$6.1, for instance, \ha (for W$z$4.4) and then \oiii (for W$z$6.1) would fall in the F356W/F444W and F277W/F356W overlaps respectively, splitting the line between two bands, so neither wide-band window admits it.
Medium-band fields, however, can reach into this gap, e.g.\ with F300M and F410M as the line bands at $z=4.89$--5.30.

\textbf{Roman W$z$2.2} ($z=2.09$--2.31).
\oiii falls in the \textit{Roman} band F158 and \ha in F213, with the intervening F184 continuum, which is line-free only over this narrow interval (\ha enters F184 from the red below, \oiii$+$\hb from the blue above).
\lya parents: HETDEX, the HSC narrow bands (NB387, NB391, NB395, NB400), the CSST $v$ band, and IFU surveys with KCWI (MUSE reaches \lya only at $z\geq2.95$).
This window is $\Delta z=0.22$ wide, little more than half the span of W$z$2.1, so NB400 no longer fits inside it: a quarter of its slice lies above $z=2.31$, where \oiii$+$\hb begins to enter the F184 continuum, and an NB400-selected emitter therefore needs an external redshift before it can be placed in the window.
Because the F184 continuum sits between the lines rather than redward of \ha, it is difficult to reject red-continuum interlopers (dusty galaxies, little red dots) at the colour stage, so W$z$2.2, like W$z$6.1, is treated as exploratory.
It is the unique wide-area window, with tens of deg$^2$ of overlapping \lya and \textit{Roman} coverage, but at the \textit{Roman} HLWAS Deep-tier depths (depth F158/F184/F213\,$=27.5/27.0/25.9$, the \ha band F213 setting the floor; \citealt{ObservationsTimeAllocationCommittee2025arXiv250510574O}) it reaches only the most massive, youngest ($f_{\rm cov}\simeq1$) events ($\gtrsim10^7\,{\rm M}_{\sun}$).
A pointed deep \textit{Roman} field would lift this mass floor, for instance, the \textit{Roman} eXtreme Deep Field \citep[RXDF; GAS-2001;][]{Yan2026arXiv260908145Y} targets $5\sigma$ depths of $30.0/29.0/28.0$ in F158/F184/F213 over a full-depth area of $\sim$700\,arcmin$^2$ inside a total footprint of $>1200$\,arcmin$^2$ near the North Ecliptic Pole, about $2$\,mag deeper than HLWAS in the \ha band F213, so a W$z$2.2 selection there would trade the very wide HLWAS area for access to far lower masses.

\textbf{Windows beyond this work.}
\label{sec:mediumband}
We build the windows of this work on the broad bands of JWST/NIRCam and \textit{Roman}/WFI to cover \oiii+\hb and \ha, and treat NIRCam medium bands as a refinement.
Where several medium bands overlap a window they sharpen its line isolation and can add further diagnostics, such as the Balmer jump and the \hei\,$\lambda10830$ line.
Medium-band imaging is increasingly available, with NIRCam medium-band programmes either completed or planned, including JEMS \citep{Williams2023ApJS..268...64W}, the JADES Origins Field \citep{Eisenstein2025ApJS..281...50E}, MegaScience \citep{Suess2024ApJ...976..101S,Bezanson2024ApJ...974...92B}, CANUCS/Technicolor \citep{Sarrouh2026ApJS..282....3S}, JUMPS \citep{Withers2024jwst.prop.5890W}, MINERVA \citep{Muzzin2025arXiv250719706M}, and SPAM \citep{Davis2025jwst.prop.8559D}.
Selection windows can equally be built from medium bands, or from any combination of bands of different widths.
The construction also extends below $z=1.92$, where \lya falls blueward of the ground-based atmospheric window.
We do not focus on those windows in this work.
Together with our main five windows, they are provided on our web application dubbed \textsc{Lated Explorer}\footnote{\url{https://lated.pop3star.com}}, which applies the construction rules of this work to a combination of JWST/NIRCam, JWST/MIRI and \textit{Roman}/WFI bands.
For each combination, it also derives selection criteria with the models and procedure shown in the following sections.


\subsection{Models and selection criteria per window}
\label{sec:models}

The windows above are purely geometric.
This section makes them quantitative.
Turning a window into a selection requires two components: a forward-model library spanning the target populations and the contaminants
and a Monte-Carlo mock with realistic photometric noise that turns both into selection probabilities.

\subsubsection{Population templates}
\label{sec:templates}

\textbf{Pop~III models.}
For Pop~III models, we use the Yggdrasil templates \citep{Zackrisson2011ApJ...740...13Z} and the Pop~III SEDs of \citet{Nakajima2022MNRAS.513.5134N}.
Yggdrasil templates combine metal-free stellar populations with self-consistent nebular continuum and line emission.
We employ three IMF assumptions (extremely top-heavy Pop~III.1, 50--500\,M$_{\sun}$, moderately top-heavy Pop~III.2, and a normal Kroupa IMF at $Z=0$), each as an \emph{instantaneous burst} sampled on the native Yggdrasil age grid up to 20\,Myr, with nebular covering fractions $f_{\rm cov}=0.5$ and 1 where the grids provide them.
We drop the constant-SFR histories because a brief burst in pristine, inflowing gas could be the physically expected Pop~III mode, and an instantaneous-burst grid gives the cleaner, more conservative selection envelope.
The underlying line predictions trace to \citet{Schaerer2002A&A...382...28S,Schaerer2003A&A...397..527S} and \citet{Raiter2010A&A...523A..64R}.
We also exclude the $f_{\rm cov}=0$ populations from the target locus, because $f_{\rm cov}=0$ is highly unlikely for Pop~III, as these are short lived and therefore they are expected to be mostly embedded in the cloud from which they formed.
These $f_{\rm cov}=0$ populations leak all of their ionising photons and so emit no recombination lines and are invisible to any emission-line selection.

As an independent cross-check of the Yggdrasil templates, we add the Pop~III SEDs of \citet{Nakajima2022MNRAS.513.5134N}, which process \citet{Schaerer2003A&A...397..527S} metal-free stellar spectra (three IMFs: Salpeter $1$--$100$, $1$--$500$, and $50$--$500\,{\rm M}_{\sun}$) through \textsc{cloudy} at $\log U=-2.0$ to $-0.5$ and $n_{\rm H}=10^3\,{\rm cm}^{-3}$.
We exclude the extreme-ionisation $\log U=-0.5$: its strong \hei\,$\lambda5876$ could fall inside continuum bands and mimic a weakened \ha excess, an effect not corroborated by any Yggdrasil template.
All three gas-metallicity grids ($Z_\mathrm{gas}=0$, $10^{-5}$, and $10^{-4}$, i.e.\ $\lesssim10^{-2}\,Z_\odot$) are Pop~III stellar populations (Schaerer metal-free stars), and all join the Pop~III target class: the $Z_\mathrm{gas}=10^{-5}$ and $10^{-4}$ grids differ from the pristine $Z=0$ case only by trace nebular-gas enrichment, not by their stellar content.
Built from a different photoionisation code and stellar library than Yggdrasil, they fall in the same extreme corner of the EW(\ha)--O/H diagram with EW(\ha)\,$\approx2700$--4150\,\AA{} and \oiii$\lambda5007$/\hb\,$\lesssim0.02$ at $Z=0$, rising to $\sim$1 only at the enriched-gas $Z_\mathrm{gas}=10^{-4}$ end, and move the derived Pop~III selection envelope by $\lesssim0.04$\,mag (Section~\ref{sec:selcrit}): the selection criteria are insensitive to the choice of Pop~III models.
The broader cross-model robustness assessment, covering every published Pop~III family, is given in Section~\ref{sec:popiiimodels}.

\textbf{Extremely metal-poor galaxies (EMPGs).}
All non-Pop~III classes, this one and the other star-forming classes below, are generated with FSPS \citep{Conroy2009ApJ...699..486C,Conroy2010ApJ...712..833C} via \textsc{python-fsps}, using the MIST isochrones, MILES spectra, and the \citet{Byler2017ApJ...840...44B} \textsc{cloudy} nebular grids, which supply self-consistent emission-line luminosities (including \lya, \hb, \oiii, \ha, and \hei\,$\lambda10830$) as a function of metallicity, age, and ionisation parameter.
The extremely metal-poor class spans $\log(Z/Z_{\sun})=-2.0$ to $-1.5$ (the MIST/\textsc{cloudy} floor), ages of 1--100\,Myr under burst and constant histories, and $\log U=-2.5$ to $-1.5$.
Yggdrasil $Z=0.0004$ models ($Z/Z_\odot=0.02$ on the Yggdrasil solar scale) are included in the same class for cross-check.
In this regime, a few low-redshift textbook targets including J0811$+$4730 ($\sim0.02\,Z_\odot$; \citealt{Izotov2018MNRAS.473.1956I}) and the EMPRESS sample \citep{Kojima2020ApJ...898..142K,Nakajima2022ApJS..262....3N} anchor the class empirically.

\textbf{Metal-poor galaxies with strong emission lines.}
This class has $\log(Z/Z_{\sun})=-1.0$ to $-0.5$ (the metallicity range where \oiii/\hb peaks), ages 1--30\,Myr, $\log U=-2.5$ to $-1.5$, low dust.
The resulting rest-frame EWs (median EW(\ha)\,$\approx1500$\,\AA{} in the class as gridded) bracket the observed JWST medium-band EELG population \citep{Boyett2024MNRAS.535.1796B,Withers2023ApJ...958L..14W}.
We label this class in the figures as MPG with its model metallicity ($Z/Z_\odot=0.1$) rather than `EELG', because the latter is an observational selection category, a high equivalent-width cut, rather than a unique physical population; observationally, many such objects would indeed be classified as EELGs.

\textbf{Normal star-forming galaxies (SFGs) and dusty star-forming galaxies (DSFGs).}
We use near-solar metallicity, constant and delayed-$\tau$ histories, ages 30\,Myr--1\,Gyr, $\log U=-3.5$ to $-2.5$, with attenuation $A_V=0.3$--2 under both the \citet{Calzetti2000ApJ...533..682C} starburst law and a steeper SMC-like law, the latter a bumpless power law $A_\lambda=A_V\,(5500\,\mbox{\AA}/\lambda)^{1.2}$ approximating the \citet{Gordon2003ApJ...594..279G} SMC bar extinction curve, the steep attenuation empirically favoured for low-mass, low-metallicity systems \citep{Salim2020ARA&A..58..529S}.

\textbf{AGN templates.}
We retrieved MPA-JHU-style emission-line measurements \citep{Brinchmann2004MNRAS.351.1151B,Tremonti2004ApJ...613..898T} for 119\,956 galaxies at $0.02\leq z\leq0.4$ from the SDSS DR17 database \citep{Abdurro'uf2022ApJS..259...35A} (S/N(\ha)\,$>5$ and S/N(\hb)\,$>3$; an unbiased whole-plate subsample of the 473\,928 spectra satisfying the cuts), classified with the standard BPT demarcations \citep{Kewley2001ApJ...556..121K,Kauffmann2003MNRAS.346.1055K} into 63\,416 star-forming, 20\,840 composite, and 13\,094 AGN spectra, the remaining 22\,606 falling outside the BPT classification.
AGN templates are represented by power-law continua carrying the \emph{measured} rest-frame EW sets subsampled from the SDSS DR17 BPT-classified set; this replaces any hand-tuned line ratios.

\textbf{Pristine black holes (PBHs).}
A metal-free or metal-poor accreting black hole is the exotic source most readily confused with a Pop~III stellar population, and discriminating the two is the express purpose of \citet{Nakajima2022MNRAS.513.5134N}.
We refer to these as pristine black holes (PBHs): massive black holes, or heavy black-hole seeds, born in an essentially zero-metallicity environment, a class likely to include direct-collapse black holes, primordial black holes, and other heavy seeds.
We include their full grid of these models, labelled DCBH by \citet{Nakajima2022MNRAS.513.5134N} (216 emergent \textsc{cloudy} SEDs: a black body of $T_{\rm bb}=5\times10^4$--$2\times10^5$\,K plus a power-law accretion continuum of slope $\alpha=-1.2$ to $-2.0$, gas metallicity $Z=0$--$10^{-3}$ from pristine to mildly enriched gas ($Z\lesssim0.07\,Z_{\sun}$), $\log U=-3.0$ to $-0.5$, $n_{\rm H}=10^3\,{\rm cm}^{-3}$).

\textbf{Little red dots (LRDs).}
The non-negligible contaminant for an \ha-strong, \oiii-weak selection is the little red dot (LRD) population: compact, abundant sources with a characteristic `v-shaped' SED, a blue rest-UV continuum and a red rest-optical continuum inflecting at the wavelength around Balmer break, now mostly understood to be a population of predominantly faint, broad-line AGN \citep{Matthee2024ApJ...963..129M,Greene2024ApJ...964...39G,Kokorev2024ApJ...968...38K,Labbe2025ApJ...978...92L,Akins2025ApJ...991...37A}.
They are the worst case for LATED because three of their properties conspire: their strong Balmer lines and red optical continuum mimic an \ha-band excess and an \oiii deficit; their blue UV slope can pass a naive UV-slope gate; and they sometimes show \lya.
We include them as an empirical contaminant class using the template library presented in \citet{Zhang2025arXiv251205180Z}, built from JWST/NIRSpec prism spectra of 44 spectroscopically confirmed LRDs at $z\approx2$--7 and binned in the $(\beta_{\rm UV},\beta_{\rm opt})$ continuum-slope plane.

\textbf{Low-$z$ interlopers in parent LAE sample.}
The most likely contaminants for parent \lya emitter selection are low-$z$ \oii emitters.
They are star-forming galaxies placed at the redshifts where \oii\,$\lambda3727$ enters each \lya selection band (e.g., $z\approx0.24$--0.41 for W$z$3.1); their parent-stage observable is the \oii-driven NB/MB excess, propagated with the same machinery as \lya.
They contaminate \emph{purely photometric} (narrow-/intermediate-band excess) \lya selection only: a spectroscopic or IFU \lya redshift identifies the line directly, by its asymmetric profile and the absence of a resolved doublet, and removes the \oii emitters at the parent stage, so we retain them in the grid to test their location in the selection planes.

\textbf{Passive quiescent galaxies (QGs).}
We take old ($\geq0.3$\,Gyr), near-solar stellar continua with red, line-free SEDs.
They fall \emph{above} the selection box (at $y\gtrsim0$, far above the $y_{\rm max}$ cut, like the red continua) and are never selected in the mock, confirming that a red passive continuum cannot mimic an \ha-strong metal-poor target.

\textbf{Synthetic photometry.}
\label{sec:synthphot}
All templates are attenuated in the rest frame, redshifted across the window grid, subjected to mean IGM transmission \citep{Inoue2014MNRAS.442.1805I}, and convolved with the measured filter transmission curves.
The convolution integral is evaluated on a wavelength grid sampled finely enough ($\lesssim5$\,\AA) to resolve the nebular lines.

\subsubsection{Per-window selection planes}
\label{sec:windowgallery}

\begin{figure*}
\includegraphics[width=\textwidth]{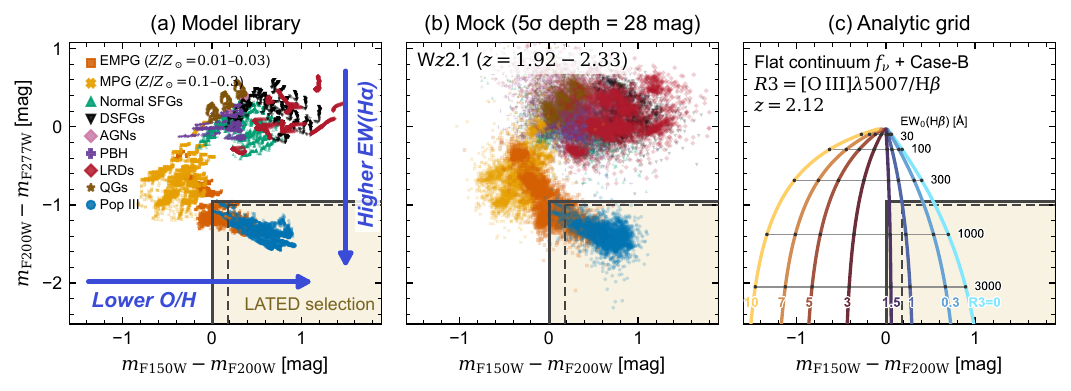}
\caption{LATED selection on the EW(\ha)--O/H plane, \textbf{W$z$2.1} ($z=1.92$--2.33; NIRCam F150W/F200W with the red-side F277W continuum).
\textbf{(a)} The noiseless model-class library, the model colour-grid points coloured by class.
All classes are introduced in Section~\ref{sec:templates}.
The LATED selection corner for Pop~III is filled in light yellow, which is the adopted target locus of Section~\ref{sec:selcrit}.
The arrows point towards lower O/H and higher EW(\ha).
In every panel the selection corner is drawn at both redshift extremes of the window: \emph{solid} = the loosest $z$-edge, \emph{dashed} = the tightest.
\textbf{(b)} Monte-Carlo realisation of the mock catalogue at a uniform 5$\sigma$ depth of 28\,mag, showing where each class scatters once photometric noise is applied.
Per-class selection probabilities are quoted in Section~\ref{sec:selfunc}.
\textbf{(c)} Analytic grid at the window-centre redshift, a flat-$f_\nu$ continuum carrying all hydrogen Balmer and Paschen lines (canonical Case B) with different EW(H$\beta$) and $R3=[\mathrm{O\,III}]\lambda5007/\mathrm{H}\beta$.
The coloured iso-$R3$ curves and grey iso-EW(H$\beta$) lines make explicit that $y$ tracks line EW while $x$ tracks the O/H.}
\label{fig:gallery_W$z$21}
\end{figure*}

\begin{figure*}
\includegraphics[width=\textwidth]{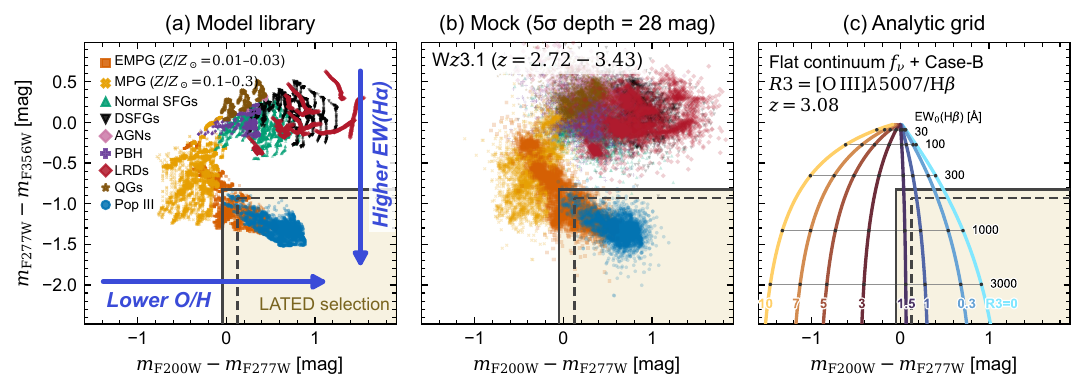}
\caption{LATED selection on the EW(\ha)--O/H plane, \textbf{W$z$3.1} ($z=2.72$--3.43; NIRCam F200W/F277W with the red-side F356W continuum), in the same format as Fig.~\ref{fig:gallery_W$z$21}.
}
\label{fig:gallery_W$z$31}
\end{figure*}

\begin{figure*}
\includegraphics[width=\textwidth]{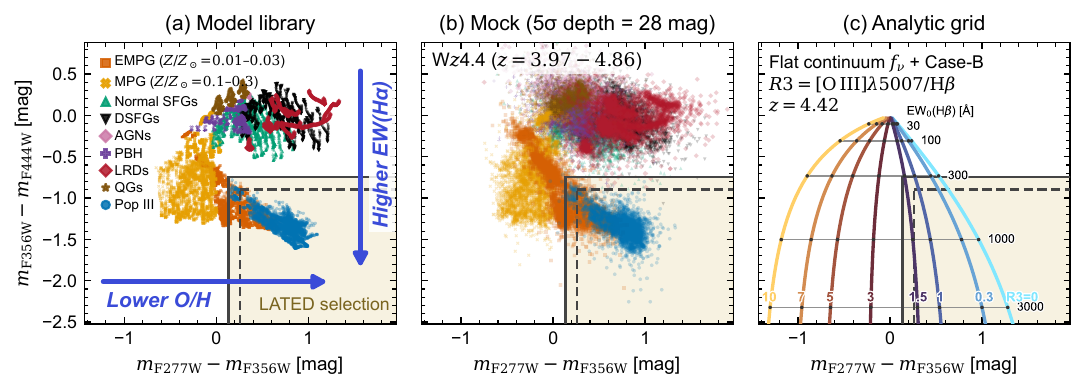}
\caption{LATED selection on the EW(\ha)--O/H plane, \textbf{W$z$4.4} ($z=3.97$--4.86; NIRCam F277W/F356W with the red-side F444W continuum), in the same format as Fig.~\ref{fig:gallery_W$z$21}.}
\label{fig:gallery_W$z$44}
\end{figure*}

\begin{figure*}
\includegraphics[width=\textwidth]{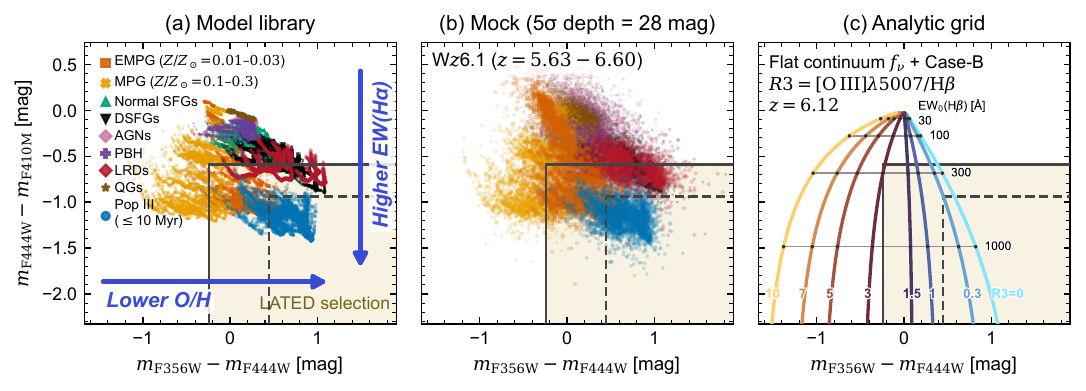}
\caption{LATED selection on the EW(\ha)--O/H plane, \textbf{W$z$6.1} ($z=5.63$--6.60; NIRCam F356W/F444W with the between-the-lines \emph{medium}-band F410M continuum), in the same format as Fig.~\ref{fig:gallery_W$z$21}.
Because F410M lies blueward of \ha, the LRD locus abuts the selection box from above: under a stricter young-age boundary (Pop~III inst $\leq$10\,Myr, Section~\ref{sec:selcrit}, as the legend notes) no LRD template enters the noiseless box per redshift, and the further LRD flag guards the noise-driven residual (Section~\ref{sec:lrdveto}).}
\label{fig:gallery_W$z$61}
\end{figure*}

\begin{figure*}
\includegraphics[width=\textwidth]{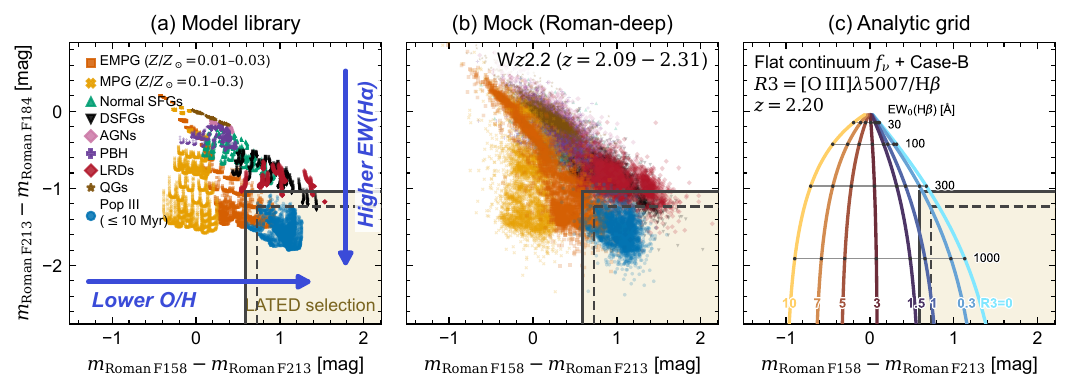}
\caption{LATED selection on the EW(\ha)--O/H plane, \textbf{W$z$2.2} ($z=2.09$--2.31; \textit{Roman} F158/F213 lines with the between-the-lines F184 continuum), in the three-panel format of Fig.~\ref{fig:gallery_W$z$21}.
Pop~III with instantaneous ages $>10$\,Myr is excluded in this window.
Here panel (b) is mocked at the \textit{Roman} HLWAS Deep-tier per-band depths (F158/F184/F213\,$=27.5/27.0/25.9$), so its colour scatter is broader and \ha(F213)-limited.
The model loci separate cleanly and the $Z{=}0.1\,Z_{\sun}$-burst leakage is zero; under the strict young-age boundary (Pop~III inst $\leq$10\,Myr, Section~\ref{sec:selcrit}, as the legend notes) no LRD template enters the colour box either, and \textit{Roman}'s binding constraint is survey depth.}
\label{fig:gallery_W$z$22}
\end{figure*}

Figures~\ref{fig:gallery_W$z$21},\ref{fig:gallery_W$z$31},\ref{fig:gallery_W$z$44},\ref{fig:gallery_W$z$61},\ref{fig:gallery_W$z$22} show the adopted EW(\ha)--O/H selection plane of each LATED window, the per-window realisation of the EW(\ha)--O/H diagram of Section~\ref{sec:contrast}.
They are shown in a three-panel format:
\emph{Panel (a)}: the noiseless model-class library, the model colour-grid points coloured by class, with the selection corner dominated by Pop~III models.
\emph{Panel (b)}: one Monte-Carlo realisation of the mock at the fiducial depth.
\emph{Panel (c)}: an analytic nebular grid at the window-centre redshift, making explicit that $y$ tracks H$\alpha$ line EW while $x$ tracks the \oiii/\hb ratio.
In every panel the selection corner is drawn at both redshift extremes of the window: \emph{solid} = the loosest $z$-edge, \emph{dashed} = the tightest.
The selected criteria shown here are derived in Section~\ref{sec:selcrit}.

Within these per-window planes the qualitative geography of the model classes is general across filter implementations; recall that strong \ha drives $y$ negative and \ha-dominance over \oiii$+$\hb drives $x$ positive (Section~\ref{sec:contrast}), so the selection corner sits at the bottom right:

\begin{itemize}
\item Pop~III galaxies occupy that bottom-right corner: a large \ha excess (strongly negative $y$) with only \hb and nebular continuum in the \oiii band (positive $x$).
\item Extremely metal-poor galaxies ($Z/Z_\odot=0.01$--$0.03$) occupy the bottom-centre region, partially overlapping the Pop~III locus.
This reflects a continuous distribution of metallicity and line EW for metal-free and extremely metal-poor galaxies across the range $Z/Z_\odot=0$--$0.03$.
It further suggests that photometric selection alone cannot confirm Pop~III detections; spectroscopic follow-up is required to determine the metallicity definitively.
\item {Metal-poor galaxies ($Z/Z_\odot=0.1$--0.3) with strong emission lines}, the most numerous strong-lined population, sit to the left of the corner: their \oiii$+$\hb band matches or outshines their \ha band, driving $x$ negative.
\item {Normal star-forming galaxies, PBHs, QGs, and AGNs} cluster near the origin with mild excesses in both colours.
\item {DSFGs and LRDs} have red optical continua, which can mimic an \ha excess.
Propagated through the bands, a continuum band redward of \ha turns this against them: it makes the continuum band bright and pushes them \emph{above} the selection box.
In W$z$2.2 and W$z$6.1, whose between-the-lines and blueward continua cannot separate a red continuum from an \ha excess.
To handle this case, we use a stricter Pop~III model boundaries with younger ages ($<10$\,Myr) to close the strip they enter through, and a dedicated rest-optical-continuum colour flags any residual leakage (Section~\ref{sec:selcrit}).

\end{itemize}

\subsubsection{Selection criteria}
\label{sec:selcrit}

Here we derive the redshift-dependent selection criteria.
The rule is geometric: at each redshift the selection region encloses the full extent of the adopted noiseless Pop~III target locus.
The locus is the combined Yggdrasil and \citet{Nakajima2022MNRAS.513.5134N} \textsc{cloudy} grids over all IMFs and covering factors $f_{\rm cov}>0$, with two cuts, both set out at the end of this section.

Concretely, at each redshift we take the noiseless colours of every template in the target locus and place the two thresholds at their extremes:
\begin{align}
x &\;\geq\; x_{\rm min}(z),
\label{eq:xcrit}\\
y &\;\leq\; y_{\rm max}(z).
\label{eq:ycrit}
\end{align}

\begin{figure}
\includegraphics[width=\linewidth]{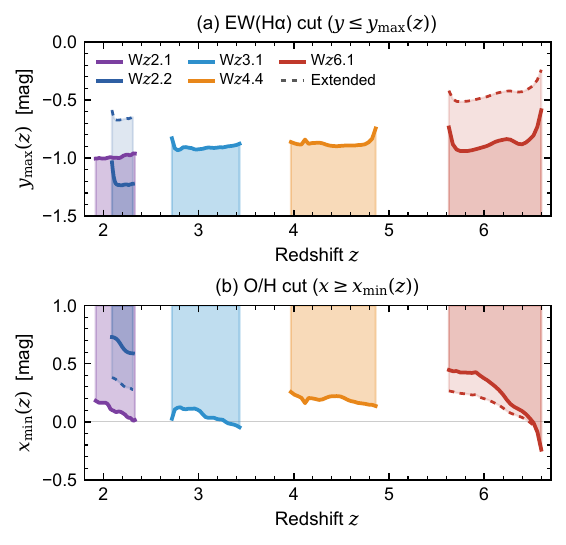}
\caption{The redshift-dependent LATED selection criteria for every window, from the released \texttt{selection\_boundary\_<W>.csv} tables where \texttt{<W>} is the window name.
\emph{(a)} the EW(\ha) cut $y\leq y_{\rm max}(z)$; \emph{(b)} the line-band contrast O/H cut $x\geq x_{\rm min}(z)$.
Both are full-extent envelopes of the combined Pop~III model locus, per-redshift extrema of the Yggdrasil \citep{Zackrisson2011ApJ...740...13Z} $+$ \citet{Nakajima2022MNRAS.513.5134N} grids; they slide smoothly with redshift as the lines move through the bands.
Solid curves are the default criteria.
The two windows whose continuum band lies between the lines, W$z$2.2 and W$z$6.1, also carry an extended variant that retains the 10--20\,Myr instantaneous bursts, drawn dashed in the same colour with the difference shaded.}
\label{fig:boundary}
\end{figure}

The \ha-strength threshold $y_{\rm max}(z)$ is the weakest \ha excess of any template, so the cut $y\leq y_{\rm max}$ keeps them all.
For example, across W$z$3.1, $y_{\rm max}=-0.93$ to $-0.83$, i.e.\ a selected source must be at least $0.83$\,mag (a factor of $\gtrsim2$) brighter in the \ha band than in the continuum band.
The line-band contrast threshold $x_{\rm min}(z)$ is likewise the smallest $x$ of any template, requiring the \ha band to be about as bright as, or brighter than, the \oiii$+$\hb band (e.g., $x_{\rm min}=-0.05$ to $+0.12$ across W$z$3.1).
Because the thresholds trace the model spectra rather than an idealised line list, they automatically allow for the nebular continuum, which contributes real \oiii-band flux in nebular-dominated ($f_{\rm cov}\simeq1$) metal-free populations and which a pure-line criterion would wrongly cut away.
The envelope is deliberately permissive: no template inside the target locus is excluded, photometric noise is handled by the Monte-Carlo mock rather than by inflating the boundary, and contaminant rejection is quantified afterwards (Section~\ref{sec:selfunc}) rather than built into the cut.
The choice of Pop~III models does not drive the result: the Yggdrasil and \textsc{cloudy} loci nest inside one another, and removing the \citet{Nakajima2022MNRAS.513.5134N} grids moves the boundaries by $\lesssim0.04$\,mag.
The two adopted cuts are shown for every window in Fig.~\ref{fig:boundary}, and the full $z$-resolved boundaries are released as machine-readable tables (\texttt{selection\_boundary\_Wz\{2.1/2.2/3.1/4.4/6.1\}.csv}, together with \texttt{selection\_boundary\_Wz\{2.2/6.1\}\_extended.csv} for the extended variant defined below).

Two cuts are applied to the locus.
First, the \textsc{cloudy} $\log U=-0.5$ corner is excluded in every window: its strong \hei\,$\lambda5876$ lands inside continuum bands and fakes a weakened \ha excess, a single-corner extreme with no Yggdrasil counterpart.
Second, in W$z$2.2 and W$z$6.1 the continuum band does not lie redward of \ha, and there instantaneous bursts older than 10\,Myr are excluded from the locus.
These snapshots exist only in Yggdrasil (the \textsc{cloudy} grids carry no age dimension), emit \lya an order of magnitude more weakly than the 2-Myr reference burst (intrinsic EW$_0$(\lya)\,$=74$--383\,\AA{}, median 167\,\AA{}, against $\approx$1600\,\AA{}), and their evolved continua occupy the same colour strip as the red-continuum contaminants that precisely these windows' band geometry is difficult to reject.
The three primary NIRCam windows, whose red-side continuum bands separate evolved bursts from red contaminants on their own, retain the full 0--20\,Myr instantaneous age range.

The trim is a statement about how many bands the observation can provide, not about Pop~III physics.
The 10--20\,Myr bursts remain legitimate targets; what changes is whether anything downstream can catch a red-continuum interloper that the colour stage has let through.
We therefore release both age-capped windows, W$z$2.2 and W$z$6.1, in two variants, and Fig.~\ref{fig:boundary} draws the extended pair dashed against the default.
We describe the trade for W$z$6.1 here (W$z$2.2 behaves in the same way, its extended criteria loosening $x_{\rm min}$ from $[0.59,0.73]$ to $[0.28,0.38]$ and $y_{\rm max}$ from $[-1.23,-1.04]$ to $[-0.67,-0.59]$).
The default, conservative criteria carry the trim and are the ones to use in a field that holds only F356W, F410M and F444W, where no further band exists to test a continuum: at $z=5.7255$ they give $x_{\rm min}=+0.432$ and $y_{\rm max}=-0.931$.
The extended criteria retain the full 0--20\,Myr locus and are the ones to use where the field carries enough bands to measure the multi-band contamination flags, especially LRDs of Section~\ref{sec:lrdveto}, which then supplies the downstream diagnostic the colour stage gives up: at the same redshift they give $x_{\rm min}=+0.248$ and $y_{\rm max}=-0.512$.
The price is paid at the colour stage and is strongly redshift dependent.
Over the noiseless W$z$6.1 model grid, the LRD templates go from a pass fraction of 0.004 under the default criteria, non-zero only above $z\approx6.5$ ($0.015$ at $z=6.53$--6.58), to 0.957 under the extended criteria, rising from 0.77 at $z=5.63$--5.68 through 0.83 at $z=5.73$--5.78 and 0.92 at $z=5.83$--5.88 to 1.00 in every sampled bin above $z\approx5.93$; dusty star-forming templates go from 0.008 to 0.668 over the same grid.
Through the full selection function the trade is far milder, because the \lya parent stage removes most dusty or red-continuum sources before the colour box ever sees them.
Every W$z$2.1 and W$z$6.1 number quoted below is computed under the default criteria unless stated otherwise.

The derived cuts slide only mildly with redshift in most windows, with $x_{\rm min}(z)$ spanning 0.12--0.17\,mag and $y_{\rm max}(z)$ 0.05--0.19\,mag across the four lower-redshift windows, but W$z$6.1 is the exception.
Running \ha to the very end of the filter set makes its criteria vary far more strongly: across $z=5.63$--6.60 the derived $x_{\rm min}(z)$ spans 0.69\,mag and changes sign (from $+0.45$ to $-0.24$, crossing zero at $z\approx6.51$) and $y_{\rm max}(z)$ spans 0.35\,mag.
Three filter-geometry effects drive this.
First, because the upper edge \emph{is} \ha's half-power exit from F444W, the F444W transmission at \ha slides from 0.98 to 0.49 of peak across the window and the \ha boost of the band falls from 1.2 to 0.8\,mag for a Pop~III template, so by $z\gtrsim6.5$ even genuine Pop~III sources no longer show an \ha band clearly brighter than their \oiii$+$\hb band, which forces $x_{\rm min}$ negative and steepens it toward the edge.
Second, lines cross the narrow F410M continuum band within the window: at the blue edge the \ha wing still enters F410M at $\approx$9 per cent of peak transmission, brightening the ``continuum'' by $\approx$0.3\,mag before decaying away by $z\approx5.8$, and \hei\,$\lambda5876$ sits inside F410M at near-full transmission up to $z=6.32$ ($\approx$0.15\,mag) before exiting mid-window.
Third, the \oiii$+$\hb band brightens toward the top of the window as \hb climbs onto the F356W flat top and H$\gamma$ and \oiii\,$\lambda4363$ enter its blue edge at $z\approx6.23$ ($+0.13$\,mag).
The per-redshift criteria absorb all three effects by construction, but the physical discriminating power of the colour pair genuinely degrades toward $z\approx6.6$, where half of the \ha flux is no longer collected, a further reason to treat W$z$6.1 as exploratory.

\subsubsection{Selection gates and flags}
\label{sec:score}

When applying LATED selection to real data we suggest a small set of selection \emph{gates} together with a set of \emph{flags} (Table~\ref{tab:score}).
The four gates are the two detections, in \lya (S/N$>$3 for spectroscopy and S/N$>$5 for photometry) and in \ha (S/N$>$5 in the H$\alpha$ band), and the two colour criteria; a source failing any of them is not a candidate.
The flags record counterpart multiplicity, a blue UV continuum, a blue UV-to-optical continuum colour, the absence of an AGN counterpart, and photometric-redshift consistency; they provide more information to warn the selection without removing, so a flagged candidate stays in the sample and stays inspectable.

The adopted flags, and of the corroborating measurements that extend them beyond the selection photometry, is as follows.
None is a selection criterion; each raises or lowers a surviving candidate's priority for spectroscopy.

\textbf{A blue ultraviolet continuum.}
Extremely metal-poor populations are intrinsically blue, but the nebular continuum reddens the composite spectrum even at $Z=0$ \citep{Schaerer2002A&A...382...28S,Raiter2010A&A...523A..64R}, so the ultraviolet slope works as a flag against red, dusty or evolved contaminants rather than as a positive metal-free indicator.
We place the flag permissively at $\beta_{\rm UV}\leq-1.5$ rather than at the bluest values a dust-free stellar population would reach.
Simulations support that placement: in MEGATRON, Pop~III-dominated galaxies span $\beta_{\rm UV}\approx-2$ to $-1.5$, \emph{redder} than their Pop~II-dominated counterparts because of the nebular continuum \citep{Storck2026MNRAS.548ag529S}, so a stricter blue cut would remove the very systems being sought.

\textbf{A blue ultraviolet-to-optical continuum.}
The second flag is an ultraviolet-to-optical continuum colour, measured as an AB magnitude difference on bands that carry no strong emission line,
\begin{equation}
C_{\rm red} = m^{\rm cont}_{\rm UV} - m^{\rm cont}_{\rm opt},
\label{eq:cred}
\end{equation}
where $m^{\rm cont}_{\rm UV}$ is the median magnitude of the bands whose rest-frame pivot wavelength lies between 1500 and 4000\,\AA{} and $m^{\rm cont}_{\rm opt}$ that of the bands between 5500 and 9000\,\AA, in both cases discarding any band whose passband contains \lya, \oii, \hb$+$\oiii or \ha$+$\nii at the source redshift.
The bands are line-free by construction, so emission-line boosting cannot redden a galaxy whose continuum is in fact blue.
A nebular-continuum-dominated spectrum is brighter in the rest ultraviolet than in the rest optical, so the flag is raised at $C_{\rm red}\leq0$, while a red-continuum interloper such as LRDs sits at $C_{\rm red}>0.7$.
One line-free colour therefore carries both diagnostics, with the target population at one end and the dominant contaminant at the other, and the intervening range left deliberately unflagged in either direction.
Its red end is what identifies little red dot candidates, a use set out with its thresholds in Section~\ref{sec:lrdveto}.
The same colour is also the robust photometric form of the Balmer jump (Section~\ref{sec:balmerjump}).

\textbf{\hei\,$\lambda10830$.}
At $z\approx3$ this line falls in F444W, and for Pop~III-like populations its predicted strength is sufficient to produce a measurable excess in deep data (Fig.~\ref{fig:sed}).
A simultaneous \ha and \hei double excess with absent \oiii substantially strengthens a candidate, because it is difficult to mimic with a redshift error.

\textbf{\heii\,$\lambda1640$.}
Where rest-ultraviolet spectra exist, nebular \heii is the classic hard-ionisation signature \citep{Schaerer2003A&A...397..527S}, and its detection elevates a candidate's priority, while its association with broad stellar features instead indicates Wolf--Rayet contamination \citep{Shirazi2012MNRAS.421.1043S}.
Narrowness alone is not conclusive either, since the slow, dense winds of metal-poor very massive stars can produce narrow \heii\,$\lambda1640$ at ${\rm FWHM}\approx300$--500\,km\,s$^{-1}$ \citep{Grafener2015A&A...578L...2G}, so line width supports but does not by itself establish a Pop~III interpretation.
However, a high EW(\heii) is typical of Pop~III in most models and, especially when combined with other diagnostics such as line ratios, can isolate Pop~III from other populations.

\textbf{Morphology, and the AGN and dust vetoes.}
High-resolution imaging from JWST and Roman distinguishes compact clumps, outskirt pockets and blends, and pristine-pocket scenarios predict \lya-bright structure offset from the host continuum \citep{Mas-Ribas2016ApJ...833...65M,Vanzella2023A&A...678A.173V}.
X-ray counterparts, radio detections, mid-infrared power-law colours and optical variability each veto a candidate or mark it for AGN-sensitive follow-up, and a far-infrared or submillimetre detection indicates obscured star formation inconsistent with a chemically primitive interpretation.

\begin{table}
\caption{LATED selection gates and flags.
A gate is a requirement: a source failing any gate is not a candidate.
A flag is a warning: it is recorded per candidate and never removes a source, so that every criterion stays individually inspectable and the sample stays reproducible from the photometry alone.
The colour thresholds $y_{\rm max}(z)$ and $x_{\rm min}(z)$ are publicly released per window.}
\label{tab:score}
\begin{tabular}{p{2.5cm}p{5.2cm}}
\hline
Gate & Requirement \\
\hline
\lya detection & ${\rm S/N}\geq5$ in the \lya narrow or medium bands; or ${\rm S/N}\geq3$ in the spectroscopy \\
\ha detection & ${\rm S/N}\geq5$ in the \ha band (\oiii and continuum bands may be upper limits) \\
EW(\ha) cut & $y = m_{{\rm H}\alpha} - m_{\rm cont} \leq y_{\rm max}(z)$ \\
O/H cut & $x = m_{\rm OIII} - m_{{\rm H}\alpha} \geq x_{\rm min}(z)$ \\
\hline
\hline
Flag & Requirement \\
\hline
Counterpart quality & all sources within $r=1$\arcsec of the \lya centre, with multiplicity flagged\\
Blue UV continuum & $\beta_{\rm UV}\leq-1.5$ ($f_\lambda\propto\lambda^{\beta}$) \\
Blue UV-optical SED & $C_{\rm red}=m^{\rm cont}_{\rm UV}-m^{\rm cont}_{\rm opt}\leq0$ on strong-emission-line-free bands \\
Non-AGN & no X-ray/radio/MIR counterpart; no variability \\
Phot-$z$ consistency & $p(z)$ overlaps the \lya slice \\
\hline
\end{tabular}
\end{table}


\subsection{Uncertainties and selection functions}
\label{sec:mockunc}

We close the methodology with the irreducible uncertainties of the models and mock, and the recovery and leakage they imply.

\subsubsection{Uncertainties}
\label{sec:uncertainties}

We state the uncertainties of the models and of the mock explicitly, because together they bound what a photometric selection can claim.
Items 1--6  are properties of the model library; item 7 are properties of the mock and of the statistics built on it.

\begin{enumerate}
\item \textbf{Nebular covering fraction dominates detectability.}
$f_{\rm cov}=0$ populations \citep{Zackrisson2011ApJ...740...13Z}, which leak all of their ionising photons, produce no recombination lines and are unselectable by construction.
We therefore quote the Pop~III recovery fraction conditional on $f_{\rm cov}>0$, and treat $f_{\rm cov}=0$ templates as a separate, unselectable floor, with a zero selection possibility in every window.
LATED statements about the Pop~III population are necessarily statements about its \emph{nebular-dominated} fraction, and every recovery fraction quoted inherits this condition.

\item \textbf{Pop~III-dominated systems are the target.}
Simulations find that Pop~III components hosted by metal-enriched galaxies power strong metal lines through their dominant Pop~II cohabitants, so the presence of \oiii does not exclude Pop~III \citep{Rusta2025ApJ...989L..32R,Venditti2026ApJ..1005..226V,Venditti2026OJAp....967811V}.
The \oiii-weakness criterion targets the Pop~III-\emph{dominated}, metal-weak phase, which semi-analytic models limit to the first $\sim$15--20\,Myr of a system's star-forming life \citep{Rusta2025ApJ...989L..32R}.
However, the populations selected by LATED can span a continuous metallicity distribution, including a `hybrid' phase.
Distinguishing them requires further \heii–based diagnostics.

\item \textbf{The Pop~III IMF is unknown.}
EW predictions differ by factors of a few between IMF assumptions \citep{Schaerer2003A&A...397..527S,Zackrisson2011ApJ...740...13Z,Nakajima2022MNRAS.513.5134N}.
Our grid spans six IMF realisations across the two synthesis codes, three in Yggdrasil and three in the \citet{Nakajima2022MNRAS.513.5134N} \textsc{cloudy} models.
Those six realisations pair up by stellar mass range across the two codes, so they bracket the IMF grids in use rather than the space of physically distinct IMF shapes.
The resulting per-variant recovery fraction is not monotonic in mass range, so our grid does not on its own fix the sign of the IMF dependence.

\item \textbf{The $Z=0$ engines can be compared but not fully bracketed.}
FSPS, which generates every enriched class, floors at $\log(Z/Z_{\sun})=-2$ and cannot represent $Z=0$.
The metal-free tier therefore rests on the two dedicated grids, Yggdrasil and the \citet{Nakajima2022MNRAS.513.5134N} \textsc{cloudy} models, which between them vary the photoionisation code, the IMF sampling and the nebular treatment, and which agree to $\lesssim0.04$\,mag in the derived selection envelope (Section~\ref{sec:selcrit}).
What they do not vary is the stellar side: both trace to \citet{Schaerer2002A&A...382...28S,Schaerer2003A&A...397..527S} metal-free atmospheres, so no independent metal-free stellar library enters the boundaries.

\item \textbf{Nebular geometry and ionisation parameter are unconstrained photometrically.}
Density-bounded nebulae depress all lines; high ionisation parameter shifts \oiii/\hb at fixed metallicity \citep{Nakajima2022MNRAS.513.5134N}.
Both act in the colours we select on, and neither is separable from metallicity by broad-band photometry alone.

\item \textbf{Photometric metallicities are indirect.}
The mapping from colours to line ratios to metallicity passes through the model library; systems flagged by LATED are candidates with model-dependent significance, and only spectroscopy (deep enough to detect or stringently limit \oiii\,$\lambda5007$, \oiii\,$\lambda4363$, and \oii) can better establish chemical abundances.

\item \textbf{Population priors are not modelled.}
Per-class selection probabilities are computed exactly, but they are conditional in two ways, and neither can be resolved with the information available.
Converting leakage into contamination \emph{fractions} requires the relative abundances of the classes in LAE samples, which are survey-specific and observationally unknown.
The Pop~III recovery fraction is likewise an average over our model grid under uniform weighting of IMF, covering fraction, age and ionisation parameter, not over a physical population, so it is a property of the criteria and the library rather than the completeness of a sample.
Both steps require a population prior that no current observation constrains, and we therefore defer them to future works.
\end{enumerate}

\subsubsection{Dependence on the choice of Pop~III model}
\label{sec:popiiimodels}

Two independent metal-free stellar-plus-nebular SED grids are ingested here, Yggdrasil \citep{Zackrisson2011ApJ...740...13Z} and the \citet{Nakajima2022MNRAS.513.5134N} \textsc{cloudy} models, and the adopted boundaries are the envelope of their union.
The cross-code test is therefore a result rather than a plan: the two loci nest inside one another, and the derived selection envelope moves by $\lesssim0.04$\,mag.
In W$z$3.1 the per-redshift extrema that set both thresholds are Yggdrasil templates in every redshift slice, so removing the \textsc{cloudy} grids leaves the boundaries of that window unchanged.

We assess how far the LATED-relevant predictions depend on these two choices by comparing against the other published Pop~III and extremely-metal-poor model families.
Of the families not ingested here, only \citet{Raiter2010A&A...523A..64R} publish a downloadable non-Yggdrasil $Z=0$ stellar-plus-nebular SED grid (eight IMFs, with the enhanced two-photon continuum), whereas \citet{Schaerer2002A&A...382...28S,Schaerer2003A&A...397..527S} and \citet{Inoue2011MNRAS.415.2920I} supply tabulated EWs and line ratios rather than SEDs.
\citet{Gessey-Jones2022MNRAS.516..841G} and \citet{Liu2025MNRAS.541.3113L} also release metal-free stellar spectra and ionising photon production rates, for non-rotating and for chemically homogeneous evolution respectively.
Neither provides nebular emission, so they bear on the ionising output of the stars rather than on the line ratios LATED selects with.
We note explicitly that BPASS \citep{Eldridge2017PASA...34...58E,Stanway2018MNRAS.479...75S} and Starburst99 \citep{Leitherer1999ApJS..123....3L} \emph{cannot} serve as Pop~III anchors, because their metallicity floors ($Z=10^{-5}$ and $\sim10^{-3}$ respectively) are not metal-free and BPASS ships no nebular lines, so they are listed only as low-metallicity stellar comparisons.

The essential point is robustness.
Every metal-free family predicts a high \ha equivalent width, with maxima $\sim$3200\,\AA{} for \citet{Schaerer2002A&A...382...28S} and 1232--4800\,\AA{} across the Yggdrasil grid measured here.
By construction of a pure H+He nebula \oiii/\hb is essentially zero, and the metal-inclusive \citet{Inoue2011MNRAS.415.2920I} grid sets the citable threshold \oiii\,$\lambda5007$/\hb\,$<0.1$ for $Z<10^{-3}\,Z_{\sun}$.
The adopted target class is slightly broader than that statement, because it also admits the \textsc{cloudy} grids with trace nebular enrichment up to $\lesssim10^{-2}\,Z_{\sun}$.
For example, over the W$z$3.1 window target locus the model \oiii\,$\lambda5007$/\hb runs 0.008--0.014 for the Yggdrasil $Z=0$ templates and 0.015--0.020 for the \textsc{cloudy} $Z=0$ templates, rising to 0.06--0.14 at $Z_{\rm gas}=10^{-5}$ and to 0.5--1.1 at $Z_{\rm gas}=10^{-4}$, so a minority of the class (12 per cent of locus templates) lies above the \citet{Inoue2011MNRAS.415.2920I} threshold and outside its stated metallicity range.
Those trace-enriched members widen the class without loosening the cut, because the per-redshift extrema that set the thresholds are pristine templates.
The LATED colour signature (strong \ha, weak \oiii) is solid and robust across different photoionisation codes.

Where the different models genuinely diverge is the rest-frame He~{\sc ii}\,$\lambda1640$ EW, a spectroscopic rather than photometric discriminant.
The cross-model compilation of \citet{Trussler2023MNRAS.525.5328T} spans $\approx$15--90\,\AA (Yggdrasil 16--50, \citealt{Nakajima2022MNRAS.513.5134N} 25--80\,\AA, \citealt{Raiter2010A&A...523A..64R} 20--90\,\AA, \citealt{Inoue2011MNRAS.415.2920I} $\sim$15\,\AA, with a ZAMS maximum of $\sim$100\,\AA{} in \citealt{Schaerer2002A&A...382...28S}), and an ageing burst dilutes it further through the nebular continuum \citep{Schaerer2003A&A...397..527S}.

The limitation that remains is narrower than a single-grid dependence, but it is not empty: the varied axes are the photoionisation code, the IMF sampling and the nebular treatment, while the metal-free stellar atmospheres underlying both grids are the same.
Ingesting the \citet{Raiter2010A&A...523A..64R} SEDs, the one public grid that would vary that axis, is the natural further cross-check and is deferred to future work.
Beyond the public grids, \citet{Lecroq2025A&A...695A..17L} derive the spectral properties of Pop~III populations with a different stellar evolution code and a different atmosphere model, and, alone among the families considered here, include binary interactions.
Their SEDs are not public, but they vary precisely the stellar axis that Yggdrasil and the \textsc{cloudy} grids hold in common, which makes them the most direct way to close this gap.

\subsubsection{The selection function: Pop~III recovery}
\label{sec:selfunc}

We now specify the Monte-Carlo mock that turns the model library and the adopted criteria into selection probabilities.
The mock catalogue draws $20\,000$ objects per class from the model grid, the latter built on a fine $\Delta z=0.001$ redshift grid so the sampling is continuous in redshift, assigns continuum magnitudes $m_{\rm cont}\sim U(22, 32)$, and perturbs the photometry with Gaussian noise set by 5$\sigma$ point-source depths (background-limited; the depth is a free parameter).
The mock focuses on the colour stage.
The preceding \lya parent stage is survey-specific and is discussed separately in Section~\ref{sec:selcompleteness}.
An object is selected when it passes, in a given noise realisation, (i) an S/N\,$\geq5$ detection in the \ha band, the \oiii$+$\hb and continuum bands being carried at their $2\sigma$ upper limits where undetected, exactly as Table~\ref{tab:score} specifies; and (ii) the adopted colour--colour criteria at its slice redshift, $y\leq y_{\rm max}(z)$ and $x\geq x_{\rm min}(z)$ (equations~\ref{eq:ydef}--\ref{eq:xdef}).
We apply no S/N gate to the \oiii$+$\hb band, since a weak or absent line there indicates high EW or low O/H rather than a failure.
The selection probability $P_{\rm sel}$ is the Monte-Carlo pass fraction.
We call the Pop~III value a \emph{recovery fraction} rather than a completeness, because it is the probability that a template drawn from our Pop~III grid under uniform weighting passes the stated gates, not the fraction of a physical population a survey would find.

One effect lies outside the mock by construction: the boundaries are evaluated at each object's true redshift, so the window-edge bias that a redshift error would induce is not modelled.
IGM attenuation is not such an effect, since it does not reach the selection bands, which lie far redward of \lya in every window (Section~\ref{sec:synthphot}).

Noiselessly, the criteria recover the target locus in full, and by construction, the boundaries are its per-redshift envelope (Section~\ref{sec:selcrit}), so that number restates the construction rather than measuring anything.
What can be measured is how much of it survives photometry.
Requiring S/N\,$\geq5$ in the \ha band, for sources at least 1\,mag brighter than the depth, the recovery fraction is $>0.99$ in every window.
This is the quantity the \textsc{Lated Explorer} reports for any band combination.
At the depths they are designed for, then, the colour criteria are not what limits a search.

\begin{figure*}
\includegraphics[width=\textwidth]{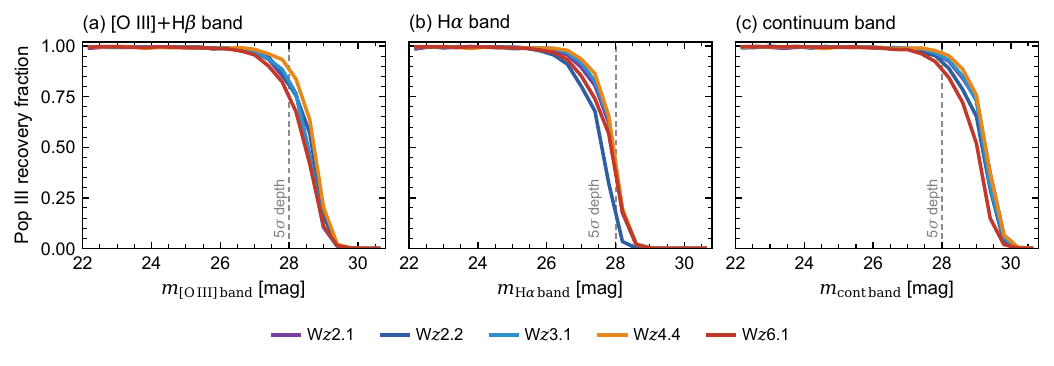}
\caption{Colour-stage recovery of the Pop~III templates against source magnitude, at a fixed $5\sigma$ depth of 28\,mag in all three bands (dashed line).
Each panel bins the same objects by their magnitude in one selection band, so each curve is a completeness curve in that band.
Only the colour stage is applied.
Recovery is complete at the bright end, $\sim1.0$ brighter than 25\,mag, and the half-recovery point falls at $27.93$ in the \ha band \emph{(b)}, at the depth itself, against $28.54$ in \oiii$+$\hb \emph{(a)} and $29.22$ in the continuum \emph{(c)}.
The offsets between the panels are the median colours of the Pop~III templates, so all three curves trace the one gate the LATED algorithm imposes, the \ha detection; the continuum and \oiii$+$\hb bands enter as $2\sigma$ upper limits where undetected.}
\label{fig:banddepth}
\end{figure*}

Figure~\ref{fig:banddepth} holds all three bands' $5\sigma$ depths at 28\,mag and bins the target locus by its magnitude in each selection band in turn, giving a completeness curve per band.
All three are flat at unity over the bright half of the range, reaching $\sim1.0$ brighter than 25\,mag in every window, and then fall steeply.
Taking W$z$3.1 for definiteness, the half-recovery point sits at $m=27.93$ in the \ha band (i.e., at the depth itself) and at $29.22$ and $28.54$ in the continuum and \oiii$+$\hb bands.
The three are not three separate limits, because the \ha detection is the only gate the algorithm imposes, and the other two panels show it displaced by the typical colours of Pop~III templates.
Over the same locus the median $m_{\rm cont}-m_{{\rm H}\alpha}$ is $1.33$\,mag and the median $m_{\rm OIII}-m_{{\rm H}\alpha}$ is $0.64$\,mag, which place the half-recovery points at $29.26$ and $28.57$ against the $29.22$ and $28.54$ measured.
A target is therefore recovered whenever its \ha is detected, however faint its continuum, because an undetected continuum enters the EW(\ha) cut as an upper limit and a faint continuum only strengthens the \ha excess.
The five windows lie on top of one another in all three panels, so at matched depth the colour stage behaves the same way everywhere.

\begin{figure*}
\includegraphics[width=0.67\textwidth]{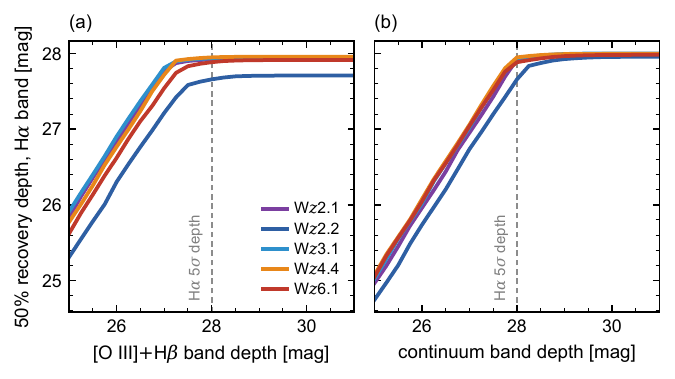}
\caption{How deep the other two bands should be, with the \ha $5\sigma$ depth fixed at 28\,mag (dashed line).
The curve is the 50 per cent recovery depth in the \ha band, the \ha magnitude at which colour-stage recovery of the Pop~III target locus ($f_{\rm cov}>0$) falls through one half.
Both panels flatten at the \ha depth, but either can hold it back if it is the shallowest.
\emph{(a)} The \oiii$+$\hb band reaches that ceiling once it is within $\approx0.5$\,mag of the \ha depth, and may be the shallowest of the three, being the brighter band for these blue targets and carrying no S/N gate.
\emph{(b)} The continuum band reaches it at $\approx28$--$28.75$.
A band that is not detected enters its colour at its $2\sigma$ upper limit.}
\label{fig:linebanddepth}
\end{figure*}

Figure~\ref{fig:linebanddepth} asks another question from the survey's side.
Holding the \ha depth at 28\,mag, it sweeps each of the other two bands and reports the 50 per cent recovery depth in the \ha band, the magnitude to which half the locus is still recovered.
Both curves rise and then flatten at $27.7$--$28.0$, the \ha depth itself, so neither band can push the selection past the H$\alpha$ depth.
Once the continuum band reaches $28.0$--$28.75$, its curve lies within $0.05$ mag of that ceiling; the same is true for the \oiii+\hb curve once that band reaches $27.25$--$28.25$.
Below those depths, the reach drops by $\approx 3$ mag and $\approx 2$ mag, respectively, over the range considered.
This suggests neither band needs to be deeper than H$\alpha$.
\oiii+\hb can even be the shallowest of the three, because it is brighter for these metal-enriched targets and does not have an S/N gate.
In practice, matching the three depths could be the best choice.
Extra depth beyond the H$\alpha$ gate is wasted, and shallower depth would affect the Pop~III recovery.

For statistical applications, the same machinery can propagate the depths of any given survey, so that candidate counts can be expressed per comoving volume once a parent selection function and a population prior are supplied. Detailed calculations and a discussion of overall completeness are left to future work.

\begin{figure*}
\includegraphics[width=\textwidth]{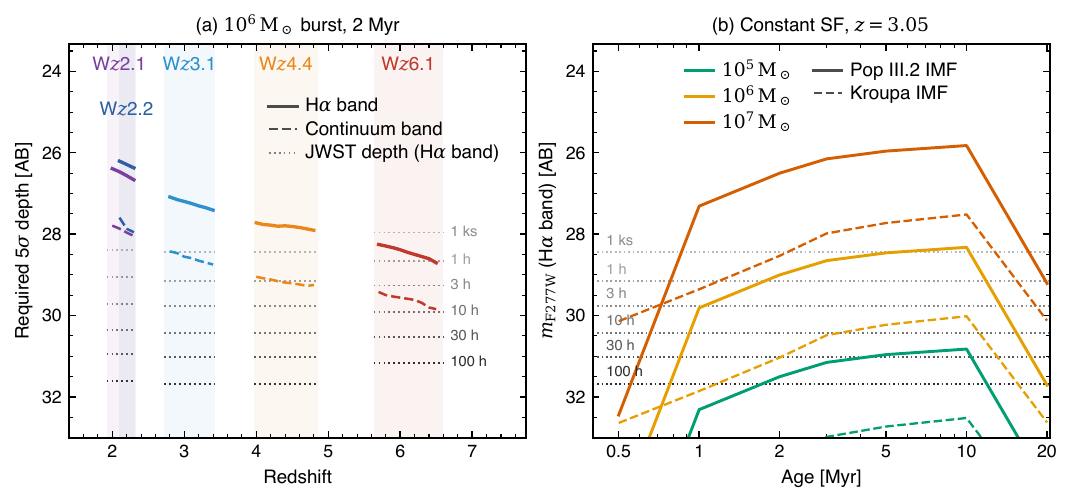}
\caption{Detectability of Pop~III / extremely metal-poor domains, referred to JWST exposure time.
\emph{(a)} The 5$\sigma$ depth required to detect the \ha band (solid) and the continuum band (dashed) of a $10^6\,{\rm M}_{\sun}$, 2-Myr Pop~III burst across the five windows.
\emph{(b)} Apparent \ha-band magnitude versus age at $z=3.05$ for constant-SF Pop~III populations of $10^5$--$10^7\,{\rm M}_{\sun}$ of gas (Yggdrasil, $f_{\rm cov}=1$; solid: moderately top-heavy IMF; dashed: Kroupa at $Z=0$), on the same magnitude scale as (a).
Dotted levels are the 5$\sigma$ point-source depths NIRCam reaches in 1\,ks to 100\,h of on-source time (\textsc{pandeia}; \citealt{Pontoppidan2016SPIE.9910E..16P}), drawn in (a) for each window's \ha band with the exposure labels written once against W$z$6.1, and in (b) for F277W; W$z$2.2 is a \textit{Roman} window and carries none.
The levels are idealised (fiducial background, default aperture, optimised readout), so achieved depths in real mosaics differ by a few tenths of a magnitude.}
\label{fig:depth}
\end{figure*}

\begin{figure*}
\includegraphics[width=\textwidth]{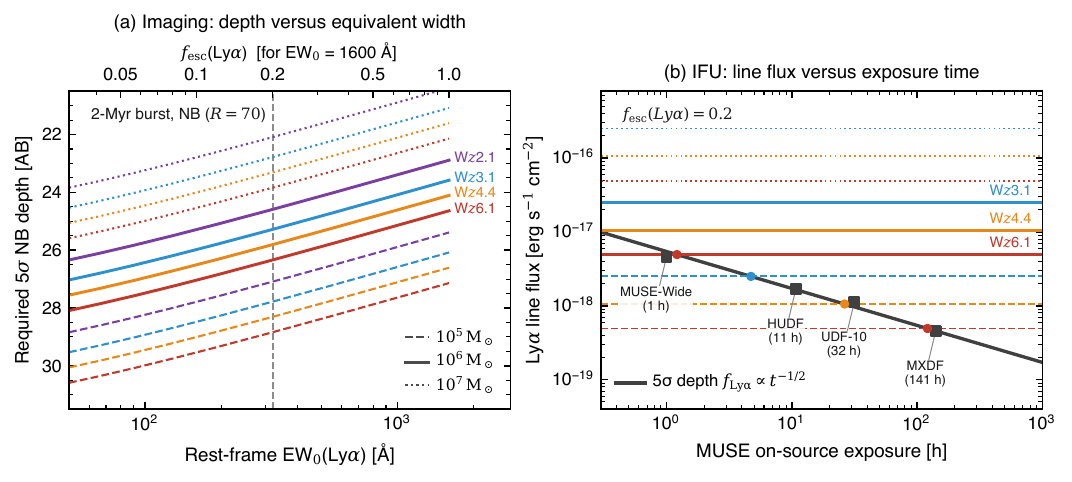}
\caption{The \lya depth requirement, for the reference 2-Myr Pop~III burst at $10^5$, $10^6$ and $10^7\,{\rm M}_{\sun}$ (dashed, solid, dotted) at the central redshift of the four NIRCam windows (W$z$2.2 shares W$z$2.1's redshift).
\emph{(a)} The $5\sigma$ depth of a narrow band of resolving power $R=70$, that of the real \lya narrow bands, must reach, against the rest-frame equivalent width $f_{\rm esc}{\rm EW}_0$ (equation~\ref{eq:nbdepth}; top axis: $f_{\rm esc}$ for the template's intrinsic ${\rm EW}_0\approx1600$\,\AA).
At fixed $R$, the line boost is redshift-independent, so the windows differ only by distance dimming; the vertical line marks the fiducial $f_{\rm esc}=0.2$.
\emph{(b)} The same requirement as a line flux at $f_{\rm esc}=0.2$, for the three windows VLT/MUSE can reach (\lya enters its bandpass only at $z\geq2.95$, excluding W$z$2.1), against the faintest \lya emitters MUSE actually recovers: the fifth percentile of the \lya flux of ${\rm S/N}\geq5$ emitters per tier (squares; MUSE-Wide, \citealt{Urrutia2019A&A...624A.141U}; HUDF mosaic, UDF-10 and MXDF, \citealt{Bacon2023A&A...670A...4B}), with a $t^{-1/2}$ fit (solid), a real selection limit for extended emission that sits a factor $2$--$5$ above published point-source limits.
Dots mark the exposure at which each burst is reached.
The \lya line is several magnitudes brighter than the continuum of the same source, and existing narrow-band imaging and integral-field spectroscopy already reach it.}
\label{fig:lyadepth}
\end{figure*}

\section{Detectability and contaminants}
\label{sec:performance}

The last section established what the model library and the mock can and cannot constrain.
Here we ask what that implies for a real survey: the depth it must reach and the false positives it will admit.

\subsection{Detectability and required depth}
\label{sec:depth}

A survey is useful to LATED only if it can reach the targets' expected apparent magnitudes.
Because the model library is absolutely normalised (Yggdrasil per $10^6\,{\rm M}_{\sun}$ of gas formed into the burst; FSPS per ${\rm M}_{\sun}$ formed), the mock can predict those magnitudes directly (Fig.~\ref{fig:depth}).

Here, we perform a rough estimate.
A $10^6\,{\rm M}_{\sun}$ Pop~III burst (Yggdrasil Pop~III.2 template) at $z=3.0$, observed at its 2-Myr brightest phase, has $m_{\rm F277W}=27.2$ in the \ha-boosted band but a continuum magnitude of $m_{\rm F356W}=28.6$: the \emph{continuum} sets the depth a clean colour measurement needs.
In NIRCam time, that gap is an order of magnitude: $2$\,min of on-source exposure to detect the line against $21$\,min to measure the continuum beside it (Fig.~\ref{fig:depth}a).
The same calculation across the five windows is shown in Fig.~\ref{fig:depth}(a), and the pattern is uniform.
At each window's central redshift the same burst requires a 5$\sigma$ depth of $26.5$ and $27.9$ in the \ha and continuum bands for W$z$2.1, $26.3$ and $27.9$ for W$z$2.2, $27.2$ and $28.6$ for W$z$3.1, $27.8$ and $29.2$ for W$z$4.4, and $28.5$ and $29.6$ for W$z$6.1.
The continuum is the harder requirement in every window, by $1.1$--$1.6$\,mag, so a survey that detects the line but not the continuum yields a limit rather than a colour.
Converted into on-source NIRCam time (Fig.~\ref{fig:depth}a), the line costs $32$\,s in W$z$2.1, $2$\,min in W$z$3.1, $5$\,min in W$z$4.4 and $41$\,min in W$z$6.1, against $6$\,min, $21$\,min, $2.7$\,h and $8.6$\,h for the respective continuum bands.
The colour therefore costs an order of magnitude more time than the detection in every window, and $31$ times more in W$z$4.4, whose F444W continuum is the least sensitive band in the set.
The requirement climbs by $\approx2$\,mag from W$z$2.1 to W$z$6.1, which is why the high-redshift windows are reserved for the deepest fields.
A $10^7\,{\rm M}_{\sun}$ event brightens to $m_{\rm F277W}\approx24.7$ and is accessible over wide fields; a $10^5\,{\rm M}_{\sun}$ pocket ($m_{\rm F277W}\approx29.7$) is only visible to those deepest fields, or when magnified by lensing clusters.
Ageing is equally punishing: between 2 and 20\,Myr a constant-SFR Pop~III population fades by $\sim$1\,mag in the \ha band, and instantaneous bursts fade much faster.

These requirements also explain why LATED could not have been carried out before JWST, because no earlier facility combined the wavelength coverage and the depth that a single LATED colour demands.
For example, \citet{Fontana2014A&A...570A..11F} describe the HAWK-I UDS and GOODS Survey (HUGS) as one of the deepest $K$-band imaging surveys, reaching $K_s=26.5$ in $31.5$\,h per pointing at $5\sigma$.
The independent ZFOURGE $K_s$ images reach $25.5$--$26.5$ ($5\sigma$) over $400$\,arcmin$^2$ \citep{Straatman2016ApJ...830...51S}, while the deepest wide-field near-infrared survey, UltraVISTA \citep{McCracken2012A&A...544A.156M}, reaches only $K_s=24.8$ over $\sim$1.9\,deg$^2$ in its sixth data release despite $171$\,h per pixel.\footnote{\url{https://irsa.ipac.caltech.edu/data/COSMOS/images/Ultra-Vista/ultravistadr6.pdf}}
Against this $\approx 26.5$\,mag ceiling, the \ha band is only marginally detected, while the continuum depth required for a LATED colour falls short by $1.4$\,mag in W$z$2.1 and W$z$2.2.
Integration alone cannot close that gap: extrapolating the HUGS depth under idealised background-limited scaling gives $\sim$400\,h per pointing for W$z$2.1 and W$z$2.2, and these are lower limits, because the depth of real coadded stacks improves more slowly than $\sqrt{t}$.
The wavelength argument is stronger still than the depth argument.
At $z=2.09$--2.31 the \ha line falls at $2.03$--$2.18\,\mu$m, inside the ground $K$ window, and the reference burst requires only $m=26.3$ in F213, the closest analogue of $K_s$, so the deepest $K$-band pointing ever obtained would just have detected it.
It is brighter still in the rest-ultraviolet, at $m\approx27.2$ in $g$, $r$ and $i$ at $z\approx2.1$, well inside HSC-SSP UltraDeep imaging \citep{Aihara2018PASJ...70S...4A}.
What was never accessible from the ground is the continuum band that turns such a detection into a colour.
At half power that band spans $2.42$--$3.13\,\mu$m in W$z$2.1, where the atmosphere is opaque, and $1.69$--$2.00\,\mu$m in W$z$2.2, straddling the telluric water band between $H$ and $K$, while in windows of higher redshifts it lies between $3.0$ and $5.0\,\mu$m, in the thermal infrared.

Space observations did not close the gap either.
Spitzer/IRAC surveyed the deep extragalactic fields for sixteen years and left a well-populated ladder in depth and area, from the wide and shallow SMUVS over $0.45$\,deg$^2$ \citep{Ashby2018ApJS..237...39A}, through SEDS at $12$\,h per pointing over $1.46$\,deg$^2$ \citep{Ashby2013ApJ...769...80A} and S-CANDELS at $50$\,h over $0.16$\,deg$^2$ \citep{Ashby2015ApJS..218...33A}, to the ultradeep IRAC Ultra Deep Field \citep{Labbe2015ApJS..221...23L} and finally GREATS, which accumulated $4260$\,h over the GOODS fields and remains the deepest mid-infrared imaging obtained before JWST \citep{Stefanon2021ApJS..257...68S}.
Their $5\sigma$ point-source depths run from $24.8$ in SMUVS and $25.5$ in SEDS to $26.8$ at $3.6\,\mu$m in the deepest GREATS coverage once source blending is accounted for.
Those are respectable depths, and they are not the reason LATED had to wait.
The reason is that IRAC's bands do not line up with the ones a LATED colour needs.
IRAC's two warm channels are close analogues of F356W and F444W, at half power $3.18$--$3.92$ and $4.00$--$5.01\,\mu$m against $3.14$--$3.98$ and $3.88$--$4.99\,\mu$m, but ch1 begins at $3.18\,\mu$m while the ground $K$ window closes near $2.4\,\mu$m.
Only W$z$6.1 can adopt the selection with these two filters.
Across the window \hb and the \oiii doublet fall in ch1 while \ha falls in ch2, so $[3.6]-[4.5]$ is the LATED $x$ colour, and it was measured well before JWST.
For example, \citet{Bowler2017MNRAS.469..448B} deblended the SPLASH imaging of CR7 at $z=6.604$ and found $[3.6]-[4.5]=-1.20\pm0.30$ for its brightest clump, part of the case against the Pop~III interpretation proposed by \citet{Sobral2015ApJ...808..139S}.
That colour lies far below our O/H threshold of $x_{\rm min}\simeq-0.24$ at $z=6.6$, so this cut alone would reject the clump.
The IRAC continuum band, however, could not reach the depth we require.
Its cryogenic $5.8\,\mu$m channel had the right placement, sitting redward of \ha throughout the window, but it reached only $24.3$ ($5\sigma$) in the deepest GOODS coverage and ceased when the cryogen ran out in 2009, more than five magnitudes short of what W$z$6.1 demands.
Neither decades of deep ground-based imaging nor sixteen years of Spitzer could therefore have revealed extremely metal-poor domains at these masses, not because the sources are undetectable but because the band that makes the measurement metallicity-sensitive was either unobservable or far too shallow.

That situation has now changed.
The facility implications divide into three statements, one about JWST, one about \textit{Roman}, and a final one about the \lya parent surveys.

\textbf{Most JWST fields can reach the depth requirements.}
JADES-class depths, with $m_{5\sigma}\gtrsim29$ \citep{Eisenstein2026ApJS..283....6E}, clear the continuum requirement by about two magnitudes and reach the $10^{5}$--$10^6\,{\rm M}_{\sun}$ regime where late-time Pop~III events are most plausible, and the same holds for the other deep NIRCam programmes at comparable depth.
Wide-and-shallow tiers such as COSMOS-Web, at 28\,mag, detect the $10^6\,{\rm M}_{\sun}$ burst in the \ha band, and since the continuum has only to enter as an upper limit the burst is still recovered, in nine of ten mock realisations, although its continuum lies $0.6$\,mag below that depth.
In exposure-time terms (Fig.~\ref{fig:depth}) the continuum requirement of every NIRCam window is met inside a few hours per pointing, for example, $6$\,min in W$z$2.1, $21$\,min in W$z$3.1, $2.7$\,h in W$z$4.4 and $8.6$\,h in W$z$6.1, so for JWST the major binding constraint is survey area, not depth.

\textbf{For \textit{Roman}, the deepest field works.}
The HLWAS deep tier reaches 5$\sigma$ depth of F158/F184/F213\,$=27.5/27.0/25.9$ \citep{ObservationsTimeAllocationCommittee2025arXiv250510574O}, short of the W$z$2.2 requirement in both the continuum band, by $0.9$\,mag, and the \ha band, by $0.4$\,mag, so over that footprint LATED selects only the most massive, youngest, $f_{\rm cov}\simeq1$ events, at $\gtrsim10^7\,{\rm M}_{\sun}$, a search for rare luminous domains rather than typical metal-poor pockets.
The \textit{Roman} eXtreme Deep Field \citep{Yan2026arXiv260908145Y} breaks that trade-off.
RXDF (Section~\ref{sec:combwin}) targets $5\sigma$ depths of $30$ in $R$, $Z$, $Y$, $J$ and $H$ (F062, F087, F106, F129, F158), $29$ in $F$ (F184) and $28$ in $K$ (F213), over a full-depth area of $\sim$700\,arcmin$^2$ inside a total footprint of $>1200$\,arcmin$^2$.
The three bands that define W$z$2.2 are exactly F158, F184 and F213, so RXDF clears the window's continuum requirement by $1.1$\,mag and its \ha requirement by $1.7$\,mag.
It is $2.0$\,mag deeper than the HLWAS deep tier in F184 and $2.1$\,mag deeper in F213, so on the same criterion that gives $\gtrsim10^7\,{\rm M}_{\sun}$ at HLWAS depth the W$z$2.2 mass floor falls by a factor of about six, to $\gtrsim1.6\times10^6\,{\rm M}_{\sun}$.
Its description as unprecedented for such a wide area is relative to depth rather than to area: $\sim$700\,arcmin$^2$ is $0.19$\,deg$^2$, some twenty times the area of a JADES Deep pointing at comparable $\approx$30\,mag depth, but a third of COSMOS-Web and four orders of magnitude below HLWAS.
What RXDF will lack is a \lya parent, since \textit{Roman} carries no band blueward of $0.5\,\mu$m, and here W$z$2.2 is fortunate: its $z=2.09$--2.31 is covered by four Subaru/HSC narrow bands, NB387, NB391, NB395 and NB400 (Table~\ref{tab:windows}), and the field, near the North Ecliptic Pole, is accessible to Subaru.
HETDEX offers a second route: its $z=1.88$--3.52 range spans the window, but its current Spring and Fall footprints do not include the pole.
A dedicated HETDEX pointing over the $\sim$700\,arcmin$^2$ RXDF footprint would deliver a spectroscopic \lya parent sample with the redshifts, and an HSC narrow-band campaign would do the same photometrically; we regard either as the natural companion to the RXDF, and call for them from the community in the upcoming semesters.

\textbf{The \lya depth for the parent sample is not the bottleneck.}
The parent stage admits a source only if it is detected at $5\sigma$ in the matched \lya band ($3\sigma$ for spectroscopy), so the depth that band must reach is the apparent magnitude of the source in it.
A filter fixes its fractional width, so for a narrow band of resolving power $R=\lambda/\Delta\lambda$ holding an unresolved line of rest-frame equivalent width ${\rm EW}_0$, escaping with fraction $f_{\rm esc}$, on a continuum of magnitude $m_{\rm c}$,
\begin{equation}
m_{\rm NB} = m_{\rm c} - 2.5\log_{10}\!\left(1+\frac{f_{\rm esc}\,{\rm EW}_0\,R}{\lambda_{\rm Ly\alpha}}\right),
\label{eq:nbdepth}
\end{equation}
where the $(1+z)$ of the observed equivalent width and of the band width have cancelled: at fixed $R$ the line boost is independent of redshift, and the only redshift dependence left is the distance dimming of the continuum.
Figure~\ref{fig:lyadepth}(a) evaluates this for the reference 2-Myr burst at $10^5$, $10^6$ and $10^7\,{\rm M}_{\sun}$ and at each window's central redshift, for $R=70$, the resolving power of the real \lya narrow bands (HSC NB387, NB816 and NB921 measure $R=68$, $73$ and $68$ from their transmission curves, and ODIN N419, N501 and N673 give $55$, $65$ and $67$).
The tracks are plotted against the rest-frame equivalent width $f_{\rm esc}{\rm EW}_0$, so every one spans the same axis from $f_{\rm esc}=0$ to $1$, and the top axis gives the corresponding $f_{\rm esc}$ for the template's intrinsic ${\rm EW}_0\approx1600$\,\AA.
We adopt $f_{\rm esc}=0.2$ as the fiducial escape fraction (the dashed vertical), representative of \lya-selected emitters at $z\approx2$--4 \citep[per-object typical values of 0.02--0.4 for LAE-selected samples;][]{Matthee2016MNRAS.458..449M,Sobral2017MNRAS.466.1242S}; any other value can be read off the top axis directly.
At $f_{\rm esc}=0.2$ the $10^6\,{\rm M}_{\sun}$ burst requires a narrow-band depth of $24.6$, $25.3$, $25.8$ and $26.3$\,mag in W$z$2.1, W$z$3.1, W$z$4.4 and W$z$6.1 (W$z$2.2 shares the W$z$2.1 value, its \lya parent coming from the same ground-based bands), and $10^5\,{\rm M}_{\sun}$ requires $27.1$, $27.8$, $28.3$ and $28.8$\,mag, in every case $3.3$\,mag brighter than the continuum requirement of the same $10^6\,{\rm M}_{\sun}$ source.
Existing ground-based narrow-band surveys reach $25$--$26$\,mag \citep{Aihara2022PASJ...74..247A,Lee2024ApJ...962...36L}, so they already recover the $10^6\,{\rm M}_{\sun}$ burst in the three lower windows and sit within $0.3$\,mag of it in W$z$6.1, whereas the $10^5\,{\rm M}_{\sun}$ burst lies beyond them everywhere.

For an IFU observation the limit is given by a line flux, and Fig.~\ref{fig:lyadepth}(b) sets it against what VLT/MUSE actually recovers.
Rather than the published point-source limits for an unresolved line, which real \lya emission, extended over several arcseconds, never reaches, we take as the depth tracer the faintest \lya emitters each survey tier detects, measured identically in every tier from the released catalogues as the fifth percentile of the \lya flux of ${\rm S/N}\geq5$ emitters.
That gives $4.6\times10^{-18}$, $1.7\times10^{-18}$, $1.1\times10^{-18}$ and $4.5\times10^{-19}$\,erg\,s$^{-1}$\,cm$^{-2}$ for MUSE-Wide at 1\,h \citep{Urrutia2019A&A...624A.141U}, the HUDF mosaic at 11\,h, UDF-10 at 32\,h and MXDF at 141\,h \citep{Bacon2023A&A...670A...4B}, from 460, 474, 76 and 267 emitters respectively.
A background-limited $t^{-1/2}$ law with its normalisation fitted to the four, 
\begin{equation}
    F=5.4\times10^{-18}\,(t/1\,{\rm h})^{-1/2}\rm \,erg\,s^{-1}\,cm^{-2},
\end{equation}
passes within $0.07$\,dex of every point, and is the $5\sigma$ depth curve drawn through the four tiers in Fig.~\ref{fig:lyadepth}(b).
MUSE covers $4800$--$9300$\,\AA, so \lya enters its range only at $z\geq2.95$ and the two $z\approx2$ windows lie beyond it; against that curve the $10^6\,{\rm M}_{\sun}$ burst at $f_{\rm esc}=0.2$ is reached within $16$\,min in W$z$3.1 and W$z$4.4, inside a 1-h wide survey, and in $1.2$\,h in W$z$6.1, at that survey's edge, whereas the $10^5\,{\rm M}_{\sun}$ burst needs $4.7$, $26$ and $121$\,h, the deep-field regime rising to MXDF depth by W$z$6.1.
The \lya line is therefore the feasible measurement of the three, in imaging and in spectroscopy alike, and what a \lya-selected parent sample is limited by is the redshift coverage of the surveys that provide it, not their depth.


\subsection{Contaminants and failure modes}
\label{sec:contaminants}

A photometric selection lives or dies by its false positives.
This section examines the classes that can enter the LATED selection region, how each enters, and the steps that reduce the contamination risk.
Two points apply throughout.
Every mitigation reduces risk without eliminating it, and no photometric argument alone can confirm a metal-free interpretation.

\subsubsection{Observed galaxies on the selection planes}
\label{sec:jadesplanes}

\begin{figure*}
\includegraphics[width=\textwidth]{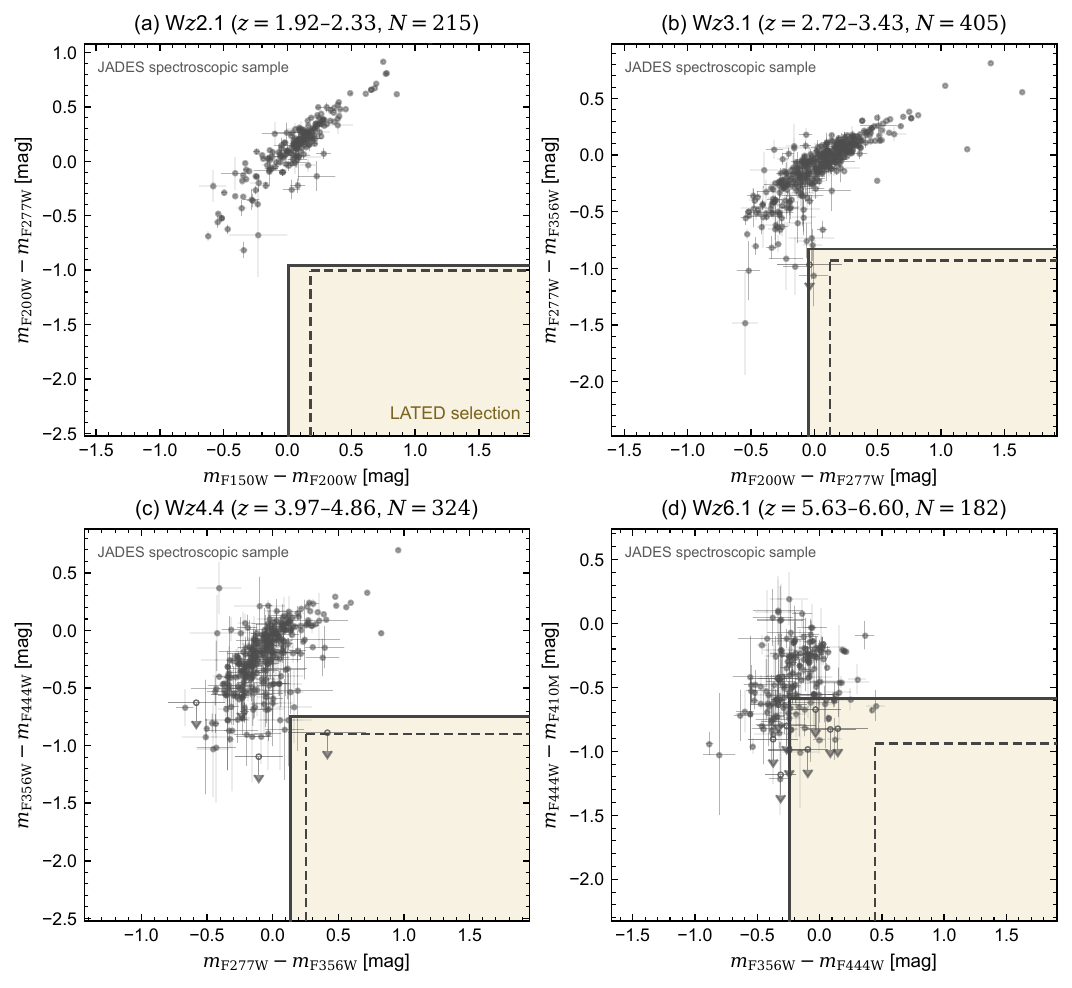}
\caption{The JADES spectroscopic sample on the LATED selection planes of four adopted NIRCam windows, each panel following Figs~\ref{fig:gallery_W$z$21}--\ref{fig:gallery_W$z$61}.
We draw every JADES source with a robust spectroscopic redshift inside the window (DR4 quality A--C) and DR5 CIRC0 photometry in the three selection bands, with the \ha band detected at ${\rm S/N}\geq5$: filled points carry $1\sigma$ error bars, and open points with arrows mark sources whose continuum band enters at its $2\sigma$ limit.
The population lies away from the LATED corner in all four windows.
The few sources overlapping the selection region cannot pass the corresponding redshift-dependent selection criteria at their spectroscopic redshifts (see Fig.~\ref{fig:boundary}).
}
\label{fig:jadesplanes}
\end{figure*}

Before examining individual contaminant classes, we first check where real observed galaxies fall on the adopted selection planes.
Figure~\ref{fig:jadesplanes} shows the JADES spectroscopic sample on the selection planes of four NIRCam windows, with every panel in the same configuration as Figs~\ref{fig:gallery_W$z$21}--\ref{fig:gallery_W$z$61}.
We take all 1126 JADES sources with a robust spectroscopic redshift (DR4 quality A--C) inside a window and DR5 photometry in the three selection bands \citep{Curtis-Lake2026MNRAS.549ag836C,Scholtz2026MNRAS.549ag939S,Johnson2026arXiv260115954J,Robertson2026arXiv260115956R}, require ${\rm S/N}\geq5$ in the \ha band, and substitute the \oiii and continuum bands by their $2\sigma$ limits when undetected.
Our aim is not to select any sources here, but to show the distribution of the real galaxy population relative to our selection corner to check contamination empirically.
We use the \texttt{CIRC0} aperture photometry.
In all four windows, the spectroscopic population forms a sequence locus separated from the target corner.
Although a few sources overlap the selection region on Fig.~\ref{fig:jadesplanes}, they cannot pass the corresponding redshift-dependent selection criteria at their spectroscopic redshifts (Fig.~\ref{fig:boundary}).

\subsubsection{Little red dots}
\label{sec:lrdveto}

Little red dots are the contaminant that most directly defeats a naive \ha-strong, \oiii-weak selection.
They are compact, abundant sources, predominantly faint and often broad-line active nuclei at $z\approx3$--8 \citep{Matthee2024ApJ...963..129M,Greene2024ApJ...964...39G,Kokorev2024ApJ...968...38K,Akins2025ApJ...991...37A}, with comoving number densities of order $10^{-5}\,{\rm Mpc^{-3}}$ at $z\approx4$--5.5 \citep{Matthee2024ApJ...963..129M}.
Their defining feature is a `V-shaped' spectral energy distribution, a blue rest-ultraviolet continuum joined to a red rest-optical continuum at an inflection near the Balmer limit.
Every part of that shape works against us.
The red optical continuum mimics an \ha-band excess, the strong Balmer lines that accompany it reinforce the same excess, and the blue ultraviolet continuum satisfies the very slope flag expected of a metal-poor candidate.

The defence operates at two stages.
Wherever the window carries a continuum band redward of \ha (Section~\ref{sec:contrule}), that band is bright for a red-continuum source and pushes LRDs out of the selection box on the colour plane alone: the noiseless template leakage is almost zero in W$z$2.1, W$z$4.4, and W$z$3.1.
W$z$2.2 and W$z$6.1 are in a much weaker position, because their continuum bands lie between the \oiii and \ha lines.
There, we adopt stricter default criteria to mitigate the LRD contamination, restricting the target locus to young Pop~III templates ($<10$\,Myr; Section~\ref{sec:selcrit}).
Over the noiseless W$z$6.1 grid, the LRD templates go from a pass fraction of $0.004$ under the default 10-Myr boundary to $0.957$ under the extended 20-Myr one.
For any LRD that does pass, a second problem follows: its red continuum does not follow the typical power-law assumption and can manufacture a spurious \oiii/\hb limit (see further discussion in Section~\ref{sec:recovery}).

We therefore raise a photometric \emph{warning} flag on candidate LRDs.
A flagged source remains a LATED candidate, an \ha-strong and \oiii-weak \lya emitter, but is marked as a likely LRD and prioritised accordingly for spectroscopy.
The discriminator is the red rest-optical continuum itself, because LRDs are red redward of the Balmer limit near $3645$\,\AA{} \citep{Setton2025ApJ...995..118S} while young metal-poor systems stay blue throughout.

We deliberately avoid using an $f_\lambda$ continuum slope $\beta_{\rm opt}$ for this purpose.
At $z\approx4$--6 the rest-optical $f_\lambda$ slope is close to zero and therefore weakly diagnostic, and, more seriously, emission-line boosting of broad and medium bands by \oiii, \hb and \ha can fake a red $\beta_{\rm opt}$ for a line-rich galaxy whose continuum is in fact blue, a systematic emphasised by \citet{Hainline2025ApJ...979..138H}.
Instead we use the line-free ultraviolet-to-optical colour $C_{\rm red}$ of equation~(\ref{eq:cred}), whose ultraviolet and optical bands are selected to exclude \lya, \oii, \hb$+$\oiii and \ha$+$\nii at the source redshift.
A warning is raised when the rest-ultraviolet slope is blue and $C_{\rm red}>0.7$\,mag, that is, when the rest-optical continuum is about $1.9$ times brighter than the rest-ultraviolet.
We refer to the colour criterion $C_{\rm red}>0.7$ alone as the \emph{red flag}; the LRD warning is the red flag joined by a blue rest-ultraviolet slope, the v-shape, while a red flag with a red ultraviolet slope marks the monotonically red continuum of a typical dusty or evolved galaxy.
This is similar to the AB-colour form of the canonical LRD selections, the red1 and red2 colour boxes of \citet{Kokorev2024ApJ...968...38K} and the $(\beta_{\rm UV},\beta_{\rm opt})$ v-shape of \citet{Hviding2025A&A...702A..57H} and \citet{Kocevski2025ApJ...986..126K}, with the threshold set below their $\approx$1\,mag cuts precisely because line-free bands carry no emission-line boost.
Its main limitation is that it is a colour-only criterion: compactness, which complements colour in the literature selections \citep{Kokorev2024ApJ...968...38K,Greene2024ApJ...964...39G}, and broad-line spectroscopy remain the decisive independent tests.

\subsubsection{Extreme Balmer-jump objects}
\label{sec:balmerjump}

A distinct and instructive contaminant emerged from the JWST GLIMPSE programme.
GLIMPSE-16043 was selected photometrically as a Pop~III candidate at $z_{\rm phot}\approx6.5$, on strong \ha, no detectable \oiii, and a pronounced Balmer jump \citep{Fujimoto2025ApJ...989...46F}, but NIRSpec follow-up placed it at $z=6.20$ with clearly detected \oiii (\oiii/\hb\,$=1.78$), refuting the metal-free interpretation \citep{Fujimoto2025arXiv251211790F}.
Its photometry is reproduced not by a stellar-plus-nebular population but by an exotic source with an extraordinary Balmer jump of $-1.66$\,mag and EW(\ha)\,$\approx3750$\,\AA{} on a near-undetected continuum, best matched by a hot blackbody in a low-temperature nebula, plausibly a tidal-disruption or microquasar-like object.

Two lessons follow.
The first is that the continuum level and the equivalent widths of a photometrically selected candidate must not be inferred through stellar-population SED fitting, because the true solution can lie far outside the priors such fits impose: the equivalent widths can be much higher, and the continuum much lower, than any template family admits.
GLIMPSE-16043 is the demonstration: its combination of an enormous equivalent width and a near-undetected continuum lies in a corner of parameter space that population synthesis does not reach.
The revised GLIMPSE criteria respond with an explicit continuum-detection floor, and their marginal case A370-z6LAE-2, whose continuum band is detected at only ${\rm S/N}\approx1.4$, too shallow to exclude a strong Balmer jump, shows how weakly the photometry itself constrains that continuum.
LATED internalises the lesson in two places.
The selection carries the continuum only as a measured band, substituted by its $2\sigma$ limit when undetected, so that the \ha excess enters as a conservative lower limit and an unconstrained continuum cannot manufacture a strong-excess detection.
And it is the direct motivation for the deliberately non-physical parametric line-ratio inference of Section~\ref{sec:recovery}, which fits the lines and the continuum freely with no population prior and therefore admits solutions of exactly this kind: applied to GLIMPSE-16043's published photometry alone, its \oiii/\hb posterior ($<2.08$ at $2\sigma$; Section~\ref{sec:recoveryval}) accommodates the spectroscopic value that the Pop~III interpretation had excluded.
The second lesson is that a photometric redshift is a load-bearing input of such a selection.
GLIMPSE-16043 was selected at $z_{\rm phot}\approx6.5$ but sits at $z_{\rm spec}=6.20$, and at these redshifts \ha rides the falling red slope of the F444W bandpass, so the redshift error matters twice.
The transmission at the line is 0.84 of peak at the true redshift but 0.71 at the photometric one, distorting any line flux or equivalent width inferred from the band excess, and the selection thresholds themselves are strongly redshift-dependent exactly where band edges are in play, swinging by up to $0.69$\,mag and changing sign across W$z$6.1 (Section~\ref{sec:selcrit}), so a redshift error evaluates the right colours against the wrong boundary.
A NIRCam-only selection must carry this uncertainty through every line assignment, transmission correction, and threshold, whereas the \lya parent stage hands every LATED candidate a spectroscopic redshift before the colour criteria are evaluated, the photometric redshift entering only as a posterior consistency flag (Table~\ref{tab:score}).
The extreme-Balmer-jump degeneracy itself, however, is shared by any \oiii-weakness selection and is broken only spectroscopically.

\begin{figure*}
\includegraphics[width=\textwidth]{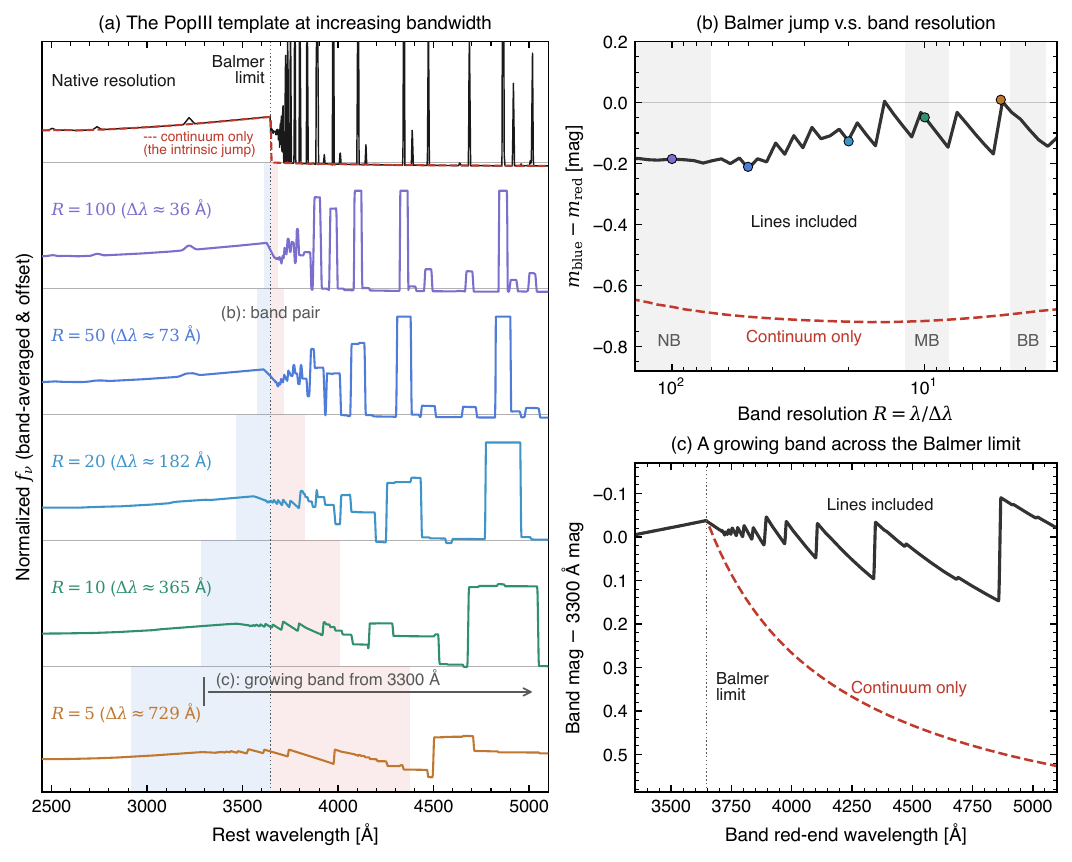}
\caption{Why band-integrated photometry cannot see the Balmer jump.
\emph{(a)} The Pop~III template of Fig.~\ref{fig:windowseds} around the Balmer limit in $f_\nu$, with the Balmer series restored beyond the grid's H14 truncation (Case~B emissivities scaled to the template's \hb; \citealt{Storey1995MNRAS.272...41S}), so orders up to $n\approx30$ remain discrete while higher orders merge toward the edge.
Rows show the native resolution and band averages at $R=100$ down to 5 (offset vertically); on each row the blue and red spans mark the abutting band pair used in (b), and the arrow marks the growing band of (c).
The dashed curve removes the emission lines, isolating the intrinsic free-bound discontinuity of a factor $\approx$1.8.
\emph{(b)} The apparent jump measured with two abutting bands touching at the limit, versus band resolution $R=\lambda/\Delta\lambda$: the continuum-only jump (dashed) stays at $\approx$0.7\,mag, while with the lines included it shrinks from $-0.19$\,mag at $R=100$, a real residual set by the sub-Saha high-order populations, to zero at broadband, where the cascade's $\approx$2.4 Balmer-line photons per Balmer-continuum photon refill the red band.
Shaded spans mark ground narrow-band and JWST medium- and wide-band resolutions.
\emph{(c)} A single band with its blue end fixed and its red end swept across the limit: the continuum-only band fades with a kink at the limit, whereas with the lines included the magnitude evolves smoothly through it.
The raw jump is thus a spectroscopic observable, which is why LATED carries it as the blue end of the wide-baseline colour $C_{\rm red}$ instead.}
\label{fig:balmerband}
\end{figure*}

\textbf{The Balmer jump} deserves a separate treatment, because it is the diagnostic most often proposed for nebular-dominated systems and the one whose photometric behaviour is least intuitive.
Recombinations to $n=2$ emit bound-free photons at $\lambda\leq3646$\,\AA, so a nebular-continuum-dominated spectrum is \emph{brighter blueward} of the limit, the Balmer jump in emission \citep{Schaerer2002A&A...382...28S,Cameron2024MNRAS.534..523C}, whereas an evolved population shows the opposite break.
It is a general recombination feature of starbursts younger than a few Myr, and medium-band imaging has been used to search for it directly over $1.5<z<8.5$ \citep[e.g.,][]{Trussler2026MNRAS.549ag788T}.
Broad-band photometry, however, is almost blind to the Balmer jump, as Fig.~\ref{fig:windowseds} shows.
The obstacle is that the jump and the Balmer emission lines are two channels of the same recombination flow.
In Case~B every recombination reaching $n=2$ emits exactly one photon adjacent to the edge, a Balmer-\emph{continuum} photon when the capture lands on $n=2$ directly and a Balmer-\emph{line} photon when it arrives by cascade from $n\geq3$, with the two channels carrying the photon budget in a ratio of about $1:2.4$ at $10^4$\,K \citep{Osterbrock2006agna.book.....O,Luridiana2015A&A...573A..42L}.
A photometric band straddling the limit therefore re-collects, on the red side, the very line photons whose absence from the continuum defines the jump.
We quantify this with a controlled experiment on the Pop~III template (Fig.~\ref{fig:balmerband}).

The template is prepared in two steps.
Because the model grid truncates the Balmer series at H14, we first restore the $n=15$--2000 orders with Case~B emissivities \citep{Storey1995MNRAS.272...41S} computed with PyNeb \citep{Luridiana2015A&A...573A..42L} and scaled to the template's measured \hb flux, at $T_e=2\times10^4$\,K and $n_e=10^{2}$\,cm$^{-3}$, the emissivity ratios shifting by less than 2 per cent over $T_e=1$--$3\times10^4$\,K.
These are deposited at their Rydberg wavelengths and convolved with a single 1.5-\AA{} kernel, so that orders up to $n\approx30$, about the highest separable in galaxy spectra before velocity merging, remain discrete while the higher orders blend into the pseudo-continuum approaching the edge.
Second, a line-free counterpart is built by interpolating the continuum across all H and He lines, isolating the intrinsic free-bound discontinuity, which is a factor of about 1.8, or 0.7\,mag.

The result is a strong, bandwidth-dependent suppression.
The continuum-only jump is about 0.7\,mag at every band resolution, while with the lines included the apparent jump is $-0.19$\,mag for a narrow-band pair ($R=100$), $-0.05$\,mag at medium-band widths ($R=10$), and $+0.01$\,mag at broad-band widths ($R=5$).
The narrow-band residual is physical rather than incomplete bookkeeping: at nebular densities the high-order populations sit below their Saha values, so the merged-line pseudo-continuum delivers only 0.6--0.9 of the free-bound emissivity at the edge, the fraction rising with $T_e$ and $n_e$, and a narrow band abutting the limit collects exactly that deficit.
At wider bandwidths the cascade's photon budget takes over, the roughly 2.4 Balmer-line photons per Balmer-continuum photon refill the red band, and the jump is erased.
A growing band shows the same physics from another angle, its magnitude evolving smoothly through the edge with no feature at 3646\,\AA{} as each Balmer line pulls it back toward its starting flux, whereas the continuum-only band kinks at the limit and fades by 0.5\,mag.

\begin{figure}
\includegraphics[width=\columnwidth]{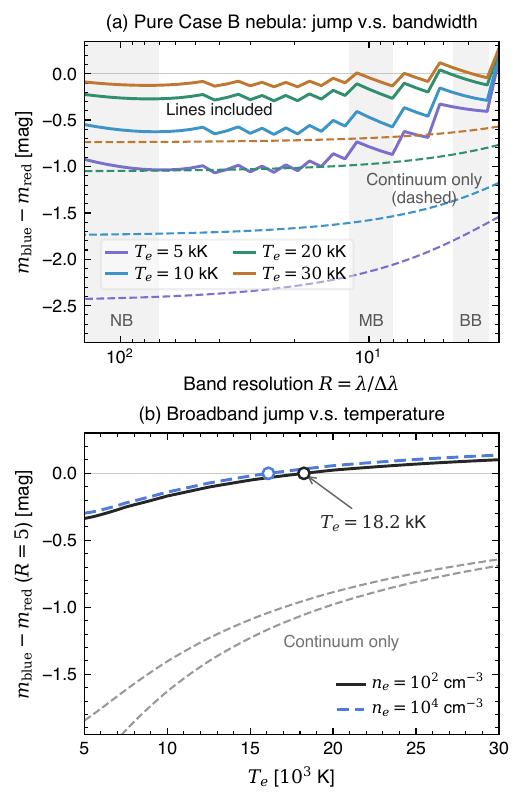}
\caption{The template-independent PyNeb Balmer-jump test: a pure-hydrogen Case~B nebula (free-bound, free-free and two-photon continua; line emissivities to $n=50$ plus the asymptotic tail; \citealt{Storey1995MNRAS.272...41S}), measured with the abutting band pair of Fig.~\ref{fig:balmerband}b.
\emph{(a)} The apparent jump versus band resolution at $T_e=5$--30\,kK ($n_e=10^{2}$\,cm$^{-3}$), lines included (solid) and continuum only (dashed): the intrinsic jump deepens as $T_e^{-3/2}$ toward low temperature, while line refilling closes it toward broadband.
\emph{(b)} The broadband ($R=5$) jump versus $T_e$: the colour crosses zero at $T_e=18.2$\,kK ($16.1$\,kK at $n_e=10^{4}$\,cm$^{-3}$), so the warm nebulae of extremely metal-poor systems show no broadband jump even before stellar continuum dilution, whereas cool, metal-rich nebulae retain a residual.}
\label{fig:balmerpyneb}
\end{figure}

A template-independent PyNeb calculation confirms this picture and fixes its temperature dependence.
For a pure-hydrogen Case~B nebula built entirely from PyNeb, with free-bound, free-free and two-photon continua and the \citet{Storey1995MNRAS.272...41S} line emissivities to $n=50$ plus the asymptotic tail beyond, the same abutting-band measurement gives a continuum-only jump growing from $-0.7$ to $-2.4$\,mag as $T_e$ falls from $3\times10^4$ to $5\times10^3$\,K, the $T_e^{-3/2}$ free-bound step that underlies the Balmer-jump temperature diagnostic, while the line emissivities fall only as about $T_e^{-0.9}$.
Line refilling therefore overtakes the shrinking step as temperature rises, and the broad-band blue-minus-red colour crosses zero at $T_e\approx1.8\times10^4$\,K for $n_e=10^{2}$\,cm$^{-3}$, or $1.6\times10^4$\,K at $10^{4}$\,cm$^{-3}$ (Fig.~\ref{fig:balmerpyneb}).
For the warm nebulae of extremely metal-poor systems, with $T_e\gtrsim1.5\times10^4$\,K, the broad-band jump is thus erased to within about 0.1\,mag by the hydrogen spectrum alone, before any stellar continuum dilutes the step further, whereas a cool, metal-rich nebula would retain a residual of $-0.34$\,mag at $5\times10^3$\,K.

The raw Balmer jump is therefore a spectroscopic observable, recoverable only by placing continuum windows \emph{between} the lines.
The robust photometric form of the criterion is not to straddle the limit at all, but to compare line-free continuum windows far from it on either side, which is exactly the colour $C_{\rm red}$ of equation~(\ref{eq:cred}): for a nebular-dominated spectrum the rest-ultraviolet continuum is brighter than the rest-optical continuum, by about 0.7\,mag in $f_\nu$ for the template of Fig.~\ref{fig:balmerband}.
This wide-baseline comparison preserves the sign of the jump while avoiding the line pileup entirely, and it is the reason the Balmer jump enters LATED only through the blue end of $C_{\rm red}$ and never as a straddling medium-band colour.
Its price is that the colour folds the ultraviolet slope and the two-photon continuum in with the jump, the degeneracy behind the GLIMPSE-16043 misclassification.
The empirical results point the same way.
In the medium-band search of \citet{Trussler2026MNRAS.549ag788T}, nebular-dominated candidates, identified through a deficit in the rest-optical continuum, make up about 10 per cent of galaxies at $z\sim6$ falling to about 3 per cent at $z\sim2$, and carry median EW(\ha)\,$=1567$\,\AA{} and EW(\oiii$+$\hb)\,$=2244$\,\AA, the extreme line emission LATED also selects on.
Yet they find that nebular-dominated galaxies do not necessarily show the largest Balmer jumps, nor the largest ionising photon production efficiencies or the reddest ultraviolet slopes, and conclude that continuum spectroscopy is ultimately required.
That is the same conclusion this section reaches from the radiative physics, and therefore LATED carries the Balmer jump as one end of a corroborating colour rather than as a criterion.

\subsubsection{Pristine black holes}

A pristine or metal-poor accreting black hole is the contaminant most readily mistaken for a metal-free stellar population, and separating the two is the express goal of \citet{Nakajima2022MNRAS.513.5134N}.
We propagate their full PBH grid, 216 \textsc{cloudy} spectra spanning $T_{\rm bb}=5\times10^4$--$2\times10^5$\,K, accretion slope $\alpha=-1.2$ to $-2.0$, $Z=0$--$10^{-3}$ and $\log U=-3.0$ to $-0.5$, through every window.
None of them enters the selection region, and the mean bright-end selection probability is zero in all five windows.
The reason is that the blue accretion continuum suppresses the \ha equivalent width and places these models near $y\approx0$, far above the \ha-strength boundary $y\leq y_{\rm max}(z)$.
LATED therefore rejects PBHs photometrically, by the same continuum-relative \ha criterion that rejects red-continuum interlopers, and so separates accreting black holes from metal-free stars without spectroscopy, the discrimination \citet{Nakajima2022MNRAS.513.5134N} draw from the line spectrum.
What remains is a PBH embedded in a host with a strong \emph{stellar} \ha equivalent width, a configuration that \heii\,$\lambda1640$ or $\lambda4686$ strength and line widths would resolve spectroscopically.

\subsubsection{Ordinary extreme emission-line galaxies}

Galaxies with EW(\oiii$+$\hb)\,$\gtrsim1000$\,\AA{} are common at high redshift \citep[e.g.,][]{Boyett2024MNRAS.535.1796B} and share the \ha strength of LATED targets, so they are the most numerous class that reaches the neighbourhood of the selection region.
They are excluded by the O/H cut, which is the criterion that separates \oiii-weak from \oiii-strong systems, and no PBH template enters the selection region in any window.
Their route in is photometric scatter in shallow data rather than any overlap of loci.
The residual risk is an extreme emission-line galaxy with an intrinsically unusual \oiii/\ha ratio lying outside the gridded parameter space.

\subsubsection{Low-redshift interlopers}

An \oii, \oiii or \ha emitter at low redshift can masquerade as \lya in the selection band.
Narrow- and intermediate-band surveys control this with broad-band colours and photometric redshifts, and integral-field profiles largely eliminate it.
The mock propagates the class end-to-end, using FSPS templates placed at the \oii-interloper redshifts and selected through their genuine \oii-driven band excess, and finds no leakage through the adopted LATED colour cuts in the noiseless limit, and an end-to-end selection probability of 0.5 per cent once fiducial photometric noise is included.
What brings these systems close to the boundary is their near-infrared recombination lines, the Paschen series, which mimic weak excesses.
The residual risk is a low-redshift system with a coincidental line pair and a pathological photometric-redshift solution.

\subsubsection{Active galactic nuclei}

Narrow-line active nuclei produce strong recombination lines and can show weak \oiii at high density or under unusual ionisation.
The non-AGN flag of Table~\ref{tab:score} records an X-ray, radio, or mid-infrared counterpart or detected variability; \oiii-strong nuclei are excluded by the EW(\ha)--O/H diagram itself, and the broad-line little red dot population is covered by the $C_{\rm red}$ warning above.
Heavily obscured or radio-quiet, X-ray-faint nuclei remain, and these have to be screened spectroscopically through line widths and high-ionisation lines.

\subsubsection{Other stellar and nebular degeneracies}

Several further classes reach the selection region through genuine physics rather than through noise, and photometry alone cannot separate them.
Shocks produce hard spectra and can enhance \heii without any metal-free stars \citep{Thuan2005ApJS..161..240T,Plat2019MNRAS.490..978P}; photometry cannot distinguish shock ionisation, so the discrimination is deferred to spectroscopic line ratios \citep{Allen2008ApJS..178...20A}, and the risk is real but small because shocked systems with weak \oiii and strong \ha are rare.
Wolf--Rayet phases produce \heii and high equivalent widths in metal-enriched systems \citep{Schaerer1998ApJ...497..618S,Crowther2007ARA&A..45..177C,Shirazi2012MNRAS.421.1043S}, but they remain \oiii-strong and are excluded by the O/H cut, and broad \heii separates them cleanly in follow-up.
Density-bounded nebulae depress low-ionisation lines \citep{Jaskot2013ApJ...766...91J,Zackrisson2013ApJ...777...39Z}, and extreme ionisation parameters shift \oiii/\hb at fixed metallicity \citep{Nakajima2014MNRAS.442..900N,Nakajima2022MNRAS.513.5134N}, so such systems can scatter toward the selection region from the metal-poor side; there is no photometric mitigation, because this is a physical degeneracy rather than a measurement failure, but these objects are interesting in their own right as extreme Lyman-continuum-leaker analogues \citep{Izotov2018MNRAS.473.1956I,Flury2022ApJS..260....1F}, so selecting them is a feature with a different label rather than a pure failure.
Metal-rich, low-excitation starbursts show weak \oiii for the opposite reason, since \oiii/\hb also declines along the high-metallicity branch of the strong-line calibrations \citep{Pagel1979MNRAS.189...95P,Maiolino2019A&ARv..27....3M}, but they fail the EW(\ha) threshold and the blue-continuum flag, are usually bright in \oii, and usually fail the \lya pre-selection as well.

\subsubsection{Measurement and systematic failure modes}

The remaining failure modes are properties of the data rather than of the sources; each is listed here with its mitigation.

\textbf{Blending with unrelated neighbours.}
A \lya source blended with an unrelated neighbour can acquire spurious colours in ground-matched apertures.
This is common in lensing-cluster fields, where foreground cluster galaxies blend with the background sources.
High-resolution morphology, optimised background subtraction, matched-aperture photometry and visual inspection address this, recorded through the counterpart-quality flag of Table~\ref{tab:score}.

\textbf{Variability between epochs.}
The \lya imaging and the rest-frame optical imaging may be separated by years or decades, over which active nuclei and transients can vary and so fabricate or erase an excess.
Multi-epoch variability flags address this, and future optical imaging such as Rubin/LSST will settle it for the ground-based parents.
Spectroscopy of the \lya line can also address this.

\textbf{\lya astrometric offsets.}
Resonant scattering shifts, broadens and spatially redistributes \lya, so offsets between the \lya and continuum centroids complicate cross-matching.
Our generous 1\arcsec\ matching radius, together with de-blending, partly addresses this.

\textbf{Redshift error at window edges.}
A source near a filter slice edge can have a \lya-inferred redshift wrong by enough to move \ha partially out of its band.
The resulting window-edge equivalent-width bias should be modelled in the selection function.

\subsubsection{Comparison with published photometric selections}
\label{sec:selcompare}

\begin{figure*}
\includegraphics[width=\columnwidth]{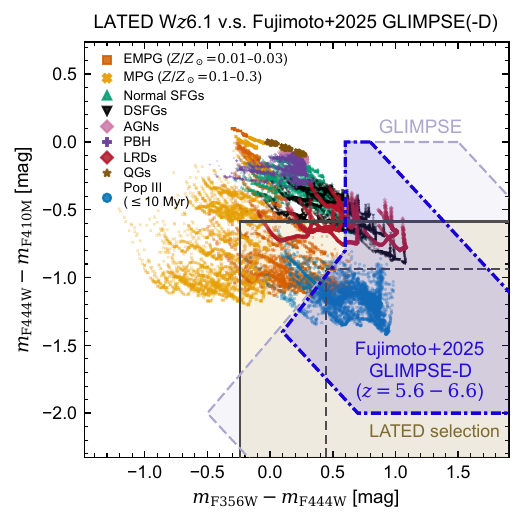}
\includegraphics[width=\columnwidth]{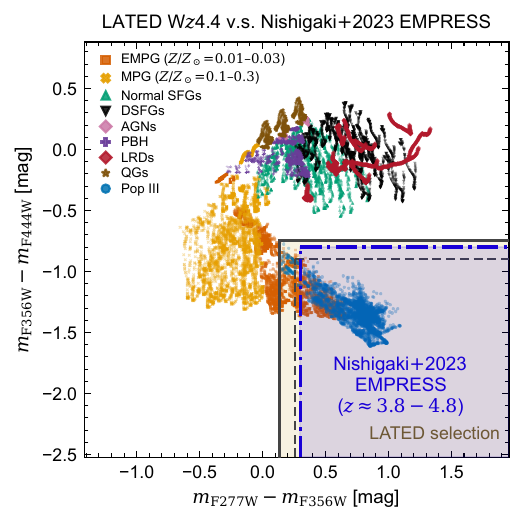}
\caption{The LATED selection compared with published photometric selections.
In both panels the model library is drawn by class (inset legend) and the LATED region is the \emph{tan shaded} corner, with a \emph{solid} edge at the loosest and a \emph{short-dashed} edge at the tightest in-window redshift of the released boundary.
\emph{Left}: the W$z$6.1 plane, $x=m_{\rm F356W}-m_{\rm F444W}$ against $y=m_{\rm F444W}-m_{\rm F410M}$ (F356W: \oiii$+$\hb; F444W: \ha; F410M: continuum; $z\approx5.6$--6.6), with both generations of the GLIMPSE Pop~III selection: \emph{lavender dashed} for GLIMPSE, equation~(4) of \citet{Fujimoto2025ApJ...989...46F}, and \emph{blue dash-dotted} for GLIMPSE-D, equation~(3) of \citet{Fujimoto2025arXiv251211790F}, each lightly shaded in its own colour.
Both are mapped exactly into this plane via $x=(m_{\rm F356W}-m_{\rm F410M})+(m_{\rm F410M}-m_{\rm F444W})$ and $y=-(m_{\rm F410M}-m_{\rm F444W})$; each GLIMPSE selection additionally applies two colour diagrams involving F200W and F277W, a Pop~III photo-$z$, and SED criteria not representable here.
The three regions agree on the Pop~III locus and on excluding the $Z{=}0.1\,Z_{\sun}$-burst locus.
\emph{Right}: the W$z$4.4 plane with the \citet{Nishigaki2023ApJ...952...11N} $z\approx3.8$--4.8 EMPG colour cuts in \emph{blue dash-dotted}: their ${\rm F356W}-{\rm F444W}<-0.8$ and ${\rm F277W}-{\rm F356W}>0.3$ (their equations 8--9) are axis-aligned in this plane with the same band roles, and their corner falls between the loosest and tightest LATED corners.
}
\label{fig:fujicomp}
\label{fig:nishicomp}
\end{figure*}

Because the same degeneracy is shared across the field, it is worth placing the LATED region beside the published selections directly.
Figure~\ref{fig:fujicomp} (left panel) does this for W$z$6.1 in the LATED EW(\ha)--O/H diagram, overlaying both generations of the GLIMPSE criteria.
Each generation intersects three colour--colour diagrams with SED-based criteria, and we draw the one whose axes are linear combinations of the LATED axes and therefore map into this plane exactly, taking the published GLIMPSE \citep{Fujimoto2025ApJ...989...46F} and GLIMPSE-D \citep{Fujimoto2025arXiv251211790F} criteria.
All three regions enclose the Pop~III target locus, holding 93, 91 and 90 per cent of it inside the LATED, GLIMPSE and GLIMPSE-D median boxes respectively and 100 per cent under LATED's per-redshift boundaries, and all three exclude the $Z{=}0.1\,Z_{\sun}$-burst locus entirely.
The constructions differ in kind.
The LATED region is a closed two-sided box read off the model envelope with per-redshift thresholds, whereas the GLIMPSE regions are hand-drawn polygons supplemented by two further colour diagrams involving F200W and F277W, a Pop~III-template photometric redshift, and $\chi^2$-based SED criteria.
Crucially, the boxes part company on the little red dot locus.
We note that both GLIMPSE third-diagram regions admit LRD templates over $z=5.6$--6.6, and their remaining criteria can reject part of these, whereas the LATED young-locus box excludes them entirely, which is the design intent of the LATED age-capped envelope.

An independent convergence check comes from the $z\approx4$--5 extremely metal-poor galaxy search of \citet{Nishigaki2023ApJ...952...11N}, which overlaps W$z$4.4 with the same band roles, \ha in F356W, \hb$+$\oiii in F277W, and the continuum in F444W.
Their two colour cuts, ${\rm F356W}-{\rm F444W}<-0.8$ and ${\rm F277W}-{\rm F356W}>0.3$, are axis-aligned cuts in exactly the adopted W$z$4.4 plane, and Fig.~\ref{fig:nishicomp} (right panel) shows their corner falling between the loosest and tightest LATED corners.
A hand-tuned search and a model-envelope construction therefore converge on the same region from independent starting points.
What LATED adds is the \lya redshift-resolved thresholds.
We note that 14 of their 17 candidates show (\hb$+$\oiii)/\ha ratios consistent with negligible \oiii at their $z_\mathrm{phot}$, which they highlight as Pop~III-like.
But their search rests on photometric redshifts alone, which exposes it to the second lesson of Section~\ref{sec:balmerjump}.
The band ladder that makes F356W the \ha band of W$z$4.4 also makes it the \oiii$+$\hb band of W$z$6.1: a strong \oiii emitter at $z\approx5.3$--6.9 reproduces the same F356W excess, its \ha moving into or beyond F444W, and so masquerades as a $z\approx4.4$ \ha emitter, a contaminant class those authors themselves discuss.
That failure mode has since materialised: the one candidate presented in their paper (their Figure~11) at $z_\mathrm{phot}=4.6$ selected from JWST photometry has been confirmed by JWST/NIRSpec as exactly such a $z_\mathrm{spec}=6.73$ \oiii emitter (priv.\ comm.), the F606W detection that anchored its low photometric redshift having come from a cross-matching error.
The \lya parent stage of LATED can remove this degeneracy by construction, because the redshift is fixed spectroscopically or given by NB/MB, before the colours are evaluated.

\section{Photometric emission line inference}
\label{sec:recovery}

The selection in Section~\ref{sec:algorithm} identifies candidates but does not measure their line fluxes or line ratios.
A source passes the O/H cut when its \oiii band is faint relative to its \ha band, which can happen either because the source is genuinely metal-poor or because its continuum is too faint for the colour to carry information.
Two candidates can therefore pass the same cuts while differing by an order of magnitude in \oiii/\hb, and the selection cannot distinguish them.

In this section we use the three selection bands to constrain \oiii/\hb quantitatively, a number that converts directly into a metallicity.
As the fiducial conversion we adopt the empirical calibration of \citet{Isobe2026arXiv260611345I}, anchored on \oiii\,$\lambda4363$ electron temperatures down to $12+\log({\rm O/H})=7.0$, or about 2 per cent solar.
At that metal-poor end the calibration reduces to a rough rule: with $R3=$ \oiii$\lambda5007$/\hb, the ratio itself reads as a metallicity in units of per cent solar, $Z/Z_{\sun}\approx R3$ per cent, so $R3=1$ corresponds to $\approx1$ per cent solar and an upper limit on the ratio becomes an upper limit on the abundance.
For each candidate the method returns either a measured ratio or an upper limit, with an uncertainty attached, so that the follow-up queue can be ranked by the strength of each constraint and a deep limit can be distinguished from a weak one (Section~\ref{sec:ranking}).
The method makes no assumption about the stellar population, age, or star-formation history, and every inferred quantity can be traced back to a band excess.

\begin{figure*}
\centering
\includegraphics[width=1.1\columnwidth]{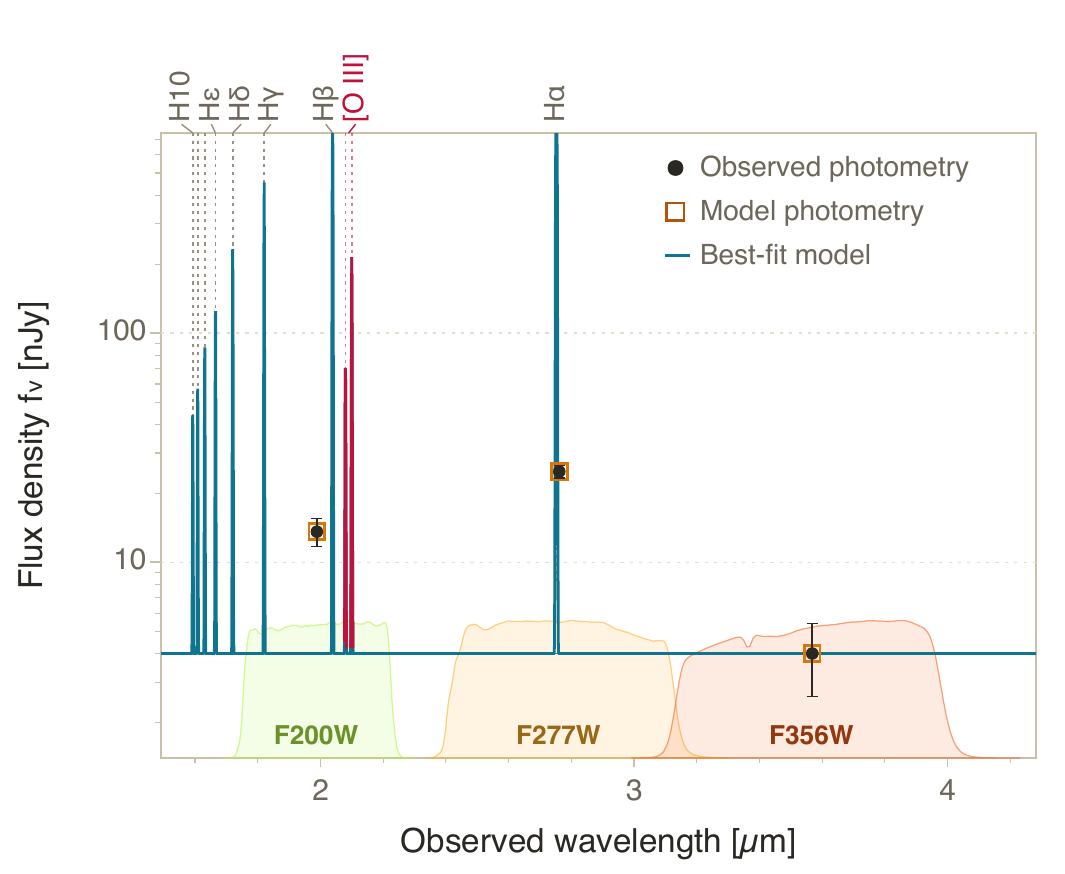}\hfill
\includegraphics[width=0.9\columnwidth]{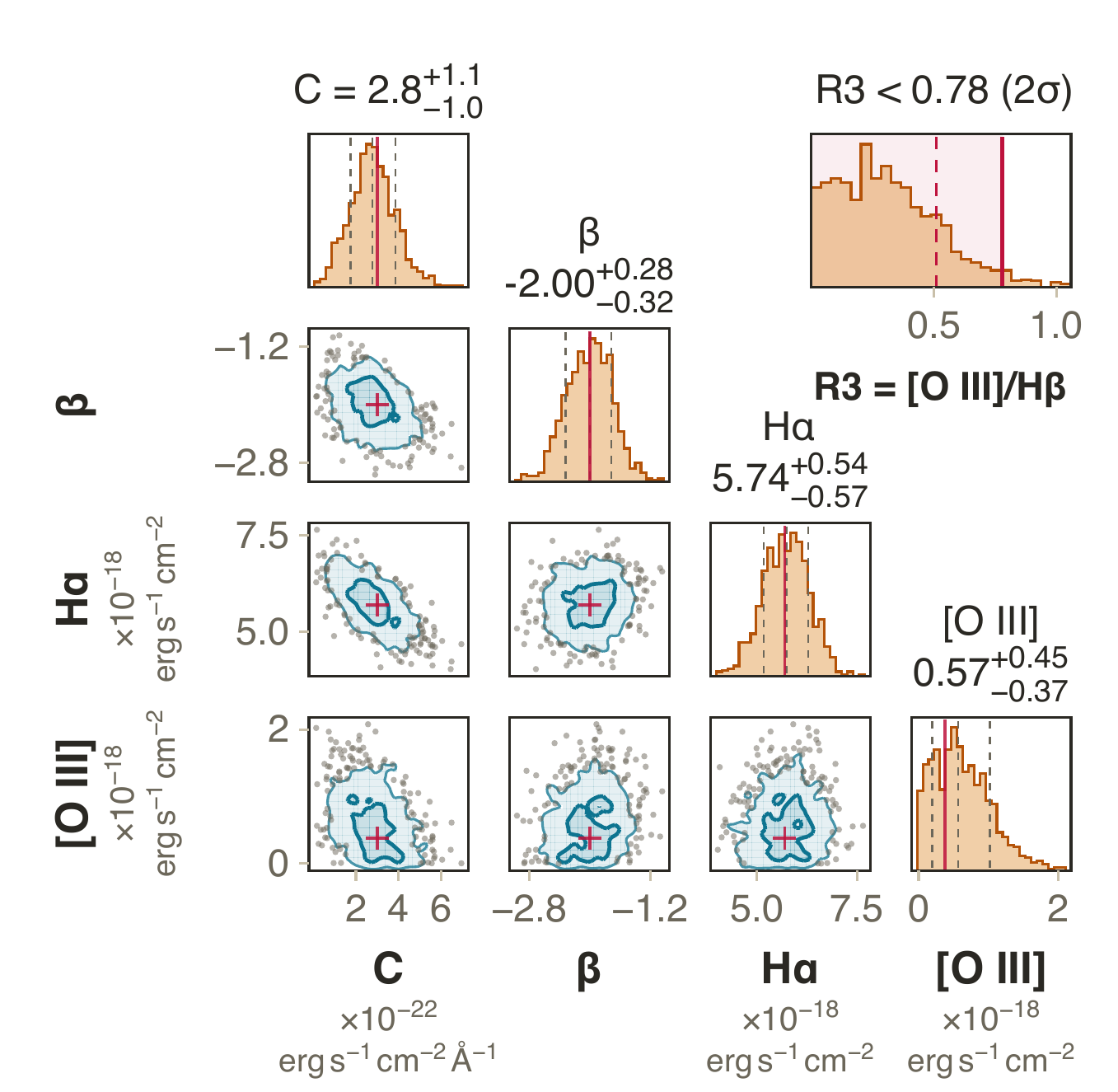}
\caption{Output of the released package \texttt{lated}, exported directly from the web application, for one of its worked examples: the published photometry of the metal-free candidate CR3 at $z=3.19$ \citep{Cai2025ApJ...993L..52C} in its three W$z$3.1 selection bands, with the slope fixed at $\beta=-2$.
\emph{Left}: the best-fit model drawn through the filter transmission curves, with the tied Balmer series at its observed wavelengths, and the model re-integrated through each band (open squares), which land on the observed photometry (black points).
The \oiii\,$\lambda\lambda4959,5007$ doublet is drawn in red.
\emph{Right}: the corner plot of the free parameters for the same fit, the continuum amplitude, the slope under its assumed prior, and the two free line fluxes, with the reported \oiii/\hb posterior inset in the top-right corner.
For a limit, the $2\sigma$ allowed range down to zero is shaded; contours enclose $1\sigma$ and $2\sigma$ in two dimensions of the Monte-Carlo points, dashed lines mark the 16th, 50th and 84th percentiles, and the red crosshairs mark the best fit.
The \oiii\ flux is undetected here, so the ratio is returned as \oiii/\hb\,$<0.78$ at $2\sigma$, the limit of Section~\ref{sec:recoveryval}, consistent with the source's spectroscopic \oiii/\hb\,$<0.89$.
}
\label{fig:recoveryapp}
\end{figure*}

The full procedure is released as a standalone, installable Python package dubbed \texttt{lated}\footnote{\url{https://github.com/lmytime/lated}}, which can be installed from PyPI\footnote{\url{https://pypi.org/project/lated/}} by \texttt{pip install lated}.
It also provides the command-line interface \texttt{lated serve}, which launches a web application for interactive inference; Fig.~\ref{fig:recoveryapp} shows its figure output for a worked example.

\subsection{The model and the fit}
\label{sec:inferencemodel}

Given three or more bands and the slice redshift, we fit
\begin{equation}
f_\lambda \;=\; C\,(\lambda/\lambda_0)^{\beta} \;+\; \textstyle\sum_l F_l\,\delta(\lambda-\lambda_l),
\label{eq:inference}
\end{equation}
propagated through the full filter curves, with a free continuum $(C,\beta)$ and non-negative line fluxes $F_{{\rm H}\alpha}$ and $F_{[{\rm O\,III}]\lambda5007}$.
The tied components matter.
\oiii\,$\lambda4959$ rides on $\lambda5007$ at the fixed atomic ratio of $2.98$, and the \emph{full Case~B Balmer series} \citep{Osterbrock2006agna.book.....O} rides on \ha: \hb at $1/2.86$, H$\gamma$ at 0.468 of \hb, H$\delta$ at 0.259, H$\epsilon$ at 0.159, and the higher orders down to H10, each placed at its own observed wavelength and attenuated differentially under a Calzetti dust law \citep{Calzetti2000ApJ...533..682C}.
The contribution from higher-order terms cannot be ignored.
For example, at $z=3$, H$\gamma$ falls \emph{inside the \oiii band} (F200W), so a model without it misattributes up to $0.47\times F_{{\rm H}\beta}$ of Balmer flux to \oiii and biases the inferred ratio low, and the same applies to H$\delta$ and H$\epsilon$ in narrower or bluer band sets.
Optional free components handle other lines including \oii, [Ne~{\sc iii}], \hei, and a tied \nii$+$[S~{\sc ii}] fraction for metal-rich contaminant fitting.

With more than three bands the slope is grid-searched.
With only the three LATED selection bands it is degenerate with the continuum and line amplitudes, three data against four unknowns, so we fix it at the metal-poor expectation $\beta=-2$ and marginalise the posterior over a Gaussian slope prior of width $0.3$, the expected spread of the targets' $\beta$, so that the fixed-slope systematic enters the error budget instead of leaving the intervals over-confident.
At a fixed slope the model of equation~(\ref{eq:inference}) is linear in its amplitudes, so the fit reduces to a single non-negative least-squares solve.
Each band's model value is the photon-weighted mean flux density of equation~(\ref{eq:inference}) through that filter's measured transmission curve $T(\lambda)$: a delta-function line at observed wavelength $\lambda_l(1+z)$ contributes $T(\lambda_l(1+z))\,\lambda_l(1+z)\,/\!\int\!T\lambda\,{\rm d}\lambda$, and the power-law continuum contributes $\int\!(\lambda/\lambda_0)^{\beta}\,T\lambda\,{\rm d}\lambda\,/\!\int\!T\lambda\,{\rm d}\lambda$, matching the synthetic-photometry convention of Section~\ref{sec:synthphot}.
The design matrix therefore carries one column per free amplitude, the continuum $C$ and one column per free line into which its tied partners are folded at their effective ratios, so that the Case~B Balmer chain scaled onto \ha and \oiii\,$\lambda4959$ scaled onto $\lambda5007$ each enter their parent's column at their own observed wavelength and differential dust factor.
The amplitudes are then obtained by error-weighted non-negative least squares, whose non-negativity constraint enforces physical line fluxes and is what returns an undetected line as an upper limit rather than a spurious negative flux.
When the slope is free it is profiled over a grid of $61$ values spanning $-3.5\le\beta\le1$, retaining the solution of smallest weighted residual; when it is fixed the grid collapses to the single value $\beta=-2$.

Uncertainties come from sampling the truncated-Gaussian posterior of the amplitudes, with a flat prior on $F\geq0$, by Gibbs sampling, with the slope drawn from its profile weight over the grid when free or from its Gaussian prior when fixed, and the slice redshift jittered by $\sigma_z$.
We quote the $2.5$, $16$, $50$, $84$ and $97.5$th percentiles of the resulting amplitude and ratio draws, so that the photometric, slope and redshift systematics enter together.
We adopt this posterior sampler rather than the bootstrap, which resamples the photometry and refits and which reproduces the legacy pipeline, because the bootstrap can collapse against the $F\geq0$ boundary and return upper limits on absent lines tighter than the photometric depth supports; the two agree for well-detected lines and differ only in the limit regime.
The reported ratio is $R3=F_{[{\rm O\,III}]\lambda5007}/F_{{\rm H}\beta}$ with $F_{{\rm H}\beta}=F_{{\rm H}\alpha}/2.86$, bounded to suppress the noise tail of a vanishing denominator, and when the \oiii amplitude is not detected we report the ratio as its $97.5$th-percentile ($2\sigma$) upper limit rather than as a detection.
A line counts as detected when the 16th percentile of its flux posterior exceeds one quarter of its 84th:
\begin{equation}
Q_{16}(F_l) \;>\; \tfrac{1}{4}\,Q_{84}(F_l),
\label{eq:detection}
\end{equation}
where $Q_{p}(F_l)$ denotes the $p$th percentile of the posterior of the line flux $F_l$.
This is a scale-free test on the posterior's shape: the $F\geq0$ draws are strictly positive, so a simple positivity cut can never fail, whereas the quantile-ratio test of equation~(\ref{eq:detection}) flips at an effective significance of $\sim1.5\sigma$ for the zero-truncated posterior ($1.7\sigma$ in the Gaussian limit).
The same test applied to \ha decides the denominator, since \hb is tied to it: a detected \oiii over an undetected \ha is instead reported as a lower limit, and when neither line passes, the ratio is still quoted as an upper limit wherever its own posterior carries information.

\begin{figure*}
\includegraphics[width=\textwidth]{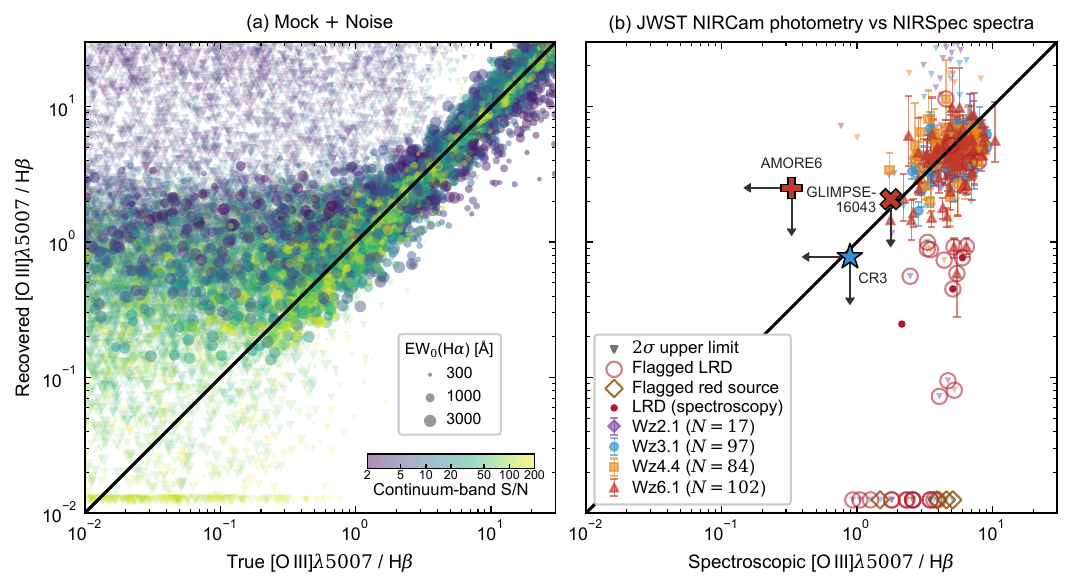}
\caption{Validation of the photometric \oiii/\hb inference.
\emph{Panel (a)}: recovered versus true ratio for the mock test; marker colour encodes the continuum-band S/N, marker size EW$_0$(\ha), and sources returned as limits are drawn faintly at their $2\sigma$ bounds.
\emph{Panel (b)}: the real-data test, the ratio inferred from the three JWST NIRCam selection bands versus the NIRSpec spectroscopic ratio, per window: 300 detections (filled symbols) and 190 upper limits (downward triangles at their $2\sigma$ bounds; 18 bounds below the frame are pinned at the floor, and 38 uninformative bounds above it are not drawn).
Large labelled markers, coloured by window, show our inference for the three literature metal-poor candidates against their spectroscopic ratios, with leftward arrows for spectroscopic upper limits and downward arrows for the inferred $2\sigma$ upper limits.
Open rings and diamonds mark the photometric little red dot and red flags, and small red dots the spectroscopically identified little red dots, discussed in Section~\ref{sec:recoverylrd}.}
\label{fig:recovery}
\end{figure*}

\subsection{Validation against mocks and against spectroscopy}
\label{sec:recoveryval}

We validate the photometric emission line inference in two independent tests, shown together in Fig.~\ref{fig:recovery}.

The first test is 30\,000 mock sources with synthetic noise across the four JWST wide-band windows, 24\,000 with \oiii\,$>0$ and 6\,000 with \oiii\,$=0$ for the test of upper limits.
Each source is drawn with a continuum-band signal-to-noise ratio log-uniform over $2$--$200$, a rest-frame EW$_0$(\ha) log-uniform over $300$--$3000$\,\AA, and a true \oiii/\hb log-uniform over $0.01$--$30$, a range chosen to sample the weak-\oiii regime densely.
The attenuation takes the values $A_V=0$ and $0.5$ with equal probability and is supplied to the fit as the known screen, whereas the true continuum slope is drawn uniformly over $-2.5$ to $-1.5$ while the fit assumes $\beta=-2$, so the test includes the fixed-slope systematic.
Where \oiii is well detected, at true \oiii/\hb\,$\gtrsim1$, the three-band recovery is unbiased with a scatter of 0.07\,dex and fewer than 1 per cent catastrophic outliers.
Refitting with a full band set with all possible NIRCam medium bands, in which $\beta$ is measured rather than assumed, reduces the scatter to 0.04\,dex, so the residual three-band scatter is dominated by the fixed-slope assumption.
At lower true ratios the sampler returns marginal \oiii as upper limits rather than as noisy detections, and the sources that remain classified as detections retain a pooled scatter of 0.08\,dex.
No misclassification occurs in the opposite direction: none of the 6926 sources with true \oiii/\hb\,$\geq3$ receives a 97.5th-percentile limit below 1.
The depth of the limits follows the continuum-band signal-to-noise ratio: among the \oiii-free sources, the $2\sigma$ bound of \oiii/\hb falls below 1 for 20 per cent of sources at S/N\,$<20$, 70 per cent at S/N\,$=20$--$50$, and 79 per cent at S/N\,$>50$, and below 0.3 for 46 per cent at S/N\,$>50$.
This mock test verifies the two properties that we will adopt for candidate ranking in Section~\ref{sec:ranking}: where \oiii is detected, the ratio is measured accurately, and where it is not, the inference returns an upper bound set by the photometric depth rather than a spurious \oiii-weak classification.

The second test uses real measurements.
We assemble 490 JADES sources with both DR4 grating spectroscopy, including the DR4 DarkHorse DR1 sibling release in GOODS-S, and DR5 photometry in all three selection bands \citep{Curtis-Lake2026MNRAS.549ag836C,Scholtz2026MNRAS.549ag939S,Johnson2026arXiv260115954J,Robertson2026arXiv260115956R}, after removing two DarkHorse sources (DarkHorse 40609 and 642456) whose NIRSpec spectra are contaminated by overlap from neighbouring traces.
The three-band inference classifies 300 of these sources as detections, and for them the photometric ratio agrees with the spectroscopic one with a bias of $-0.04$\,dex and a scatter of 0.12\,dex.
Refitting the same 300 sources with the slope free and the full DR5 band set gives a scatter of 0.11\,dex, and the two fits agree to 0.02\,dex per source, so the fixed-slope assumption costs about 0.02\,dex on real data, consistent with the mock estimate.
The cost is small because every source in this comparison is a relatively strong \oiii emitter, with spectroscopic \oiii/\hb\,$\geq1.7$, so the band excesses are dominated by the lines and the assumed slope enters only weakly.
The spectroscopic comparison therefore tests the strong-\oiii regime on real measurements, and the weak-\oiii regime is covered by the mock.
This real-data test shows that the three selection bands alone recover the spectroscopic \oiii/\hb to 0.12\,dex, essentially as well as a full multi-band fit.

We use three metal-poor candidates from the literature with spectroscopic constraints to anchor the weak-\oiii regime on real data; they are plotted as large markers in Fig.~\ref{fig:recovery}(b).
For CR3, a metal-free candidate at $z=3.19$ with a spectroscopic \oiii/\hb\,$<0.89$ \citep{Cai2025ApJ...993L..52C}, the \oiii amplitude is undetected and the three-band inference gives a $2\sigma$ upper limit of \oiii/\hb\,$<0.78$ (Fig.~\ref{fig:recoveryapp}), tighter than the spectroscopic bound.
Deeper NIRSpec/MSA observations with higher spectral resolution from JWST GO-9496 (PI: M.~Li) will test this limit.
For AMORE6, at $z=5.73$ with a spectroscopic \oiii/\hb\,$<0.33$ \citep{Morishita2025arXiv250710521M}, the same procedure applied to the published photometry gives \oiii/\hb\,$<2.52$ at $2\sigma$: the limit is weak because the continuum signal-to-noise ratio is low, but it is consistent with the metal-poor classification, and our own deblended photometry of the two lensed images gives a substantially tighter constraint (Section~\ref{sec:a2744cand}).
For the Pop~III candidate GLIMPSE-16043 \citep{Fujimoto2025ApJ...989...46F}, for which NIRSpec follow-up later measured \oiii/\hb\,$=1.78\pm0.18$ \citep{Fujimoto2025arXiv251211790F}, we follow the worked example of the released tool and fit the four NIRCam bands of the follow-up photometry, the three selection bands plus F480M, which also covers \ha at this redshift.
The \oiii amplitude then falls below the detection threshold of equation~(\ref{eq:detection}) and the inference returns \oiii/\hb\,$<2.08$ at $2\sigma$ (posterior median $0.88$), consistent with the spectroscopic value, despite a continuum band detected at only S/N\,$\approx0.8$.

\subsection{The failure mode of the inference}
\label{sec:recoverylrd}

We note the three-band emission-line inference has catastrophic failures for red-continuum sources, and these failures share a single mechanism.
A continuum that is red across the Balmer break violates the power-law assumption of equation~(\ref{eq:inference}), and the fit absorbs the mismatch by suppressing the \oiii amplitude, so a spectroscopically \oiii-strong source is inferred as \oiii-weak, either a low detection or a confidently low bound.
The mock isolates the same signature: of the $34$ sources in $30\,000$ whose true \oiii/\hb\,$\geq1$ is inferred as zero, $97$ per cent carry $A_V>0.3$.
We therefore flag the vulnerable population directly from the photometry, with the two-tier colour flag of Section~\ref{sec:lrdveto}: the line-free colour of equation~(\ref{eq:cred}) raises a \emph{red flag} at $C_{\rm red}>0.7$, and a red flag joined by a blue rest-ultraviolet slope is the v-shaped LRD warning, while a red flag alone marks a typical dusty or evolved galaxy.
In Fig.~\ref{fig:recovery}(b), the open red rings mark the LRD flag and the open diamonds the remaining red flags, drawn wherever the photometry makes an \oiii-weak claim, a detection, or a $2\sigma$ bound below 1.
The flags account for the failures almost exactly.
Every detection with inferred \oiii/\hb\,$<1$ carries the LRD flag, $21$ of the $24$ are spectroscopically \oiii-strong ($R3>1.5$) sources bounded as \oiii-weak, and all four diamond-marked red flags are dusty W$z$3.1 galaxies whose bounds collapse to zero; the LRD flags include two spectroscopically confirmed LRDs of \citet{Geris2026arXiv260621614G} in the validation sample, \oiii-strong at \oiii/\hb\,$\approx5$ yet bounded at $<0.45$ and $<0.77$.
A third spectroscopic LRD, DarkHorse\,642396 with broad \ha (priv.\ comm.), cannot be photometrically flagged at all: at $z=6.554$ no line-free rest-optical band exists in its coverage, so its low bound rests on a continuum colour the data cannot check, a caution that applies generally at the top of W$z$6.1 without the reddest medium bands.
This shows that the red flag is carried even though the colour cut already removes most LRDs, and it matters most in the blueward-continuum W$z$6.1 window, where the colour cut alone cannot reject them.

\subsection{Candidate ranking for spectroscopic follow-up}
\label{sec:ranking}

Within the gate-passing sample of Section~\ref{sec:score}, the three-band inference described above sets spectroscopic priority by returning, for each candidate, a measured \oiii/\hb or its $2\sigma$ upper limit from the selection photometry alone, so the ranking is homogeneous across the sample by construction.
Candidates can be ranked in four classes, with \oiii counted as detected or undetected by the quantile test of equation~(\ref{eq:detection}), ordered within each class by the bound or the measured value:
\begin{enumerate}
\item \oiii-undetected candidates with a $2\sigma$ bound \oiii/\hb\,$<1$ (the deepest limits, the most metallicity-constraining outcome broad-band photometry can deliver);
\item \oiii-detected candidates with \oiii/\hb\,$<1$;
\item \oiii-undetected candidates whose $2\sigma$ bound exceeds 1 (consistent with zero but weakly constrained);
\item \oiii-detected candidates with \oiii/\hb\,$\geq1$.
\end{enumerate}
We note that the ranking only orders the queue for spectroscopic follow-up; it does not reject any source.
This restraint reflects the interpretation ceiling of broad-band data: photometry cannot distinguish $Z=0$ from $Z\lesssim10^{-2}\,Z_{\sun}$, so even the deepest bound is not a direct metallicity measurement.
We apply the ranking to the following Abell~2744 demonstration field in Section~\ref{sec:a2744}.


\section{Demonstration in the Abell 2744 field}
\label{sec:a2744}

Before turning to survey-scale applications in future works of the LATED series, we demonstrate the complete LATED selection framework on one legacy field whose archival data exercise every stage of the pipeline: the deep VLT/MUSE mosaic of the lensing cluster Abell 2744 with JWST/NIRCam photometry.
We deliberately chose the field as a hard test.
MUSE provides a secure spectroscopic LAE parent sample, the NIRCam imaging is among the deepest available, and the cluster core supplies exactly the bright-neighbour blending, intracluster light, and extreme magnification gradients that a lensed-field application must survive.
Abell 2744 has been observed by extensive JWST programs and explored by many teams worldwide.
The field also contains two spectroscopically established, extremely metal-poor systems that a credible selection should confront: AMORE6 at $z=5.725$ \citep{Morishita2025arXiv250710521M} and LAP2 at $z=4.19$ \citep{Vanzella2026A&A...705L..12V}.

\subsection{Parent LAE sample, source detection, and photometry}
\label{sec:a2744data}

The \lya parent sample is provided by the public MUSE redshift catalogue of \citet{Richard2021A&A...646A..83R}.
We keep sources with a detected \lya emission line ($\mathrm{S/N}\geq3$ and flux $>0$) and a secure redshift ($\mathrm{zconf}\geq2$), which yields 138 \lya emitters at $2.72\leq z\leq6.60$.
Of these, 85 fall inside the adopted windows (23 in W$z$3.1, 45 in W$z$4.4, and 17 in W$z$6.1) and 53 fall in the inter-window redshift gaps.
Lensing magnification does not enter the selection: all criteria are flux ratios, so lensing cancels in the colours and $\mu$ is carried for reference only.

Rather than matching to a public photometric catalogue, whose source-detection strategy and deblending choices can fail exactly where lensed metal-poor systems live, we run source detection and photometry on the NIRCam imaging ourselves.
We refer to \citet{Fu2025ApJ...987..186F} and \citet{Li2026arXiv260411892L} for the details of the NIRCam imaging and its reduction.
For every in-window \lya emitter we define a detection region as the union of a 1-arcsec radius circle region around the \lya peak position and the 3$\sigma$ contour of the continuum-subtracted MUSE \lya narrow-band map (registered to the JWST frame; truncated at 3~arcsec), so that sources anywhere under the \lya emission are screened to avoid missing any counterparts.
Every NIRCam band is background-subtracted with a $0.85$-arcsec running-median surface, which tracks the bright-neighbour and intracluster-light gradients of the cluster core; sources are detected on a noise-normalised, matched-filtered stack of the wide bands, where the per-band noise is estimated with the median absolute deviation.
Peaks above $2.5\sigma$ inside the region are sources, and peaks separated by more than $0.15$~arcsec are treated as distinct (peak-based deblending); if nothing is detected, photometry is forced at the \lya position so that every emitter carries at least upper limits.
Each source is then measured in all bands with forced 50-per-cent encircled-energy apertures (point-source aperture correction of 2), with uncertainties from the blank-aperture scatter of the background-subtracted image, and the gates of Section~\ref{sec:score} are applied verbatim with the standard $2\sigma$ substitution for undetected bands.
We have sufficient bands in this field.
The extended criteria of Section~\ref{sec:selcrit} apply wherever the multi-band LRD diagnostic of Section~\ref{sec:lrdveto} can be measured, and that diagnostic needs a red-side continuum band that only wide coverage supplies.
Abell~2744 carries it, so we adopt the extended W$z$6.1 criteria throughout this section and quote the conservative default alongside wherever the two disagree.
W$z$3.1 and W$z$4.4 carry the default criteria and are unaffected.
The EE50 choice is deliberate for crowded cluster cores: re-measuring the candidates with a fixed $r=0.1$-arcsec aperture (per-band aperture corrections) leaves the long-wavelength colours essentially unchanged but re-admits neighbour flux into the short-wavelength bands (doubling the aperture area at $r_{50}\simeq0.05$~arcsec), and 80-per-cent encircled-energy apertures degrade both the signal-to-noise and the blend isolation enough to lose the faintest candidates; the EE50 apertures maximise point-source signal-to-noise and minimise blended light in every band.
Across the 85 LAE regions this yields 518 screened sources (a median of six per region), each carrying its colour uncertainties and limit flags.
For every LAE, we generate a standard visual-inspection figure combining the MUSE \lya spectrum, the continuum-subtracted \lya narrow-band contours registered onto a NIRCam colour image with every detected source marked, and the position in the cluster field.

\begin{figure*}
\includegraphics[width=\textwidth]{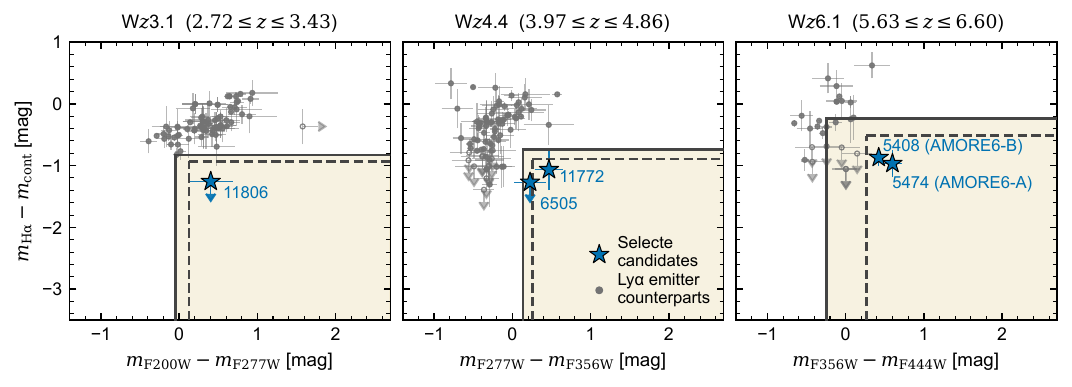}
\caption{The Abell 2744 selection in the three windows (left to right: W$z$3.1, W$z$4.4, W$z$6.1), each panel labelled with its window's bands.
Grey points are the detected sources inside the \lya regions; blue stars are the five candidates, labelled by MUSE ID \citep{Richard2021A&A...646A..83R}.
Open symbols with arrows are $2\sigma$ limits: an undetected \oiii$+$\hb\ band makes $x$ a lower limit (rightward arrow) and an undetected continuum band makes $y$ an upper limit (downward arrow), so the plotted positions are conservative.
The selection corner is drawn solid at the loosest and dashed at the tightest in-window redshift; decisions are evaluated at each source's redshift.}
\label{fig:a2744sel}
\end{figure*}

\begin{figure*}
\includegraphics[width=\textwidth]{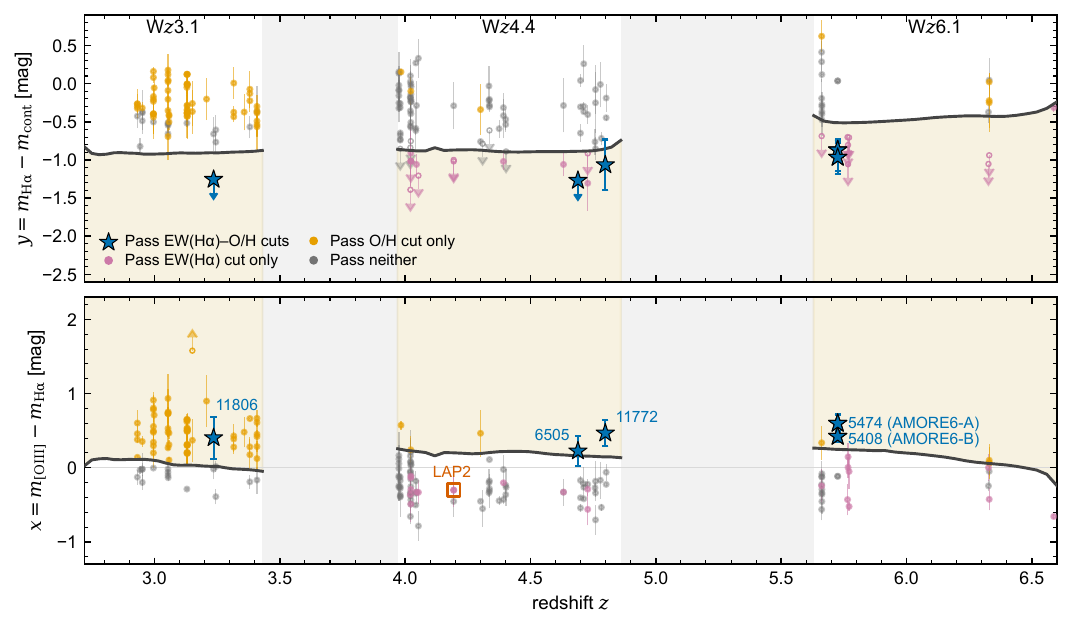}
\caption{The same measurements against the redshift-dependent criteria of Fig.~\ref{fig:boundary}; the two panels share the redshift axis, and light grey bands mark the inter-window gaps where no selection is defined.
Every \ha-detected source is plotted at its MUSE 
\lya redshift.
\emph{Top}: the EW(\ha) cut, with the selection region (shaded) below the $y_{\rm max}(z)$ curve.
\emph{Bottom}: the O/H cut, with the selection region above $x_{\rm min}(z)$.
Arrows are $2\sigma$ limits, always pointing into or along the selection region.
Colours encode the outcome: blue stars pass both cuts (the candidates), pink points pass only the EW(\ha) cut, orange points only the O/H cut, grey points neither.
The annotated LAP2 knot fails the selection, and its undetected continuum lets it pass the EW(\ha) cut only as a conservative limit; the deeper published photometry fails it physically on that cut too (Section~\ref{sec:a2744limit}).}
\label{fig:a2744zcrit}
\end{figure*}

\subsection{Selection and candidates}
\label{sec:a2744cand}

Each source detected inside a \lya region is evaluated against its window's two criteria at the emitter's MUSE redshift: the EW(\ha) cut $y\leq y_{\rm max}(z)$ and the O/H cut $x\geq x_{\rm min}(z)$, with undetected bands entering through the $2\sigma$ substitution.
Figure~\ref{fig:a2744sel} shows the resulting colour-colour planes for the three windows, with each selection corner drawn at its loosest and tightest in-window redshift, and Fig.~\ref{fig:a2744zcrit} shows the same decisions as a function of redshift, each source plotted at its MUSE redshift against the running $x_{\rm min}(z)$ and $y_{\rm max}(z)$ curves.
The 518 screened sources yield five candidates, four physical sources once the two lensed images of AMORE6 are counted together.
A source lying inside more than one \lya region is measured once per region and then merged.
At window W$z$6.1, two lensed images of the known near-pristine system AMORE6 at $z=5.73$ are selected \citep[strong \hb\ with undetected \oiii;][]{Morishita2025arXiv250710521M}.
Each image has its own \lya emitter in the parent catalogue, so we name the candidates by that emitter and note the image: 5474 is AMORE6-A and 5408 is AMORE6-B.
Each identifier therefore belongs to its own image's \lya region rather than to the pair as a whole, and the two are measured, gated, and ranked independently.
Our pipeline detects both.
5474 (AMORE6-A) measures $x=0.60\pm0.11$ and $y=-0.97\pm0.22$ (detection at $35\sigma$, \ha-band $\mathrm{S/N}=16$ in its own region), and 5408 (AMORE6-B) measures $x=0.42\pm0.08$ and $y=-0.87\pm0.12$ ($50\sigma$, \ha-band $\mathrm{S/N}=26$ in its own region).
Both images pass the adopted W$z$6.1 thresholds at this redshift, $x_{\rm min}=+0.248$ and $y_{\rm max}=-0.512$.
Selecting both images also supplies a consistency check that no single-image case can provide: lensing is achromatic, so two images of one source are expected to carry the same colours, and the measured offsets $\Delta x=0.175$ and $\Delta y=0.096$\,mag are consistent within $1.3\sigma$ and $0.4\sigma$ of the combined uncertainties.

The remaining three candidate sources are new, and each enters through a single \lya region.
6505 at $z=4.69$ ($x=0.23\pm0.21$, $y\leq-1.27$ with the continuum band at its $2\sigma$ limit, \ha-band $\mathrm{S/N}=8.9$) and 11772 at $z=4.80$ ($x=0.47\pm0.17$, $y=-1.07\pm0.33$, \ha-band $\mathrm{S/N}=12.5$) are the W$z$4.4 candidates.
We note that for both sources the UNCOVER catalogue \citep{Bezanson2024ApJ...974...92B,Weaver2024ApJS..270....7W} lists the same MUSE redshift for the matched source, which further secures the association independently.
11806 at $z=3.24$ is the W$z$3.1 candidate, and it enters only through our own detection: its nearest counterpart ($0.06$~arcsec) fails the colour cut at $x=-0.31$, but the knot our detection isolates at $0.25$~arcsec passes at $x=0.40\pm0.28$ and $y\leq-1.26$ (detection $9.5\sigma$, \ha-band $\mathrm{S/N}=6.0$).
The full MUSE spectra of these sources all show \lya as the only detected feature.

Finally, the \texttt{lated} emission-line inference of Section~\ref{sec:recovery} converts each candidate's photometry into the homogeneous \oiii/\hb ranking, run on the window's three selection bands with the continuum slope fixed at $\beta=-2$ under the released tool's slope-prior treatment (Table~\ref{tab:a2744rank}).
Every band enters the fit as its measured forced flux with its uncertainty (negative fluxes included); no limit substitution or post-hoc floor modifies the quoted posteriors.
The two AMORE6 images rank first and second.
AMORE6-A is \oiii-undetected with a $2\sigma$ bound $R3\leq0.82$, the deepest class, while AMORE6-B returns a weak \oiii\ detection at $R3=0.57^{+0.25}_{-0.33}$ (97.5th percentile $1.11$), whose 16th percentile of $0.24$ leaves it consistent at $1\sigma$ with the spectroscopic \oiii/\hb\,$<0.33$.
That two independent measurements of one source, along different light paths and from different \lya regions, land in adjacent classes is the ranking's own version of the achromatic consistency check.
The small difference is consistent with the photometric uncertainties and the different measurement regions.
Neither AMORE6 image raises the multi-band LRD warning, and neither is flagged on its UV continuum.
6505 ranks last with a measured $R3=1.13^{+0.89}_{-1.03}$ (97.5th percentile 3.4), and it illustrates the intended division of labour between the gates and the ranking: its colour passes the selection boundary only marginally ($x=0.23\pm0.21$), and because its continuum band is itself a $2\sigma$ limit the inference can trade continuum against line flux, so the \oiii\ deficit is not established.
11772 and 11806 sit between the AMORE6 pair and 6505: both are \oiii-undetected with $2\sigma$ bounds above unity, $R3\leq1.54$ and $R3\leq2.65$, the weakly constrained third class, ordered within the class by their bounds.
Both fits return strong \ha\ with \oiii\ consistent with zero, and the bounds track the depth of the photometry.
The candidates therefore await spectroscopy in the order 5474 (AMORE6-A), 5408 (AMORE6-B), 11772, 11806, 6505.
The rank and results from the \texttt{lated} inference are shown in Table~\ref{tab:a2744rank}.

\begin{table*}
\caption{The LATED selected extremely metal-poor and metal-free candidates in A2744 field ranked by the photometric \oiii/\hb inference on the window's three selection bands.
Limits are quoted at the 97.5th percentile; measurements as the median with the 16th--84th percentile range; the ranking uses the 97.5th percentile of $R3$ throughout.
AMORE6 appears twice, as its two lensed images measured independently from their own \lya regions.
Line fluxes are in $10^{-19}$\,erg\,s$^{-1}$\,cm$^{-2}$ and rest-frame equivalent widths in \AA; \hb is tied to \ha by case~B (2.86) in the model, so its flux and equivalent width are derived rather than independent.
$f_{\rm cont}$ is the fitted continuum flux density at rest-frame 5000\,\AA\ in nJy (median and 16th--84th range of its marginalised posterior).
Coordinates (J2000, decimal degrees) are those of the NIRCam source our detection selects and the forced photometry measures, not the MUSE \lya centroid nor the nearest catalogue match.
All values are observed, uncorrected for magnification, and are EE50 aperture fluxes with a point-source aperture correction: for marginally resolved sources they underestimate the total flux, and no total-flux correction is attempted.}
\label{tab:a2744rank}
\setlength{\tabcolsep}{2.5pt}
\begin{tabular}{lccccccccccc}
\hline
MUSE ID & RA & Dec. & $z$ & $R3$ & $F$(\ha) & $F$(\hb) & $F$(\oiii$\lambda5007$) & EW$_0$(\ha) & EW$_0$(\hb) & EW$_0$(\oiii$\lambda5007$) & $f_{\rm cont}$ \\
\hline
5474 (AMORE6-A) & 3.584699 & $-30.403132$ & 5.73 & $\leq0.82$ & $26.8^{+4.4}_{-4.1}$ & $9.4^{+1.5}_{-1.4}$ & $\leq8.6$ & $2400^{+400}_{-360}$ & $460^{+80}_{-70}$ & $\leq450$ & $10.6^{+1.7}_{-2.0}$ \\
5408 (AMORE6-B) & 3.584390 & $-30.403374$ & 5.73 & $0.57^{+0.25}_{-0.33}$ & $35.8^{+4.0}_{-3.6}$ & $12.5^{+1.4}_{-1.3}$ & $7.1^{+3.5}_{-4.1}$ & $2020^{+230}_{-210}$ & $390^{+40}_{-40}$ & $230^{+120}_{-140}$ & $17.2^{+1.9}_{-2.1}$ \\
11772           & 3.595913 & $-30.386383$ & 4.80 & $\leq1.54$ & $11.6^{+2.7}_{-2.2}$ & $4.1^{+0.9}_{-0.8}$ & $\leq6.3$ & $2470^{+570}_{-470}$ & $470^{+110}_{-90}$ & $\leq790$ & $3.9^{+1.1}_{-1.2}$ \\
11806           & 3.600358 & $-30.386817$ & 3.24 & $\leq2.65$ & $7.9^{+3.3}_{-2.8}$ & $2.8^{+1.2}_{-1.0}$ & $\leq4.9$ & $2400^{+1000}_{-850}$ & $460^{+190}_{-160}$ & $\leq870$ & $1.9^{+0.7}_{-0.8}$ \\
6505            & 3.571375 & $-30.400104$ & 4.69 & $1.13^{+0.89}_{-1.03}$ & $9.6^{+3.4}_{-3.0}$ & $3.4^{+1.2}_{-1.1}$ & $3.9^{+2.9}_{-3.7}$ & $1910^{+680}_{-590}$ & $370^{+130}_{-110}$ & $450^{+340}_{-420}$ & $4.2^{+1.5}_{-1.5}$ \\
\hline
\end{tabular}
\end{table*}

\subsection{A physical limit: the equivalent-width floor}
\label{sec:a2744limit}

LAP2 is not selected, which marks the opposite, equally instructive outcome.
LAP2 is in the parent sample and inside W$z$4.4; the pipeline detects seven sources inside its \lya region, including the source knot itself, yet none passes, and the reason resolves into three well-separated layers once its published measurements are used \citep{Vanzella2026A&A...705L..12V}.
At the catalogue level, the image LAP2-b is a very faint source with $m_{\rm F090W}\simeq30$\,mag and the \ha\ band still reaches only $\mathrm{S/N}\simeq1$--4 depending on the aperture centring, below the detection gate, and the long-wavelength apertures remain contaminated by the blend at any fixed-aperture scale.
The counter-image LAP2-a behaves identically: its UV knot is recovered at $7\sigma$ in F090W at the position of a catalogued UNCOVER source, but its \ha band likewise reaches only $\mathrm{S/N}\simeq2$; since the colours are magnification-invariant, both images sit at the same ideal position outside the selection corner.
These conclusions are reduction-independent: repeating the forced photometry on an independent set of $0.03$~arcsec full-field mosaics reproduces the knot fluxes of both images (the two frames agree astrometrically to $0.02$~arcsec) and improves the \ha-band significance only to $\mathrm{S/N}\simeq2$--3, still below the detection gate.
At the level of ideal, PSF-scale photometry, the published fluxes (\ha\,$=1.68\times10^{-19}$, \hb\,$=0.72\times10^{-19}$, \oiii$\lambda5007<0.5\times10^{-19}$\,erg\,s$^{-1}$\,cm$^{-2}$, $m_{\rm F277W}=30.7$) imply $x\simeq+0.15$ against a boundary of $x_{\rm min}=0.20$ and $y\simeq-0.79$ against $y_{\rm max}=-0.87$: the \oiii\ deficit becomes visible, but the system still sits narrowly outside the LATED selection box.
This is physical rather than observational.
The $y$ cut at this redshift corresponds to rest-frame EW(\ha)\,$\gtrsim1.9\times10^{3}$\,\AA\ given by Pop~III templates, the high-EW regime the mock calibrates for maximally young, high-covering-fraction star formation, whereas LAP2 has EW(\ha)\,$\simeq650$\,\AA\ \citep{Vanzella2026A&A...705L..12V}.
LAP2 therefore presents the equivalent-width floor of our selection: systems this faint and this far below the high-EW regime require spectroscopy to constrain the metallicity.

That floor is a property of the template library, and the grid identifies its origin.
At $z=4.19$ the Pop~III target locus spans rest-frame EW(\ha)\,$=1.2$--$4.8\times10^{3}$\,\AA, so no template in the locus reaches LAP2's $\simeq650$\,\AA, and the lower envelope belongs to an instantaneous burst of age $\simeq20$\,Myr at $f_{\rm cov}=1$.
The high-ionisation models excluded from the locus are not responsible, since the \citet{Nakajima2022MNRAS.513.5134N} grid at $\log U=-0.5$ gives EW(\ha)\,$=2.7$--$2.8\times10^{3}$\,\AA, above the envelope the Yggdrasil tracks already set, and restoring them would leave the floor unchanged.
A reduced covering fraction is the natural remaining candidate, but it does not scale the equivalent width linearly.
Over the 57 model combinations for which the grid holds both covering fractions at fixed IMF, star formation history and age, EW(\ha) at $f_{\rm cov}=0.5$ is a median $1.14$ times its value at $f_{\rm cov}=1$, and falls below unity only beyond an age of $\simeq8$\,Myr.
The Pop~III continuum under \ha\ is itself largely nebular \citep{Schaerer2002A&A...382...28S,Raiter2010A&A...523A..64R}, so halving the covering fraction removes the line and its underlying continuum together, and the linear scaling is approached only once the stellar continuum dominates.
Matching LAP2 therefore requires $f_{\rm cov}$ well below the value a linear estimate implies, in the interval between $f_{\rm cov}=0.5$ and $0$ that the present grid does not sample.
Extending the grid there is straightforward and would relax both cuts on its own, because the boundaries are the envelope of the target locus.
What such an extension costs is not leakage from enriched systems, since relaxing the W$z$4.4 box to LAP2's position leaves every contaminant class unchanged except extremely metal-poor Pop~II, whose pass rate rises from $39.6$ to $41.7$ per cent.
That is the degeneracy the floor marks, since at this equivalent width a pristine population and an extremely metal-poor one occupy the same colours, and it is why we read LAP2 as the limit of what broadband photometry can decide rather than as a template the library merely lacks.

The demonstration in this section establishes what the selection workflow delivers end-to-end.
The verbatim gates run on source detection and forced photometry of real, deep, crowded cluster imaging and return a small, fully vetted candidate list with uncertainties and limits; the \lya-guided regions and deblending recover what public catalogues structurally miss.


\section{Future prospects}
\label{sec:future}

The preceding sections built LATED selection with a quantified selection function at each stage, attached a line-ratio inference validated against mocks and spectroscopy, and ran the whole chain on one archival field.
This section looks forward.
We assemble the completeness budget a future survey application must carry, describe the statistical constraints a candidate sample supports and their limitations, invite the community to apply the released machinery to archival data that already overlap JWST imaging, and close with what spectroscopy follow-up can and cannot confirm.

\subsection{Factors of selection completeness}
\label{sec:selcompleteness}

The completeness of a LATED-selected sample is a product of factors that arise at different stages, and only the colour stage is fixed by the released machinery.
So the numbers below are a budget to be filled in per survey, not a completeness this paper can quote.
We assemble the budget here because any survey application has to carry them.

The first factor is the parent \lya selection.
\lya escape is stochastic and sightline-dependent, the parent surveys have their own selection functions, and beyond $z\approx5.5$ the increasingly neutral intergalactic medium attenuates the line itself \citep{Ouchi2020ARA&A..58..617O}.
This factor is survey-specific and enters as the parent-stage completeness each application must supply (Section~\ref{sec:parent}).

The second factor is the colour stage, the only one the released machinery fixes.
It is also the mildest: the criteria recover $\approx0.99$ of the Pop~III target locus once the photometry reaches S/N\,$\geq5$, in every window (Section~\ref{sec:selfunc}), because the boundaries are the envelope of that locus by construction.
The recovery is governed by the margin between the continuum magnitude and the $5\sigma$ depth rather than by the depth itself, and the per-redshift boundaries absorb the band-edge effects.
Combined with a fiducial parent stage the end-to-end probability falls to $0.83$--$0.88$ in the four lower-redshift windows and $0.28$ in W$z$6.1, but that figure inherits the parent survey it assumes and should be recomputed, not adopted.
The criteria variant adds a measured cost of conservatism: under the default W$z$6.1 criteria the better measured AMORE6 image misses both cuts, by $0.009$ and $0.061$\,mag, and only the extended criteria retain it, so which variant a field earns, set by its band inventory, changes the completeness a survey should adopt.

The third factor is the `duty cycle'.
The selection reaches only the nebular-dominated phase, roughly $3$\,Myr of the $\lesssim$15--20\,Myr Pop~III-dominated window \citep{Rusta2025ApJ...989L..32R}, and only above the equivalent-width floor the cut demands.
A measured number density is therefore a density of \emph{phases}, and converting it to a rate requires dividing by a duty cycle that the models, not the data, supply.

The fourth factor is the Pop~III population itself.
Every recovery fraction in this paper averages over our model grid with uniform weights in IMF, covering fraction, age and ionisation parameter.
The real population is not distributed that way, and nothing observed constrains how it is distributed, so a recovery fraction cannot be inverted into a completeness without a prior the data do not supply.
This is the factor we can least bound, and further work is needed.

The fifth factor is measurement.
Catalogue photometry could fail preferentially for the compact, high-equivalent-width (faint continuum) knots we targets.
They are also the sources most likely to be blended with nearby objects.
For example, both AMORE6 images are absent from the public UNCOVER catalogues (Section~\ref{sec:a2744}).
Colours of resolved sources could also shift by several tenths of a magnitude between aperture systems, depending on aperture sizes and PSF matching algorithm.

These factors also vary across the windows.
For example, at matched depth, distance dimming alone makes the reference burst a factor $2.4$ harder to detect at the W$z$4.4 midpoint and $5.1$ at the W$z$6.1 midpoint than at W$z$3.1, raising the mass floor by $+0.37$ and $+0.70$\,dex.
The practical consequence is that every application must supply all these terms itself.
We do not multiply them here because three of the five are survey-specific and one is unconstrained, and quoting their product as a completeness would imply a precision that does not exist.

\subsection{From candidates to constraints}
\label{sec:statistics}

A candidate sample with a quantified selection function supports statistical analyses that single discoveries cannot.
Examples are the luminosity functions of extremely metal-poor candidates per slice, their number density relative to the parent \lya emitters, and their clustering and environments.
The large spectroscopic samples of high-$z$ galaxies sharpen what such statistics measure: no JADES spec-$z$ source passes the redshift-dependent criteria at its measured redshift, so the target corner is empty of every observed population (see Fig.~\ref{fig:jadesplanes}), and a candidate density there probes a regime spectroscopy rarely reaches.
The A2744 demonstration sets the working scale: five candidates from 85 in-window emitters in one of the most deeply observed lensing clusters, over the $\sim4$\,arcmin$^2$ MUSE mosaic (Table~\ref{tab:a2744rank}).

The statistical analysis itself is standard and can be applied per slice.
The effective volume of a window is the geometric volume of its \lya slices weighted by the factors of Section~\ref{sec:selcompleteness}, evaluated as a function of source magnitude, and a candidate count over that volume becomes a number density with small-number confidence intervals \citep{Gehrels1986ApJ...303..336G}.
Supplying those weights is beyond this paper: it needs the survey's own parent selection function and a prior on the Pop~III population, so we set out the machinery here and leave the weighting to a future survey application.
A null result carries the same weight: zero candidates over an effective volume $V_{\rm eff}$ bound the density at $n<3.0/V_{\rm eff}$ at 95 per cent confidence.
Because the parent emitters are counted in the same slices, the candidate density can also be quoted as a fraction of the \lya-emitter population, which cancels the volume and much of the parent-stage completeness.
Stacking the imaging of candidates ranked by the inference can test population-mean diagnostics that are too faint to measure individually.
The environments of those candidates test the pristine-pocket picture directly, since the simulations place late Pop III in under-dense, late-assembling regions.

\subsection{Call for community surveys}
\label{sec:communitycall}

The archival \lya data that can feed LATED already exceed what any single team will process.
For example, HETDEX is measuring \lya redshifts for millions of emitters over $540$\,deg$^2$ without preselection and is still observing \citep{Gebhardt2021ApJ...923..217G}.
The MUSE archive holds deep integral-field mosaics of the blank fields \citep[e.g.,][]{Bacon2017A&A...608A...1B,Urrutia2019A&A...624A.141U,Bacon2023A&A...670A...4B} and of tens of lensing clusters \citep[e.g.,][]{Richard2021A&A...646A..83R}.
Narrow- and intermediate-band surveys supply wide-area parents at discrete slices for decades, including SILVERRUSH \citep{Ouchi2018PASJ...70S..13O,Shibuya2018PASJ...70S..14S}, SC4K \citep{Sobral2018MNRAS.476.4725S}, and ODIN \citep{Lee2024ApJ...962...36L}.
A growing fraction of this \lya sky already carries the JWST imaging the colour stage needs: HETDEX overlaps GOODS-N, the EGS, and COSMOS; the MUSE deep fields sit inside GOODS-S; and the cluster mosaics overlap many JWST lensing surveys including GLASS, PEARLS, UNCOVER, CANUCS, MAGNIF, GLIMPSE, VENUS, and more in the future.
The expected density of genuine metal-poor systems is low, and the predicted yield per deep field spans orders of magnitude because the Pop~III star-formation-rate density and cluster mass function are uncertain by that much, so the search must cover all available volume before spectroscopic time is spent.
That is more sky than one group can screen.
We make all necessary material public: the windows and their per-redshift boundaries as machine-readable tables, the gates and flags of Table~\ref{tab:score}, and the line inference as the installable \texttt{lated} package.
We invite teams holding \lya catalogues with JWST-covered footprints to run the gates, apply the flags, and rank their candidates with the released inference.
Candidates found this way can then be observed by dedicated follow-up programmes or as fillers in other community surveys.

\subsection{From candidates to confirmation}
\label{sec:followup}

Follow-up proceeds down the ranking in Section~\ref{sec:ranking}, and its first task is to remeasure the ratio the inference bounded.
For CR3, the photometric $2\sigma$ limit is already tighter than the published spectroscopic bound, and JWST GO-9496 will test it (Section~\ref{sec:recoveryval}); the A2744 queue of Table~\ref{tab:a2744rank} awaits the same measurement.
Deep rest-ultraviolet spectroscopy can add the ionising-spectrum diagnostics, \heii\,$\lambda1640$ together with C~{\sc iii}], O~{\sc iii}], and C~{\sc iv}, and a clear narrow \heii\ detection with all metal ultraviolet lines undetected would constitute the strongest pre-ELT evidence for a chemically primitive system.
Depending on redshift, these lines fall to ground-based optical spectrographs or to NIRSpec, and the \lya\ profile itself, often already in hand from the parent survey or further spectroscopic observations, constrains the radiative transfer and rejects low-redshift interlopers.
We regard a candidate as a spectroscopically confirmed extremely metal-poor system when strong \lya\ and Balmer emission share one redshift, metal-line limits translate to $12+\log({\rm O/H})\lesssim7$, no convincing AGN indicator survives, and the spectrum is consistent with extremely metal-poor or Pop~III models.
A stronger Pop~III designation also requires \oiii/\hb\,$<0.1$, ideally with nebular \heii\ detected.

Even that measurement establishes a candidate, not a proof.
Theory defines Pop~III by formation below the critical metallicity, $Z_{\rm crit}\lesssim10^{-3.5}\,Z_{\sun}$, whereas observation constrains metallicity only from above.
The deepest currently achievable gas-phase metallicity limits on faint high-redshift systems could reach $\sim0.1$ per cent solar, but would still sit above the critical-metallicity range.
A spectrum can therefore establish that no metal line is detected down to some bound; it cannot establish that the gas is metal-free.
Nor is nebular \heii\ with undetected metal lines decisive on its own, because models produce the same signature without metal-free stars, through very massive stars at low but non-zero metallicity, stripped binary products, X-ray binaries, and shocks \citep{Grafener2015A&A...578L...2G,Schaerer2019A&A...622L..10S,Plat2019MNRAS.490..978P}.
The designation Pop~III is consequently asymptotic.
Observations can approach it through deeper abundance limits and harder ionising spectra, but the final step from no metal lines detected to metal-free is taken by models on both sides of the comparison.
What observation can deliver, and what this framework is built to deliver, is the census of star formation approaching the critical-metallicity regime: a measured number density, or a defensible bound, for \ha-luminous systems with $12+\log({\rm O/H})\lesssim7$ from the end of reionisation to cosmic noon.
A sample of such systems at $z\approx2$--6 with a measured density would move the late-time Pop~III question from speculative to quantitative, whatever name the individual objects finally earn.

\section{Summary}
\label{sec:summary}

We present LATED, a photometric framework for selecting metal-free and extremely metal-poor candidate systems from \lya-emitter samples, along with the forward model, selection function, and line-ratio inference needed to use it quantitatively.
Our main results are as follows.

\begin{enumerate}

\item \emph{The signature is strong hydrogen emissions with weak or absent oxygen emissions, measured in three bands.}
LATED starts from a \lya-selected parent sample with a known redshift slice and requires only that \ha be strong relative to the continuum and that \oiii be weak relative to \ha.
The two colours $y=m_{{\rm H}\alpha}-m_{\rm cont}$ and $x=m_{\rm OIII}-m_{{\rm H}\alpha}$ track the \ha equivalent width and the oxygen abundance respectively, and the EW(\ha)--O/H diagram they span is the whole diagnostic.

\item \emph{The usable redshift range is derived as selection windows.}
Requiring that real filters hold \oiii$+$\hb and \ha simultaneously, and that a \lya parent route exists, yields five windows spanning $z=1.92$--6.60, the upper edge set by \ha leaving JWST/NIRCam.
Four are JWST/NIRCam windows, and one is an all-\textit{Roman} window.
Where a continuum band exists redward of \ha (W$z$2.1, W$z$3.1, W$z$4.4) the selection is at its cleanest; W$z$2.2 and W$z$6.1 lack that band and are exploratory.

\item \emph{The criteria are read off the model envelope and come with a selection function.}
At each redshift the two thresholds sit at the extremes of the noiseless Pop~III target locus, including the nebular continuum, and the boundaries are released as machine-readable tables per window\footnote{\url{https://lated.pop3star.com}}.
Because the boundary is that envelope, the colour stage recovers 99 per cent of the locus in every window once the photometry reaches S/N\,$\geq5$, and the criteria are therefore not what limits a search.
Populations with zero covering fraction emit no recombination lines and are unselectable by construction, so every recovery fraction is conditional on $f_{\rm cov}>0$ and is quoted that way throughout.
These are recovery fractions of the model library, not population completeness: the relative abundances of the classes and the distribution of Pop~III properties are both unknown, so we quote no contamination fraction and no yield, and leave that step to a future survey application.
The target corner contains no observed population: no source in the JADES spectroscopic sample passes the redshift-dependent criteria in any of the four NIRCam windows.

\item \emph{Contaminants including little red dots are flagged.}
A red rest-optical continuum mimics an \ha excess, and a blue rest-ultraviolet continuum satisfies the slope flag, so the population attacks the selection from both sides.
The colour stage removes it wherever a red-side continuum band exists, and the young-locus boundary does so in the two windows that do not use a red-side continuum band.
The residual is annotated by a two-tier flag on the line-free ultraviolet-to-optical colour $C_{\rm red}$, measured only on bands free of strong lines so that line boosting cannot fake it: the colour alone raises a red flag, marking ordinary dusty galaxies, and joined by a blue ultraviolet slope it becomes the LRD warning.
Applied to the JADES validation sample, these flags account for the catastrophic failures of the photometric inference almost exactly.
Pristine black holes, by contrast, are rejected outright: their accretion continuum suppresses the \ha equivalent width, and none of the PBH models enter the selection in any window.

\item \emph{The same three bands yield a line ratio of R3.}
A forward fit, with a power-law continuum and non-negative line amplitudes carrying the full Case~B Balmer series and the \oiii\ doublet, converts the selection photometry into \oiii/\hb with an explicit status.
At the metal-poor end the ratio reads directly as a metallicity, $Z/Z_{\sun}\approx R3$ per cent under the \citet{Isobe2026arXiv260611345I} calibration.
On mocks, it is unbiased with 0.07\,dex scatter where \oiii is well detected, and it turns marginal \oiii into upper limits rather than noisy detections.
Against JADES DR4 spectroscopy of the same DR5 sources, it reproduces the spectroscopic ratio with $-0.04$\,dex bias and 0.12\,dex scatter.
The method is released as an installable Python package \texttt{lated} with a web application.

\item \emph{The \ha line band sets the required survey depth.}
A $10^6\,{\rm M}_{\sun}$, 2-Myr burst at $z=3$ requires $m=27.2$ in the \ha band but $28.6$ in the continuum beside it, and with NIRCam that difference is an order of magnitude in exposure time, 2\,min against 21\,min.
The gap widens with redshift, reaching 41\,min against 8.6\,h in W$z$6.1.
Selection itself needs only the \ha detection, because an undetected continuum enters the EW(\ha) cut as a limit.
The continuum depth is what a clean colour costs, and so what the line-ratio inference requires to achieve better constraints.
The \lya parent stage is never the bottleneck: at the fiducial escape fraction of $0.2$ the required narrow-band depth is already reached by existing imaging, and MUSE reaches the $10^{6}\,{\rm M}_{\sun}$ burst within minutes to about an hour in the windows it covers.

\item \emph{The workflow runs end to end on real archival data.}
Applied verbatim to the Abell~2744 field, with custom detection and forced photometry, LATED recovers both lensed images of a spectroscopically confirmed metal-poor emitter AMORE6, and reveals three more candidates with uncertainties and limits.
We also identify the one class the method cannot reach on current imaging suggested by LAP2: the equivalent-width floor below which a blue continuum dilutes every band excess.

\item \emph{The output is a ranked target list.}
LATED delivers candidates for spectroscopy, and confirmation requires deep metal-line limits, exclusion of an active nucleus, and consistency with extremely metal-poor model spectra.
Each candidate is an extremely metal-poor first, and every Pop~III statement a limit, because observation bounds metallicity only from above.
Because each stage carries a quantified selection function, a null result over a known volume constrains the number density of \lya-visible, nebular-bright, oxygen-weak domains, and hence the mixing efficiency and critical metallicity that govern the late-time Pop~III tail, once the survey supplies its own parent selection function.

\end{enumerate}

LATED turns \lya surveys and photometric imaging over them into a framework for finding the most chemically primitive star formation at $z=1.92$--6.60.
The parent side draws on two decades of archival \lya data and keeps growing as surveys continue observing and as new ones are planned; the imaging side is JWST today, with \textit{Roman} on the way.
We have released every stage of the framework: the windows, the per-redshift boundaries, the gates and flags, the per-class selection probabilities, and the line inference tool.
The archival data already in hand cover more sky than any one group can screen, and the expected candidates are rare, so we invite the community to run the selection wherever a \lya catalogue meets JWST imaging.
Whether the outcome is a population or a bound, the result is the same kind of number: a measured density of star formation approaching the critical-metallicity regime, at the epoch where the survival of the late Pop~III tail is most in question.

\section*{Acknowledgements}
ZC acknowledges support from National Key R\&D
Program of China (grant No. 2023YFA1605600), National Natural Science Foundation of China (\#12525303), Tsinghua University Initiative Scientific Research Program, and New Cornerstone Science Foundation through the XPLORER PRIZE. 
RM acknowledges support from the Science and Technology Facilities Council (STFC), the European Research Council (ERC) through Advanced Grant 695671 ``QUENCH'', and the UK Research and Innovation (UKRI) Frontier Research grant RISEandFALL.
RM also acknowledges support from a Royal Society Research Professorship grant. 
BL acknowledges financial support from the German Excellence Strategy via the Heidelberg Cluster of Excellence (EXC 2181 -- 390900948) STRUCTURES. 
H\"U acknowledges support by the Max Planck Society through the Lise Meitner Excellence Program. H\"U acknowledges funding by the European Union (ERC APEX, 101164796). Views and opinions expressed are however those of the authors only and do not necessarily reflect those of the European Union or the European Research Council Executive Agency. Neither the European Union nor the granting authority can be held responsible for them.
This work is based in part on observations made with the NASA/ESA/CSA James Webb Space Telescope. The data were obtained from the Mikulski Archive for Space Telescopes at the Space Telescope Science Institute, which is operated by the Association of Universities for Research in Astronomy, Inc., under NASA contract NAS 5-03127 for JWST. These observations are associated with program \#1324, \#2561, \#2756, \#2883, \#3516, \#3538 and \#4111.
The authors acknowledge the teams of JWST programs for developing their observing program with a zero-exclusive-access period.
Based on observations collected at the European Organisation for Astronomical Research in the Southern Hemisphere under ESO programme 094.A-0115.
Generative AI (OpenAI ChatGPT and Anthropic Claude) assisted with draft language editing, coding, and literature searching. The authors reviewed and approved all scientific content, interpretations, references, and final wording and take full responsibility for the manuscript.

\section*{Data availability}
\label{sec:data}
The redshift-dependent LATED selection criteria for the five adopted windows are released with this paper and can be downloaded from the project website, \url{https://lated.pop3star.com}.
The same site hosts the \textsc{Lated Explorer}, which reproduces these tables and derives the same criteria for any other band combination.
The line-inference tool is publicly released on PyPI as a standalone package with an interactive web application (\texttt{lated}).
No new observational data were generated for this methodology paper.

\bibliographystyle{mnras}
\bibliography{main}

\vspace{4mm}
\printaffils

\bsp
\label{lastpage}
\end{document}